\documentclass[notitlepage,onecolumn,10pt,nofootinbib]{revtex4-1}
\usepackage[english]{babel}
\usepackage[latin1]{inputenc}
\usepackage{amsmath}
\usepackage{amssymb}
\usepackage{bbm}
\usepackage{graphics}
\begin{document}
\title{Quantum $D = 3$ Euclidean and Poincar\'{e} symmetries from contraction limits}

\author{Jerzy Kowalski-Glikman}
\affiliation{Institute for Theoretical Physics, University of Wroc\l{}aw, pl.\ M.\ Borna 9, 50-204 Wroc\l{}aw, Poland}
\affiliation{National Centre for Nuclear Research, ul.\ Pasteura 7, 02-093 Warsaw, Poland}

\author{Jerzy Lukierski}
\affiliation{Institute for Theoretical Physics, University of Wroc\l{}aw, pl.\ M.\ Borna 9, 50-204 Wroc\l{}aw, Poland}
\affiliation{Bogolubov Laboratory of Theoretical Physics, Joint Institute for Nuclear Research, 141980 Dubna (Moscow Region), Russia}

\author{Tomasz Trze\'{s}niewski}
\affiliation{Institute of Theoretical Physics, Jagiellonian University, ul.\ S.\ {\L}ojasiewicza 11, 30-348 Krak\'{o}w, Poland}
%\email{tbwbt@ift.uni.wroc.pl}

\date{\today}

\begin{abstract}
Following the recently obtained complete classification of quantum-deformed $\mathfrak{o}(4)$, $\mathfrak{o}(3,1)$ and $\mathfrak{o}(2,2)$ algebras, characterized by classical $r$-matrices, we study their inhomogeneous $D = 3$ quantum IW contractions (i.e. the limit of vanishing cosmological constant), with Euclidean or Lorentzian signature. Subsequently, we compare our results with the complete list of $D = 3$ inhomogeneous Euclidean and $D = 3$ Poincar\'{e} quantum deformations obtained by P.~Stachura. It turns out that the IW contractions allow us to recover all Stachura deformations. We further discuss the applicability of our results in the models of 3D quantum gravity in the Chern-Simons formulation (both with and without the cosmological constant), where it is known that the relevant quantum deformations should satisfy the Fock-Rosly conditions. The latter deformations in part of the cases are associated with the Drinfeld double structures, which also have been recently investigated in detail.
\end{abstract}

\maketitle

\section{Introduction} \label{sec:1.0}
The emergence of non-commutative quantum spacetime at ultra-short distances, comparable with the Planck length $\lambda_P \approx 10^{-35} m$, is a recurring theme, constantly re-surfacing during decades of efforts to construct quantum gravity models (see e.g. \cite{Majid:1988he,Doplicher:1995ts,Ashtekar:2004bt,Thiemann:2007my,Oriti:2009ay}). There are also good reasons to believe that a direct consequence of the quantization of gravity is a necessity to replace the classical symmetries with their quantum counterparts, where the quantum versions of Lie algebras and groups are given by the algebraic groups having the structure of Hopf algebras \cite{Majid:1995fy}. It was realized in late 1980s and 1990s that of particular importance in such a context are the studies and classification of quantum spacetime symmetries, described in the language of Hopf algebras, especially the quantum Poincar\'{e} algebras and quantum Poincar\'{e} groups, as well as quantum versions of the (anti-)de Sitter and conformal symmetries.

It has also been known since the 1980s \cite{Drinfeld:1983ts,Majid:1995fy} that the quantum deformations of symmetries can be characterized by the classical $r$-matrices, which are solutions of the classical Yang-Baxter equation (CYBE). A classical $r$-matrix $r$ is linear in the deformation parameters $q_i$ and determines the coboundary Lie bialgebra structure of the algebra $\mathfrak{g}$, with the coproduct of algebra elements $g \in \mathfrak{g}$ given by the perturbative formula
\begin{align}\label{eq:10.00}
\Delta(g;q_i) = \Delta_0(g) + [r,\Delta_0(g)] + {\cal O}(q_i^2)\,,
\end{align}
where $\Delta_0$ denotes the primitive (i.e. undeformed) coproduct and the second term provides the Lie cobracket of $g$. Such an expansion leads to the quantization of Poisson-Lie structure and emergence of the corresponding quantum-deformed Hopf algebra. The procedure of generating associative and coassociative quantum Hopf algebras from classical $r$-matrices is called the quantization of bialgebras \cite{Etingof:1996qi}. If a bialgebra is coboundary, then there is a one-to-one correspondence between $r$-matrices and possible Hopf algebraic structures describing quantum symmetries.

To be more precise, the (classical, antisymmetric) $r$-matrices of a given Lie algebra $\mathfrak{g}$ are skew-symmetric elements $r \in \mathfrak{g} \wedge \mathfrak{g}$ that satisfy the classical (in general, inhomogeneous or modified if $t \neq 0$) Yang-Baxter equation \cite{Drinfeld:1983ts,Majid:1995fy}
\begin{align}\label{eq:10.01}
[[r,r]] = t\, \Omega\,, \quad t \in \mathbbm{C}\,,
\end{align}
with $[[.]]$ denoting the Schouten bracket
\begin{align}\label{eq:10.02}
[[r,r]] := [r_{12},r_{13} + r_{23}] + [r_{13},r_{23}]\,,
\end{align}
where $r_{12} \equiv r^{(1)} \otimes r^{(2)} \otimes \mathbbm{1}$ etc. (in Sweedler notation $r^{(1)} \otimes r^{(2)} \equiv \sum_i r^{(1)}_i \otimes r^{(2)}_i$), and $\Omega$ is a $\mathfrak{g}$-invariant three-form:
\begin{align}\label{eq:10.03}
\forall g \in  \mathfrak{g}: {\rm ad}_g \Omega = [g \otimes \mathbbm{1} \otimes \mathbbm{1} + \mathbbm{1} \otimes g \otimes \mathbbm{1} + \mathbbm{1} \otimes \mathbbm{1} \otimes g,\Omega] = 0\,.
\end{align}
If $t = 0$, the Yang-Baxter equation (\ref{eq:10.01}) is homogeneous and describes the twist quantizations of algebra $\mathfrak{g}$, while its solutions are called the triangular $r$-matrices; if $t \neq 0$, the Yang-Baxter equation is inhomogeneous or modified, and its solutions are known as the quasitriangular $r$-matrices.

For semisimple Lie algebras, e.g. complex orthogonal algebras $\mathfrak{o}(n;\mathbbm{C})$ and their real forms, all bialgebras are coboundary and therefore the classification of classical $r$-matrices completes the task of classifying quantum deformations. In the case of $D = 4$ spacetime symmetries, all deformations of the Lorentz algebra have been described using four multiparameter $r$-matrices more than twenty years ago \cite{Zakrzewski:1994pp}. Recently in a similar fashion, by finding all classical $r$-matrices, all deformations of $\mathfrak{o}(4;\mathbbm{C})$ and of its real forms -- the Euclidean $\mathfrak{o}(4)$, Lorentzian $\mathfrak{o}(3,1)$, Kleinian $\mathfrak{o}(2,2)$ and quaternionic $\mathfrak{o}^*(4;\mathbbm{C}) \cong \mathfrak{o}(2;\mathbbm{H})$ -- were classified \cite{Borowiec:2017bs}. On the other hand, the complete classification of all deformations of the $D = 4$ Poincar\'{e} algebra (at the Poisson structure level) has not been obtained so far \cite{Zakrzewski:1997pp}. However, the complete classification of deformations for $D = 3$ inhomogeneous Euclidean and $D = 3$ Poincar\'{e} algebras has been accomplished in \cite{Stachura:1998ps}.

In what follows we denote by $\mathfrak{o}(D+1)$ the $D$-dimensional Euclidean de Sitter algebra and by $\mathfrak{o}(D-1,2)$ the $D$-dimensional Lorentzian anti-de Sitter algebra, while we regard $\mathfrak{o}(D,1)$ either as the $D$-dimensional Lorentzian de Sitter algebra or $D$-dimensional Euclidean anti-de Sitter algebra\footnote{$\mathfrak{o}(D,1)$ describes as well the $D+1$-dimensional Lorentz algebra and $\mathfrak{o}(D-1,2)$ is also the $D-1$-dimensional conformal algebra.}. For $D = 3$, all of these real rotation algebras are described by the (pseudo-)orthogonal real forms of the complex algebra $\mathfrak{o}(4;\mathbbm{C})$. The main task considered in this paper is to obtain all quantum inhomogeneous $D = 3$ Euclidean and Lorentzian (i.e. Poincar\'{e}) algebras that can be derived via the quantum version of \.{I}n\"{o}n\"{u}-Wigner (IW) contraction, applied to the $D = 3$ quantum rotation algebras.

A standard, classical IW contraction is applied to Lie algebras that can be decomposed into $\mathfrak{g} = \mathfrak{h} \oplus \mathfrak{n}$, where
\begin{align}
[\mathfrak{h},\mathfrak{h}] \in \mathfrak{h}\,, \qquad [\mathfrak{h},\mathfrak{n}] \in \mathfrak{n}\,, \qquad [\mathfrak{n},\mathfrak{n}] \in \mathfrak{h}\,.
\end{align}
Rescaling $\mathfrak{n} \mapsto R\,\tilde{\mathfrak{n}}$ and taking the limit of the contraction parameter $R \rightarrow \infty$, we obtain a semidirect product $\tilde{\mathfrak{g}} = \mathfrak{h} \vartriangleright\!\!< \tilde{\mathfrak{n}}$, with $[\tilde{\mathfrak{n}},\tilde{\mathfrak{n}}] = 0$ and $\mathfrak{h}$ unmodified in the contraction limit. Meanwhile, the quantum deformation parameters $q_i$ enter linearly into the definition of a classical $r$-matrix
\begin{align}
r = \sum q_i\, r_{(i)}^{AB}\, g_A \wedge g_B\,.
\end{align}
In the quantum IW contraction procedure, $q_i$ will usually depend on the powers of $R$, in a way that permits to obtain the finite contraction limit of a classical $r$-matrix as well as the finite form of the classical Yang-Baxter equation (\ref{eq:10.01}). If one appropriately rescales the $\mathfrak{g}$-invariant three-form $\Omega$, it can answer which contracted classical $r$-matrices satisfy the modified Yang-Baxter equation -- without recalculating the Schouten bracket after the contraction.

As we already mentioned by showing (\ref{eq:10.00}), classical $r$-matrices describe the infinitesimal version of quantum groups, which take the form of noncommutative Hopf algebras. The quantum IW contraction procedure applied to such a quantum Hopf algebra of $\mathfrak{g}$ leads to quantum groups associated with the inhomogeneous algebra $\tilde{\mathfrak{g}}$.

The aim of the current paper is actually three-fold.
\begin{itemize}
\item Using the complete classification of classical $r$-matrices for the $\mathfrak{o}(4;\mathbbm{C})$ algebra and hence for the $\mathfrak{o}(4)$, $\mathfrak{o}(3,1)$ and $\mathfrak{o}(2,2)$ algebras \cite{Borowiec:2017bs}, we perform all possible quantum IW contractions that lead to the inequivalent quantum deformations of $D = 3$ inhomogeneous Euclidean or $D = 3$ Poincar\'{e} algebra. Subsequently, we compare our results with the complete list of $D = 3$ quantum inhomogeneous algebras presented in \cite{Stachura:1998ps}. In particular, we show that all such quantum algebras -- without certain ${\cal T} \wedge {\cal T}$ terms (see (\ref{eq:10.04}) below) in the $r$-matrices -- can be obtained as some quantum IW contraction of a $D = 4$ quantum (pseudo-)orthogonal algebra.
\item An arbitrary classical $r$-matrix for the inhomogeneous (Euclidean or Lorentzian) Lie algebra ${\cal L} \vartriangleright\!\!< {\cal T}$ can be decomposed as follows
\begin{align}\label{eq:10.04}
r = a + b + c\,, \qquad a \in {\cal T} \wedge {\cal T}\,, \quad b \in {\cal L} \wedge {\cal T}\,, \quad c \in {\cal L} \wedge {\cal L}\,,
\end{align}
where ${\cal L}$ denotes the homogeneous subalgebra and ${\cal T}$ is the momentum subalgebra, generating Abelian translations. A classical $r$-matrix with $c = 0$ but $b \neq 0$ is usually said to describe a generalized $\kappa$-deformation (the best known example is the standard $\kappa$-deformation in timelike direction, introduced in \cite{Lukierski:1991qa,Lukierski:1992ny}). Apart from one case, all non-Abelian $r$-matrices classified in \cite{Stachura:1998ps} and derived by us via the quantum IW contractions belong to the $D = 3$ generalized $\kappa$-deformations. A distinguishing feature of such deformations in the physical context is their characterization by the dimensionful deformation parameter, which allows for the geometrization of the Planck mass $m_P = \hbar c^{-1}/\lambda_P$. This suggests the potential applicability of generalized $\kappa$-deformations in the context of quantum gravity.
\item On the other hand, in the context of the Chern-Simons formulation of 3D gravity it has been shown \cite{Fock:1999px} that 3D quantum gravity models can only be constructed with quantum symmetries satisfying the Fock-Rosly compatibility conditions (see Sec.~\ref{sec:6.0}). The Fock-Rosly compatible $r$-matrices are claimed to be completely classified in \cite{Osei:2017ybk}, although they are presented there in a quite complicated way. Meanwhile, some authors argue that the relevant class of symmetries is even smaller and given by the so-called Drinfeld doubles. It is known that there exist three such $r$-matrices associated with Drinfeld doubles on the $D = 3$ de Sitter algebra and three on the $D = 3$ anti-de Sitter algebra \cite{Ballesteros:2013dy}, while there are eight for the $D = 3$ Poincar\'{e} algebra \cite{Ballesteros:2018te}. We point out if and where all of them arise in our calculations. In particular, it turns out that four Poincar\'{e} Drinfeld doubles can not be obtained via quantum IW contractions.
\end{itemize}

The plan of our paper is the following. In the next Sec.~\ref{sec:2.0} we recall all complex $\mathfrak{o}(4;\mathbbm{C})$ $r$-matrices and perform their contractions leading to quantum-deformed inhomogeneous $\mathfrak{o}(3;\mathbbm{C})$ algebras, i.e. quantum $\mathfrak{o}(3;\mathbbm{C}) \vartriangleright\!\!< {\cal T}^3(\mathbbm{C})$ algebras. Sec.~\ref{sec:3.0} is devoted to the description of real forms of the complex algebras $\mathfrak{o}(3;\mathbbm{C})$, $\mathfrak{o}(3;\mathbbm{C}) \vartriangleright\!\!< {\cal T}^3(\mathbbm{C})$ and $\mathfrak{o}(4;\mathbbm{C})$. The focal point of our paper is Sec.~\ref{sec:4.0}, where we analyze all possible inequivalent quantum IW contractions of the deformed real forms of $\mathfrak{o}(4;\mathbbm{C})$: in Subsec.~\ref{sec:4.1} we obtain $r$-matrices of the quantum $D = 3$ inhomogeneous Euclidean algebra from contractions of the $\mathfrak{o}(4)$ $r$-matrices; then we study contractions of the $\mathfrak{o}(3,1)$ $r$-matrices, leading to $D = 3$ inhomogeneous Euclidean (in Subsec.~\ref{sec:4.2}) or $D = 3$ Poincar\'{e} (in Subsec.~\ref{sec:4.2a}) quantum algebras; in Subsec.~\ref{sec:4.3} we consider contractions of $r$-matrices of three different $\mathfrak{o}(2,2)$ real forms, which provide another set of quantum $D = 3$ Poincar\'{e} algebras; all our results are collected in Subsec.~\ref{sec:4.4}. Subsequently, in Sec.~\ref{sec:5.0} we recall the complete list of $D = 3$ inhomogeneous Euclidean and $D = 3$ Poincar\'{e} $r$-matrices given in \cite{Stachura:1998ps} and compare them with the ones derived by us via the quantum IW contractions and listed in Subsec.~\ref{sec:4.4}. Finally, Sec.~\ref{sec:6.0} contains physically most interesting results. It is devoted to the analysis of relevance of Poincar\'{e} and (anti-)de Sitter quantum algebras in 3D quantum gravity models. We conclude the paper with a summary and a short list of open problems in Sec.~\ref{sec:7.0}.

\section{Deformed $\mathfrak{o}(4;\mathbbm{C})$ algebras and their inhomogeneous $\mathfrak{o}(3;\mathbbm{C})$ contractions} \label{sec:2.0}
As we mentioned in the Introduction, the $\mathfrak{o}(4;\mathbbm{C})$ algebra provides a unified setting that we can use to describe all (deformed) rotational symmetries of Euclidean, Lorentzian or Kleinian space(time) in $D = 4$. Therefore, we will first consider quantum deformations of $\mathfrak{o}(4;\mathbbm{C})$, described by five families of classical $r$-matrices, and study the existence of their inhomogeneous $\mathfrak{o}(3;\mathbbm{C})$ quantum IW contraction limits. We subsequently pass to the $\mathfrak{o}(4;\mathbbm{C})$ real forms, whose quantum IW contraction limits lead to the deformed $D = 3$ inhomogeneous Euclidean and $D = 3$ Poincar\'{e} algebras. A convenient framework for the description of the corresponding $r$-matrices is obtained by expressing the $\mathfrak{o}(4;\mathbbm{C})$ algebra in the chiral Cartan-Weyl basis (left $\{{\bf H},{\bf E}_\pm\}$ and right $\{\bar {\bf H},\bar {\bf E}_\pm\}$), where the algebra generators satisfy\footnote{We will use here the bold font to denote generators of the complex algebras.}
\begin{align}\label{eq:20.01}
[{\bf H},{\bf E}_\pm] = \pm {\bf E}_\pm\,, \qquad [{\bf E}_+,{\bf E}_-] = 2{\bf H}\,, \qquad [\bar {\bf H},\bar {\bf E}_\pm] = \pm \bar {\bf E}_\pm\,, \qquad [\bar {\bf E}_+,\bar {\bf E}_-] = 2\bar {\bf H}\,,
\end{align}
while the rest of brackets vanish. This makes explicit the chiral decomposition $\mathfrak{o}(4;\mathbbm{C}) = \mathfrak{o}(3;\mathbbm{C}) \oplus \bar{\mathfrak{o}}(3;\mathbbm{C})$. Transforming both copies of $\mathfrak{o}(3;\mathbbm{C}) \cong \mathfrak{sl}(2;\mathbbm{C})$ to the three-dimensional Cartesian bases via the formulae (this is only one of the possible transformations, cf. (\ref{eq:33.12}-\ref{eq:30.01b}))
\begin{align}\label{eq:20.02}
{\bf H} = -i {\bf X}_3\,, \qquad {\bf E}_\pm = -i {\bf X}_1 \mp {\bf X}_2\,, \qquad
\bar {\bf H} = i \bar {\bf X}_3\,, \qquad \bar {\bf E}_\pm = i \bar {\bf X}_1 \mp \bar {\bf X}_2\,,
\end{align}
we arrive at another chiral basis (left $\{{\bf X}_i\}$ and right $\{\bar {\bf X}_i\}$, $i = 1,2,3$), such that
\begin{align}\label{eq:20.03}
[{\bf X}_i,{\bf X}_j] = \epsilon_{ijk} {\bf X}_k\,, \qquad [\bar {\bf X}_i,\bar {\bf X}_j] = \epsilon_{ijk} \bar {\bf X}_k\,, \qquad [{\bf X}_i,\bar {\bf X}_j] = 0\,.
\end{align}
For our aim of studying the contraction limits we introduce the following orthogonal basis
\begin{align}\label{eq:20.04}
[{\bf J}_i,{\bf J}_j] = \epsilon_{ijk} {\bf J}_k\,, \qquad [{\bf J}_i,{\bf K}_j] = \epsilon_{ijk} {\bf K}_k\,, \qquad [{\bf K}_i,{\bf K}_j] = \epsilon_{ijk} {\bf J}_k\,,
\end{align}
which can be obtained by taking
\begin{align}\label{eq:20.05}
{\bf X}_i = \frac{1}{2} ({\bf J}_i + {\bf K}_i)\,, \qquad \bar {\bf X}_i = \frac{1}{2} ({\bf J}_i - {\bf K}_i)\,.
\end{align}
The basis (\ref{eq:20.04}) provides a three-dimensional decomposition of the standard orthogonal basis
\begin{align}\label{eq:20.06}
[{\bf M}_{AB},{\bf M}_{CD}] = \delta_{AC} {\bf M}_{BD} - \delta_{BC} {\bf M}_{AD} + \delta_{BD} {\bf M}_{AC} - \delta_{AD} {\bf M}_{BC}\,,
\end{align}
where ${\bf M}_{AB} = -{\bf M}_{BA}$, $A,B = 1,\ldots,4$ and in this Section we choose the transformation (cf. (\ref{eq:20.17}))
\begin{align}\label{eq:20.06a}
{\bf J}_i = \frac{1}{2} \epsilon_{ijk} {\bf M}_{jk}\,, \qquad {\bf K}_i = {\bf M}_{i4}\,.
\end{align}

${\bf K}_i$ are the generators of curved translations on the coset group ${\rm O}(4;\mathbbm{C}) / {\rm O}(3;\mathbbm{C})$. Therefore, in order to examine the existence of inhomogeneous $\mathfrak{o}(3;\mathbbm{C})$ quantum contraction limits, we will perform the rescaling of ${\bf K}_i$ to
\begin{align}\label{eq:20.07}
\tilde {\bf K}_i \equiv {\cal R}^{-1} {\bf K}_i\,.
\end{align}
The contraction parameter ${\cal R}$, which is chosen to be real and positive, for the real forms $\mathfrak{o}(4)$, $\mathfrak{o}(3,1)$ and $\mathfrak{o}(2,2)$ (more precisely, as we will describe in Sec.~\ref{sec:3.0}, there are three different $\mathfrak{o}(2,2)$ real forms) in the context of 3d gravity will be interpreted as the de Sitter (or respectively anti-de Sitter) radius ${\cal R}$, related to the cosmological constant $\Lambda$ via the formula ${\cal R}^2 = |\Lambda|^{-1}$. In the present complex case, in the contraction limit ${\cal R} \rightarrow \infty$ we obtain
\begin{align}\label{eq:20.08}
[\tilde {\bf K}_i,\tilde {\bf K}_j] = {\cal R}^{-2} \epsilon_{ijk} {\bf J}_k \longrightarrow [{\bf P}_i,{\bf P}_j] = 0
\end{align}
and hence (\ref{eq:20.04}) reduces then to the inhomogeneous $\mathfrak{o}(3;\mathbbm{C})$ algebra. The generators ${\bf P}_i \equiv \lim_{{\cal R} \rightarrow \infty} \tilde {\bf K}_i$ can be interpreted as the complexified Abelian three-momenta, generating commutative complex translations.

There exist five families of $r$-matrices, each depending on some parameters, that determine all possible deformations of the $\mathfrak{o}(4;\mathbbm{C})$ algebra. We will denote them as $r_I$, $r_{II}$, $r_{III}$, $r_{IV}$ and $r_V$ (see \cite{Borowiec:2017bs}, eqs. (4.5-4.9)). Let us first consider the cases of $r_I$, $r_{II}$ and $r_V$. In the chiral Cartan-Weyl basis (\ref{eq:20.01}) they are given by
\begin{align}\label{eq:20.09}
r_I(\chi) &= \chi \left({\bf E}_+ + \bar {\bf E}_+\right) \wedge \left({\bf H} + \bar {\bf H}\right)\,, \nonumber\\
r_{II}(\chi,\bar\chi,\varsigma) &= \chi\, {\bf E}_+ \wedge {\bf H} + \bar\chi\, \bar {\bf E}_+ \wedge \bar {\bf H} + \varsigma\, {\bf E}_+ \wedge \bar {\bf E}_+\,, \nonumber\\
r_V(\gamma,\bar\chi,\rho) &= \gamma\, {\bf E}_+ \wedge {\bf E}_- + \bar\chi\, \bar {\bf E}_+ \wedge \bar {\bf H} + \rho\, {\bf H} \wedge \bar {\bf E}_+\,.
\end{align}
We transform them to the Cartesian basis using (\ref{eq:20.02}) and (\ref{eq:20.05}) and subsequently rescale the generators ${\bf K}_i$ according to (\ref{eq:20.07}). This leads to
\begin{align}\label{eq:20.10}
r_I(\chi;{\cal R}) &= {\cal R}\, \chi \left(i {\bf J}_2 - {\cal R}\, \tilde {\bf K}_1\right) \wedge \tilde {\bf K}_3\,, \nonumber\\
r_{II}(\chi,\bar\chi,\varsigma;{\cal R}) &= \frac{\chi + \bar\chi}{4} \left({\bf J}_3 \wedge \left({\bf J}_1 - i {\cal R}\, \tilde {\bf K}_2\right) + {\cal R} \left(i {\bf J}_2 - {\cal R}\, \tilde {\bf K}_1\right) \wedge \tilde {\bf K}_3\right) \nonumber\\
&- \frac{\chi - \bar\chi}{4} \left({\bf J}_3 \wedge \left(i {\bf J}_2 - {\cal R}\, \tilde {\bf K}_1\right) + {\cal R} \left({\bf J}_1 - i {\cal R}\, \tilde {\bf K}_2\right) \wedge \tilde {\bf K}_3\right) + \frac{\varsigma}{2} \left({\bf J}_1 - i {\cal R}\, \tilde {\bf K}_2\right) \wedge \left(i {\bf J}_2 - {\cal R}\, \tilde {\bf K}_1\right)\,, \nonumber\\
%r_{II}(\chi,\bar\chi,\varsigma;{\cal R}) &= \frac{\chi}{4} \left(-{\bf J}_1 + i {\bf J}_2 + {\cal R}\, (-\tilde {\bf K}_1 + i \tilde {\bf K}_2)\right) \wedge \left({\bf J}_3 + {\cal R}\, \tilde {\bf K}_3\right) \nonumber\\
%&+ \frac{\bar\chi}{4} \left(-{\bf J}_1 - i {\bf J}_2 + {\cal R}\, (\tilde {\bf K}_1 + i \tilde {\bf K}_2)\right) \wedge \left({\bf J}_3 - {\cal R}\, \tilde {\bf K}_3\right) \nonumber\\
%&+ \frac{i \varsigma}{2} \left({\bf J}_1 \wedge {\bf J}_2 + i {\cal R}\, {\bf J}_1 \wedge \tilde {\bf K}_1 + i {\cal R}\, {\bf J}_2 \wedge \tilde {\bf K}_2 - {\cal R}^2 \tilde {\bf K}_1 \wedge \tilde {\bf K}_2\right)\,, \nonumber\\
r_V(\gamma,\bar\chi,\rho;{\cal R}) &= -\frac{i \gamma}{2} \left({\bf J}_1 + {\cal R}\, \tilde {\bf K}_1\right) \wedge \left({\bf J}_2 + {\cal R}\, \tilde {\bf K}_2\right) \nonumber\\
&+ \left(\frac{\bar\chi + \rho}{4}\, {\bf J}_3 - {\cal R}\, \frac{\bar\chi - \rho}{4}\, \tilde {\bf K}_3\right) \wedge \left({\bf J}_1 + i {\bf J}_2 - {\cal R}\, (\tilde {\bf K}_1 + i \tilde {\bf K}_2)\right)\,.
%r_V(\gamma,\bar\chi,\rho;{\cal R}) &= \frac{i \gamma}{2} \left(-{\bf J}_1 \wedge {\bf J}_2 - {\cal R}\, {\bf J}_1 \wedge \tilde {\bf K}_2 + {\cal R}\, {\bf J}_2 \wedge \tilde {\bf K}_1 - {\cal R}^2 \tilde {\bf K}_1 \wedge \tilde {\bf K}_2\right) \nonumber\\
%&+ \frac{\bar\chi}{4} \left(-{\bf J}_1 - i {\bf J}_2 + {\cal R}\, (\tilde {\bf K}_1 + i \tilde {\bf K}_2)\right) \wedge \left({\bf J}_3 - {\cal R}\, \tilde {\bf K}_3\right) \nonumber\\
%&+ \frac{\rho}{4} \left(-{\bf J}_1 - i {\bf J}_2 + {\cal R}\, (\tilde {\bf K}_1 + i \tilde {\bf K}_2)\right) \wedge \left({\bf J}_3 + {\cal R}\, \tilde {\bf K}_3\right)\,.
\end{align}
In order to get the finite result in the ${\cal R} \rightarrow \infty$ limit of these $r$-matrices, we should rescale the deformation parameters in the following way
\begin{align}\label{eq:20.10a}
\tilde\chi \equiv {\cal R}^2 \chi\,, \qquad \tilde{\bar\chi} \equiv {\cal R}^2 \bar\chi\,, \qquad \tilde\varsigma \equiv {\cal R}^2 \varsigma\,, \qquad \tilde\gamma \equiv {\cal R}^2 \gamma\,, \qquad \tilde\rho \equiv {\cal R}^2 \rho\,.
\end{align}
Then in the contraction limit we ultimately obtain
\begin{align}\label{eq:20.10b}
\tilde r_I(\tilde\chi) &= -\tilde\chi\, {\bf P}_1 \wedge {\bf P}_3\,, \nonumber\\
\tilde r_{II}(\tilde\chi,\tilde{\bar\chi},\tilde\varsigma) &= -\frac{\tilde\chi}{4} \left({\bf P}_1 - i {\bf P}_2\right) \wedge {\bf P}_3 - \frac{\tilde{\bar\chi}}{4} \left({\bf P}_1 + i {\bf P}_2\right) \wedge {\bf P}_3 - \frac{i \tilde\varsigma}{2}\, {\bf P}_1 \wedge {\bf P}_2\,, \nonumber\\
\tilde r_V(\tilde\gamma,\tilde{\bar\chi},\tilde\rho) &= -\frac{i \tilde\gamma}{2}\, {\bf P}_1 \wedge {\bf P}_2 - \frac{\tilde{\bar\chi} - \tilde\rho}{4} \left({\bf P}_1 + i {\bf P}_2\right) \wedge {\bf P}_3\,.
\end{align}
As one can notice, the expression for $\tilde r_V$ in (\ref{eq:20.10b}) vanishes if the original parameters satisfy the relations $\rho = \bar\chi$ and $\gamma = 0$. On the other hand, for such values of $\gamma,\bar\chi,\rho$ in (\ref{eq:20.10}) we may use the alternative rescaling of the remaining free parameter to $\hat{\bar\chi} \equiv {\cal R}\, \bar\chi$ and in the ${\cal R} \rightarrow \infty$ limit it gives
\begin{align}\label{eq:20.10c}
\hat r_V(\hat{\bar\chi}) = -\frac{\hat{\bar\chi}}{2}\, {\bf J}_3 \wedge \left({\bf P}_1 - i {\bf P}_2\right)\,.
\end{align}
Both types of quantum IW contractions possible for $r_V$ can also be performed simultaneously, by appropriately rescaling certain combinations of the deformation parameters. Namely, taking $\hat{\bar\chi} + \hat\rho \equiv {\cal R} (\bar\chi + \rho)$, $\tilde{\bar\chi} - \tilde\rho \equiv {\cal R}^2 (\bar\chi - \rho)$ and $\tilde\gamma \equiv {\cal R}^2 \gamma$ leads to the combined contraction limit
\begin{align}\label{eq:20.10ca}
\hat r_V(\hat{\bar\chi} + \hat\rho) + \tilde r_V(\tilde\gamma,\tilde{\bar\chi} - \tilde\rho) = -\frac{\hat{\bar\chi} + \hat\rho}{4}\, {\bf J}_3 \wedge \left({\bf P}_1 - i {\bf P}_2\right) - \frac{i \tilde\gamma}{2}\, {\bf P}_1 \wedge {\bf P}_2 - \frac{\tilde{\bar\chi} - \tilde\rho}{4} \left({\bf P}_1 + i {\bf P}_2\right) \wedge {\bf P}_3\,.
\end{align}

We now turn to the $r$-matrix $r_{IV}$, which has the form
\begin{align}\label{eq:20.10d}
r_{IV}(\gamma,\varsigma) &= \gamma \left({\bf E}_+ \wedge {\bf E}_- - \bar {\bf E}_+ \wedge \bar {\bf E}_- - 2 {\bf H} \wedge \bar {\bf H}\right) + \varsigma\, {\bf E}_+ \wedge \bar {\bf E}_+\,.
\end{align}
Changing the basis via (\ref{eq:20.02}) and (\ref{eq:20.05}), and subsequently performing the rescaling (\ref{eq:20.07}), we arrive at
\begin{align}\label{eq:20.10e}
r_{IV}(\gamma,\varsigma;{\cal R}) = -\gamma \left(i {\bf J}_1 \wedge {\bf J}_2 - {\cal R}\, {\bf J}_3 \wedge \tilde {\bf K}_3 + i {\cal R}^2 \tilde {\bf K}_1 \wedge \tilde {\bf K}_2\right) + \frac{\varsigma}{2} \left({\bf J}_1 - i {\cal R}\, \tilde {\bf K}_2\right) \wedge \left(i {\bf J}_2 - {\cal R}\, \tilde {\bf K}_1\right)
%r_{IV}(\gamma,\varsigma;{\cal R}) &= i \gamma \left(-{\bf J}_1 \wedge {\bf J}_2 - i {\cal R}\, {\bf J}_3 \wedge \tilde {\bf K}_3 - {\cal R}^2 \tilde {\bf K}_1 \wedge \tilde {\bf K}_2\right) \nonumber\\
%&+ \frac{i \varsigma}{2} \left({\bf J}_1 \wedge {\bf J}_2 + i {\cal R}\, {\bf J}_1 \wedge \tilde {\bf K}_1 + i {\cal R}\, {\bf J}_2 \wedge \tilde {\bf K}_2 - {\cal R}^2 \tilde {\bf K}_1 \wedge \tilde {\bf K}_2\right)\,.
\end{align}
Again, the finite ${\cal R} \rightarrow \infty$ limit can be obtained after the rescaling of parameters
\begin{align}\label{eq:20.10f}
\tilde\gamma \equiv {\cal R}^2 \gamma\,, \qquad \tilde\varsigma \equiv {\cal R}^2 \varsigma\,.
\end{align}
This leads to
\begin{align}\label{eq:20.10g}
\tilde r_{IV}(\tilde\gamma,\tilde\varsigma) = -i \frac{2\tilde\gamma + \tilde\varsigma}{2}\, {\bf P}_1 \wedge {\bf P}_2\,.
\end{align}
The above expression vanishes if the original parameter $\varsigma = -2\gamma$. However, for (\ref{eq:20.10e}) with such a fixed $\varsigma$, the alternative rescaling $\hat\gamma \equiv {\cal R}\, \gamma$ allows us to obtain in the contraction limit
\begin{align}\label{eq:20.10h}
\hat r_{IV}(\hat\gamma) = \hat\gamma \left({\bf J}_1 \wedge {\bf P}_1 + {\bf J}_2 \wedge {\bf P}_2 + {\bf J}_3 \wedge {\bf P}_3\right)\,.
\end{align}
%which is actually the $r$-matrix describing the $\kappa$-deformation of $\mathfrak{o}(3;\mathbbm{C}) \vartriangleright\!\!< {\cal T}^3(\mathbbm{C})$.

Finally, we analyze the $r$-matrix $r_{III}$, given by
\begin{align}\label{eq:20.11}
r_{III}(\gamma,\bar\gamma,\eta) = \gamma\, {\bf E}_+ \wedge {\bf E}_- + \bar\gamma\, \bar {\bf E}_+ \wedge \bar {\bf E}_- + \eta\, {\bf H} \wedge \bar {\bf H}
\end{align}
This is the only $r$-matrix compatible with all possible real forms of $\mathfrak{o}(4;\mathbbm{C})$ (see Table I). Performing the same basis transformation as in (\ref{eq:20.10b}) and (\ref{eq:20.10e}), we obtain
\begin{align}\label{eq:20.12}
r_{III}(\gamma,\bar\gamma,\eta;{\cal R}) &= -i \frac{\gamma - \bar\gamma}{2} \left({\bf J}_1 \wedge {\bf J}_2 + {\cal R}^2 \tilde {\bf K}_1 \wedge \tilde {\bf K}_2\right) - i {\cal R}\, \frac{\gamma + \bar\gamma}{2} \left({\bf J}_1 \wedge \tilde {\bf K}_2 - {\bf J}_2 \wedge \tilde {\bf K}_1\right) - {\cal R}\, \frac{\eta}{2} {\bf J}_3 \wedge \tilde {\bf K}_3\,,
%\frac{i \gamma}{2} \left(-{\bf J}_1 \wedge {\bf J}_2 - {\cal R}\, {\bf J}_1 \wedge \tilde {\bf K}_2 + {\cal R}\, {\bf J}_2 \wedge \tilde {\bf K}_1 - {\cal R}^2 \tilde {\bf K}_1 \wedge \tilde {\bf K}_2\right) \nonumber\\
%&+ \frac{i \bar\gamma}{2} \left({\bf J}_1 \wedge {\bf J}_2 - {\cal R}\, {\bf J}_1 \wedge \tilde {\bf K}_2 + {\cal R}\, {\bf J}_2 \wedge \tilde {\bf K}_1 + {\cal R}^2 \tilde {\bf K}_1 \wedge \tilde {\bf K}_2\right) - {\cal R}\, \frac{\eta}{2} {\bf J}_3 \wedge \tilde {\bf K}_3\,,
\end{align}
which has the finite ${\cal R} \rightarrow \infty$ limit if we supplement it by the rescaling of parameters to
\begin{align}\label{eq:20.13}
\tilde\gamma \equiv {\cal R}^2 \gamma\,, \qquad \tilde{\bar\gamma} \equiv {\cal R}^2 \bar\gamma\,, \qquad \tilde\eta \equiv {\cal R}^2 \eta\,.
\end{align}
The contraction limit has the form
\begin{align}\label{eq:20.14}
\tilde r_{III}(\tilde\gamma,\tilde{\bar\gamma}) = -i \frac{\tilde\gamma - \tilde{\bar\gamma}}{2}\, {\bf P}_1 \wedge {\bf P}_2\,.
\end{align}
Similarly to $\tilde r_{IV}$ and $\tilde r_V$, the above expression vanishes if the original parameter $\bar\gamma = \gamma$ but applying the alternative rescaling
\begin{align}\label{eq:20.15}
\hat\gamma \equiv {\cal R}\, \gamma\,, \qquad \hat\eta \equiv {\cal R}\, \eta
\end{align}
in (\ref{eq:20.12}) leads to another contraction limit
\begin{align}\label{eq:20.16}
\hat r_{III}(\hat\gamma,\hat\eta) = -i \hat\gamma \left({\bf J}_1 \wedge {\bf P}_2 - {\bf J}_2 \wedge {\bf P}_1\right) - \frac{\hat\eta}{2}\, {\bf J}_3 \wedge {\bf P}_3\,.
\end{align}
%Furthermore, with $\bar\gamma = \gamma$ and after the rescaling (\ref{eq:20.15}), the $r$-matrix (\ref{eq:20.12}) actually does not depend on ${\cal R}$.
We can also use a procedure analogous to (\ref{eq:20.10ca}), taking $\hat\eta \equiv {\cal R}\, \eta$, $\hat\gamma + \hat{\bar\gamma} \equiv {\cal R} (\gamma + \bar\gamma)$ and $\tilde\gamma - \tilde{\bar\gamma} \equiv {\cal R}^2 (\gamma - \bar\gamma)$, which gives
\begin{align}\label{eq:20.16a}
\hat r_{III}(\hat\gamma + \hat{\bar\gamma},\hat\eta) + \tilde r_{III}(\tilde\gamma - \tilde{\bar\gamma}) = -i \frac{\hat\gamma + \hat{\bar\gamma}}{2}\, \left({\bf J}_1 \wedge {\bf P}_2 - {\bf J}_2 \wedge {\bf P}_1\right) - \frac{\hat\eta}{2}\, {\bf J}_3 \wedge {\bf P}_3 - i \frac{\tilde\gamma - \tilde{\bar\gamma}}{2}\, {\bf P}_1 \wedge {\bf P}_2\,.
\end{align}

The result of the IW contraction $\mathfrak{o}(4;\mathbbm{C}) \mapsto \mathfrak{o}(3;\mathbbm{C}) \vartriangleright\!\!< {\cal T}^3(\mathbbm{C})$ does not depend on a coordinate axis in $\mathbbm{C}^4$ along which the rescaling and contraction is performed (this is no longer true for the pseudo-orthogonal real forms of $\mathfrak{o}(4;\mathbbm{C})$, cf. Subsec.~\ref{sec:3.3}). In (\ref{eq:20.07}) we picked the fourth axis (i.e. $A = 4$) but this is only a matter of convention. However, since $r$-matrices are not symmetric under ${\rm O}(4)$ rotations, their quantum IW contraction limits obtained with the rescaling of different axes are not always equivalent under automorphisms of $\mathfrak{o}(3;\mathbbm{C}) \vartriangleright\!\!< {\cal T}^3(\mathbbm{C})$. To see this explicitly, let us pick the first axis, associated with the corresponding orthogonal basis:
\begin{align}\label{eq:20.17}
{\bf J}'_p = \frac{1}{2} \epsilon_{pqr} {\bf M}_{qr}\,, \qquad {\bf K}'_p = {\bf M}_{p1}\,,
\end{align}
where now $p,q,r = 2,3,4$ (cf. (\ref{eq:20.06})). The generators ${\bf J}'_p$ and ${\bf K}'_p$ satisfy the same commutation relations as (\ref{eq:20.04}), modulo the substitution $1 \mapsto 4$ (due to the Euclidean metric we have $\epsilon_{234} = 1$). The quantum IW contraction along the first axis is performed by taking the rescaling
\begin{align}\label{eq:20.18}
\tilde {\bf K}'_p \equiv {\cal R}^{-1} {\bf K}'_p
\end{align}
and subsequently the limit ${\cal R} \rightarrow \infty$, $\tilde {\bf K}'_p \rightarrow {\bf P}'_p$. In order to pass from the Cartan-Weyl basis (\ref{eq:20.01}) to the basis $\{{\bf J}'_p,\tilde {\bf K}'_p\}$, one has to substitute the formulae (\ref{eq:20.05}) into (\ref{eq:20.02}), with ${\bf J}_i$ and ${\bf K}_i$ expressed in terms of ${\bf J}'_p$ and $\tilde {\bf K}'_p$. The obtained transformation between the bases has the form
\begin{align}\label{eq:20.19}
{\bf H} &= -\frac{i}{2} \left({\bf J}'_2 - {\cal R}\, \tilde {\bf K}'_2\right)\,, & {\bf E}_\pm &= \frac{1}{2} \left(\pm {\bf J}'_3 - i {\bf J}'_4 + {\cal R}\, (\mp\tilde {\bf K}'_3 + i \tilde {\bf K}'_4)\right)\,, \nonumber\\
\bar {\bf H} &= -\frac{i}{2} \left({\bf J}'_2 + {\cal R}\, \tilde {\bf K}'_2\right)\,, & \bar {\bf E}_\pm &= \frac{1}{2} \left(\mp {\bf J}'_3 + i {\bf J}'_4 + {\cal R}\, (\mp\tilde {\bf K}'_3 + i \tilde {\bf K}'_4)\right)\,.
\end{align}
Applying (\ref{eq:20.19}) to the $r$-matrices (\ref{eq:20.09}), (\ref{eq:20.10d}) and (\ref{eq:20.11}), one can express them in the orthogonal basis rescaled along the first axis. Let us consider the $r$-matrices $r_I$, $r_{II}$ and $r_{IV}$ as examples\footnote{Due to the form (\ref{eq:20.02}) of the relation between the chiral Cartan-Weyl and Cartesian bases, the contraction along the first or second axis leads to modified or even completely new contraction limits of these $r$-matrices with respect to the ones given by (\ref{eq:20.10b}) and (\ref{eq:20.10g}-\ref{eq:20.10h}).}. The procedure described above gives
\begin{align}\label{eq:20.20}
r_I^a(\chi;{\cal R}) &= -{\cal R}\, \chi\, {\bf J}'_2 \wedge \left(\tilde {\bf K}'_4 + i \tilde {\bf K}'_3\right)\,, \nonumber\\
r_{II}^a(\chi,\bar\chi,\varsigma;{\cal R}) &= {\cal R}\, \frac{\chi + \bar\chi}{4} \left(\left({\bf J}'_4 + i {\bf J}'_3\right) \wedge \tilde {\bf K}'_2 - {\bf J}'_2 \wedge \left(\tilde {\bf K}'_4 + i \tilde {\bf K}'_3\right)\right) \nonumber\\
&- \frac{\chi - \bar\chi}{4} \left(\left({\bf J}'_4 + i {\bf J}'_3\right) \wedge {\bf J}'_2 - {\cal R}^2 \tilde {\bf K}'_2 \wedge \left(\tilde {\bf K}'_4 + i \tilde {\bf K}'_3\right)\right) + {\cal R}\, \frac{\varsigma}{2} \left({\bf J}'_4 + i {\bf J}'_3\right) \wedge \left(\tilde {\bf K}'_4 + i \tilde {\bf K}'_3\right)\,, \nonumber\\
r_{IV}^a(\gamma,\varsigma;{\cal R}) &= -{\cal R}\, \gamma \left(i {\bf J}'_4 \wedge \tilde {\bf K}'_3 - i {\bf J}'_3 \wedge \tilde {\bf K}'_4 - {\bf J}'_2 \wedge \tilde {\bf K}'_2\right) + {\cal R}\, \frac{\varsigma}{2} \left({\bf J}'_4 + i {\bf J}'_3\right) \wedge \left(\tilde {\bf K}'_4 + i \tilde {\bf K}'_3\right)\,.
%r_V^a(\gamma,\bar\chi,\rho;{\cal R}) &= \frac{i \gamma}{2} \left({\bf J}'_4 - {\cal R}\, \tilde {\bf K}'_4\right) \wedge \left({\bf J}'_3 - {\cal R}\, \tilde {\bf K}'_3\right) \nonumber\\
%&- \left(\frac{\bar\chi - \rho}{4}\, {\bf J}'_2 + {\cal R}\, \frac{\bar\chi + \rho}{4}\, \tilde {\bf K}'_2\right) \wedge \left({\bf J}'_4 + i {\bf J}'_3 + {\cal R}\, (\tilde {\bf K}'_4 + i \tilde {\bf K}'_3)\right)\,,
\end{align}
which are respectively equivalent to (\ref{eq:20.10}) and (\ref{eq:20.10e}) under the $\mathfrak{o}(4;\mathbbm{C})$ automorphism $({\bf J}'_{2/3} \mapsto \pm {\cal R}\, \tilde {\bf K}_{3/2},{\bf J}'_4 \mapsto {\bf J}_1,\tilde {\bf K}'_{2/3} \mapsto \mp {\cal R}^{-1} {\bf J}_{3/2},\tilde {\bf K}'_4 \mapsto -\tilde {\bf K}_1)$. However, rescaling the deformation parameters to (under the condition $\bar\chi = \chi$ in the case of $r_{II}$)
\begin{align}\label{eq:20.20a}
\hat\chi \equiv {\cal R}\, \chi\,, \qquad \hat\varsigma \equiv {\cal R}\, \varsigma\,, \qquad \hat\gamma \equiv {\cal R}\, \gamma\,,
\end{align}
we find the contraction limits that are inequivalent to (\ref{eq:20.10b}) and (\ref{eq:20.10g}-\ref{eq:20.10h}):
\begin{align}\label{eq:20.21}
\hat r_I^a(\hat\chi) &= -\hat\chi\, {\bf J}'_2 \wedge \left({\bf P}'_4 + i {\bf P}'_3\right)\,, \nonumber\\
\hat r_{II}^a(\hat\chi,\hat\varsigma) &= \frac{\hat\chi}{2} \left(({\bf J}'_4 + i {\bf J}'_3) \wedge {\bf P}'_2 - {\bf J}'_2 \wedge ({\bf P}'_4 + i {\bf P}'_3)\right) + \frac{\hat\varsigma}{2} \left({\bf J}'_4 + i {\bf J}'_3\right) \wedge \left({\bf P}'_4 + i {\bf P}'_3\right)\,, \nonumber\\
\hat r_{IV}^a(\hat\gamma,\hat\varsigma) &= -\hat\gamma \left(i {\bf J}'_4 \wedge {\bf P}'_3 - i {\bf J}'_3 \wedge {\bf P}'_4 - {\bf J}'_2 \wedge {\bf P}'_2\right) + \frac{\hat\varsigma}{2} \left({\bf J}'_4 + i {\bf J}'_3\right) \wedge \left({\bf P}'_4 + i {\bf P}'_3\right)\,.
%\hat r_V^a(\hat{\bar\chi}) &= -\frac{\hat{\bar\chi}}{2}\, {\bf J}'_2 \wedge \left({\bf P}'_4 + i {\bf P}'_3\right)\,.
\end{align}
Analogously to (\ref{eq:20.10ca}) and (\ref{eq:20.16a}), for $r_{II}^a$ we may also use the rescaling $\hat\chi + \hat{\bar\chi} \equiv {\cal R} (\chi + \bar\chi)$, $\tilde\chi - \tilde{\bar\chi} \equiv {\cal R}^2 (\chi - \bar\chi)$, $\hat\gamma \equiv {\cal R}\, \gamma$, which leads to the more general contraction limit
\begin{align}\label{eq:20.22}
\hat r_{II}^a(\hat\chi + \hat{\bar\chi},\hat\varsigma) + \tilde r_{II}^a(\tilde\chi - \tilde{\bar\chi}) &= \frac{\hat\chi + \hat{\bar\chi}}{4} \left(({\bf J}'_4 + i {\bf J}'_3) \wedge {\bf P}'_2 - {\bf J}'_2 \wedge ({\bf P}'_4 + i {\bf P}'_3)\right) + \frac{\hat\varsigma}{2} \left({\bf J}'_4 + i {\bf J}'_3\right) \wedge \left({\bf P}'_4 + i {\bf P}'_3\right) \nonumber\\
&+ \frac{\tilde\chi - \tilde{\bar\chi}}{4}\, {\bf P}'_2 \wedge \left({\bf P}'_4 + i {\bf P}'_3\right)\,.
\end{align}
This dependence on a contraction axis will be discussed in detail for deformations of the $\mathfrak{o}(4;\mathbbm{C})$ real forms in Sec.~\ref{sec:4.0}.

\section{Real forms of the $\mathfrak{o}(3;\mathbbm{C})$, $\mathfrak{o}(3;\mathbbm{C}) \vartriangleright\!\!< {\cal T}^3(\mathbbm{C})$ and $\mathfrak{o}(4;\mathbbm{C})$ algebras} \label{sec:3.0}
\subsection{Real algebras, bialgebras and $\divideontimes$-Hopf algebras} \label{sec:3.1}

A real Lie algebra structure (a real form) $(\mathfrak{g}, \divideontimes)$ is introduced on a complex Lie algebra $\mathfrak{g}$ by defining an involutive antilinear antiautomorphism (called a $\divideontimes$-conjugation or $\divideontimes$-operation) $\divideontimes: \mathfrak{g} \mapsto \mathfrak{g}$. Then one can find such a basis of the algebra that its structure constants are real and the $\divideontimes$-conjugation is anti-Hermitian (i.e. $\forall g \in \mathfrak{g}: g^\divideontimes = -g$)\footnote{When the algebra is represented in the Hilbert space this leads to operators with the imaginary spectrum. In order to avoid this, one can instead introduce a real Lie algebra with the imaginary structure constants and Hermitian $\divideontimes$-conjugation.}. However, in an arbitrary basis, a given real form is characterized by certain nontrivial reality conditions that have to be satisfied by the algebra generators under the action of the $\divideontimes$-conjugation.

A real coboundary Lie bialgebra (see (\ref{eq:10.00})) is introduced as a triple $(\mathfrak{g}, \divideontimes, r)$, with the classical $r$-matrix $r$ assumed to be anti-Hermitian, namely
\begin{align}\label{eq:3a.01}
r^{\divideontimes \otimes \divideontimes} = -r = r^\tau\,,
\end{align}
where $\tau$ is the flip map, $\tau: a \otimes b \mapsto b \otimes a$. This leads to the appropriate reality conditions for the deformation parameters, which is especially important for the physical description of quantum deformed symmetries. Furthermore, due to the antiautomorphism property (i.e. $\forall g,h \in \mathfrak{g}: (g h)^\divideontimes = h^\divideontimes g^\divideontimes$), the $\divideontimes$-conjugation extends to the universal enveloping algebra $U(\mathfrak{g})$ of a Lie algebra $\mathfrak{g}$, as well as its quantum deformations $U_q(\mathfrak{g})$, making each of them an associative $\divideontimes$-algebra. $U_q(\mathfrak{g})$ is defined as a Hopf algebra $(U_q(\mathfrak{g}),\varepsilon,\Delta,S)$, where $\epsilon$ denotes a counit, $\Delta$ a coproduct and $S$ an antipode \cite{Majid:1995fy}. The $\divideontimes$-conjugation has to preserve the Hopf-algebraic structure of $U_q(\mathfrak{g})$ by satisfying the following compatibility conditions\footnote{I.e. a real Hopf algebra is defined as a $\divideontimes$-Hopf algebra, which is a complex Hopf algebra equipped with a $\divideontimes$-conjugation. The presence of this conjugation turns the algebraic sector into a $\divideontimes$-algebra, while the coalgebraic sector becomes a $\divideontimes$-coalgebra, satisfying (\ref{eq:3a.02}).} (here and elsewhere $*$ denotes the complex conjugation)
\begin{align}\label{eq:3a.02}
\forall a \in U_q(\mathfrak{g}): \qquad \epsilon(a^\divideontimes) = \left(\epsilon(a)\right)^*\,, \quad \Delta(a^\divideontimes) = \left(\Delta(a)\right)^\divideontimes\,, \quad S\left(S(a^\divideontimes)^\divideontimes\right) = a\,,
\end{align}
where the $\divideontimes$-conjugation is assumed to act on tensor products as
\begin{align}\label{eq:3a.03}
(a \otimes b)^\divideontimes = a^\divideontimes \otimes b^\divideontimes\,.
\end{align}
Let us note that sometimes the alternative rule $(a \otimes b)^\divideontimes = b^\divideontimes \otimes a^\divideontimes$ is used instead.

The finite quantum counterpart of a classical $r$-matrix $r$ is the universal $R$-matrix $R$, determining the Hopf algebra structure $U_q(\mathfrak{g})$ of quantum deformations of $U(\mathfrak{g})$. An invertible element $R \in U_q(\mathfrak{g}) \wedge U_q(\mathfrak{g})$ provides the flip of the coproduct $\Delta^\tau = R \Delta R^{-1}$. $U_q(\mathfrak{g})$ together with $R$ satisfying certain additional conditions is called a quasitriangular Hopf algebra. In such a case, $R$ allows us to define a quasitriangular $\divideontimes$-Hopf algebra if \cite{Majid:1995fy} it is either real, i.e. $R^{\divideontimes\otimes\divideontimes} = R^\tau$, or antireal, i.e. $R^{\divideontimes\otimes\divideontimes} = R^{-1}$ and the corresponding quantum $R$-matrix is $\divideontimes$-unitary. For the triangular $R$-matrix, i.e. $R^\tau = R^{-1}$, the above two reality conditions are identical. In the non-triangular case, there exist two universal $R$-matrices and the second of them $(R^\tau)^{-1}$ satisfies the same reality conditions as $R$. Any $R$ can be expanded as $R = \mathbbm{1} \otimes \mathbbm{1} + r + \ldots$ and there are two possibilities for the element $r$. Firstly, if $r$ is skew-symmetric, it is a classical $r$-matrix, which should also be anti-Hermitian, i.e. satisfy the relation (\ref{eq:3a.01}). Secondly, if the element $r$ is not skew-symmetric, we have a non-triangular case and then $r^{\divideontimes \otimes \divideontimes} = r^\tau$ for the real $R$, while $r^{\divideontimes \otimes \divideontimes} = -r$ for the antireal $R$.

\subsection{Real forms of $\mathfrak{o}(3;\mathbbm{C})$ and of $\mathfrak{o}(3;\mathbbm{C}) \vartriangleright\!\!< {\cal T}^3(\mathbbm{C})$} \label{sec:3.2}

The complex algebra $\mathfrak{o}(3;\mathbbm{C}) \cong \mathfrak{sl}(2;\mathbbm{C})$ in the Cartan-Weyl basis corresponds to the $\{{\bf H},{\bf E}_\pm\}$ sector of (\ref{eq:20.01}). There are three real forms of $\mathfrak{o}(3;\mathbbm{C})$, with the following reality conditions:
\begin{align}\label{eq:3b.01}
H^\dagger &= H\,, & E_\pm^\dagger &= E_\mp && {\rm for} \quad \mathfrak{su}(2)\,, \nonumber\\
H^\ddagger &= -H\,, & E_\pm^\ddagger &= -E_\pm && {\rm for} \quad \mathfrak{sl}(2;\mathbbm{R})\,, \nonumber\\
H^\# &= H\,, & E_\pm^\# &= -E_\mp && {\rm for} \quad \mathfrak{su}(1,1)\,,
\end{align}
where $\mathfrak{su}(2) \cong \mathfrak{o}(3;\mathbbm{R})$ and $\mathfrak{sl}(2;\mathbbm{R}) \cong \mathfrak{su}(1,1) \cong \mathfrak{o}(2,1;\mathbbm{R})$. The isomorphism between the $\mathfrak{su}(1,1)$ and $\mathfrak{sl}(2;\mathbbm{R})$ algebras can be realized as an automorphism of $\mathfrak{o}(3;\mathbbm{C})$ given by (here we denote the Cartan-Weyl basis in the $\mathfrak{su}(1,1)$ case as $\{H',E'_\pm\}$)
\begin{align}\label{eq:3b.02}
H' = -\frac{i}{2} (E_+ - E_-)\,, \qquad E'_\pm = \mp i H + \frac{1}{2} (E_+ + E_-)\,.
\end{align}
In the Cartesian basis, corresponding to the $\{{\bf X}_i\}$ sector of (\ref{eq:20.03}), we have only two real forms, with the reality conditions
\begin{align}\label{eq:3b.03}
X_i^\dagger = -X_i \quad {\rm for} \ \mathfrak{su}(2)\,, \qquad
X_i^\ddagger = (-1)^{i-1} X_i \quad {\rm for} \ \mathfrak{sl}(2;\mathbbm{R}) \cong \mathfrak{su}(1,1)\,.
\end{align}
However, the relation between the Cartan-Weyl and Cartesian bases does not look the same in the case of $\mathfrak{su}(1,1)$ as in $\mathfrak{sl}(2;\mathbbm{R})$. From (\ref{eq:3b.01}) one can observe that $H$ is compact for $\mathfrak{su}(1,1)$, while noncompact for $\mathfrak{sl}(2;\mathbbm{R})$. In general, the relations between the above mentioned bases can be chosen as\footnote{The formulae (\ref{eq:3b.04}) are not unique, e.g. for the left chiral sector $\{H,E_\pm\}$ in (\ref{eq:20.02}) they have been transformed by the $\mathfrak{o}(3)$ automorphism $X_i \mapsto (-1)^i X_i$.}
\begin{align}\label{eq:3b.04}
H &= -i X_3\,, & E_\pm &= -i X_1 \mp X_2 && {\rm for} \quad \mathfrak{su}(2) \quad {\rm or} \quad \mathfrak{sl}(2;\mathbbm{R})\,, \nonumber\\
H' &= -i X_2\,, & E'_\pm &= -i X_1 \pm X_3 && {\rm for} \quad \mathfrak{su}(1,1)\,.
\end{align}
It means that if we consider e.g. the Drinfeld-Jimbo deformation of $\mathfrak{o}(2,1)$ in the Cartan-Weyl basis with the first or second set of relations (\ref{eq:3b.04}), in the Cartesian basis we obtain two different types of nonlinearities (see \cite{Lukierski:2017qy}, Sec.~5 for more details).

In order to enlarge the $\mathfrak{o}(3;\mathbbm{C})$ algebra to the inhomogeneous algebra $\mathfrak{o}(3;\mathbbm{C}) \vartriangleright\!\!< {\cal T}^3(\mathbbm{C})$, one should extend the Cartan-Weyl basis by the generators ${\bf P}_i \in {\cal T}^3(\mathbbm{C})$, $i = 1,2,3$ (commuting complex momenta). Denoting ${\bf P}_\pm \equiv {\bf P}_1 \pm i {\bf P}_2$, we then write the cross brackets of $\mathfrak{o}(3;\mathbbm{C}) \vartriangleright\!\!< {\cal T}^3(\mathbbm{C})$ as
\begin{align}\label{eq:3b.05}
[{\bf H},{\bf P}_\pm] &= \mp {\bf P}_\pm\,, & [{\bf H},{\bf P}_3] &= 0\,, && \nonumber\\
[{\bf E}_\pm,{\bf P}_\pm] &= \pm 2 {\bf P}_3\,, & [{\bf E}_\pm,{\bf P}_\mp] &= 0\,, & [{\bf E}_\pm,{\bf P}_3] &= \mp {\bf P}_\mp\,.
\end{align}
In terms of the Cartesian basis they take the familiar simple form
\begin{align}\label{eq:3b.06}
[{\bf X}_i,{\bf P}_j] = \epsilon_{ijk} {\bf P}_k\,, \qquad [{\bf P}_i,{\bf P}_j] = 0\,.
\end{align}
The real $D = 3$ inhomogeneous Euclidean ($\mathfrak{o}(3) \vartriangleright\!\!< {\cal T}^3$) or Poincar\'{e} ($\mathfrak{o}(2,1) \vartriangleright\!\!< {\cal T}^{2,1}$) algebra can be obtained by imposing the appropriate conditions (\ref{eq:3b.01}) together with the consistent proper reality conditions for $P_i \in {\cal T}^3$ or $P_i \in {\cal T}^{2,1}$, respectively. The latter conditions define real momenta
\begin{align}\label{eq:3b.07}
P_\pm^\dagger &= -P_\mp\,, & P_3^\dagger &= -P_3 && {\rm for} \quad \mathfrak{su}(2) \vartriangleright\!\!< {\cal T}^3\,, \nonumber\\
P_\pm^\ddagger &= P_\pm\,, & P_3^\ddagger &= P_3 && {\rm for} \quad \mathfrak{sl}(2;\mathbbm{R}) \vartriangleright\!\!< {\cal T}^{2,1}\,.
\end{align}
It is easy to see that the brackets (\ref{eq:3b.05}) are invariant under either of the conjugations (\ref{eq:3b.07}). On the other hand, the Poincar\'{e} algebra can also be introduced as the real form of (\ref{eq:3b.05}) invariant under the $\mathfrak{su}(1,1)$ conjugation (i.e. $\mathfrak{su}(1,1) \vartriangleright\!\!< {\cal T}^{2,1}$), obtained by extending the reality conditions from the last line of (\ref{eq:3b.01}) by
\begin{align}\label{eq:3b.08}
P_\pm^\# = P_\mp\,, \qquad P_3^\# = -P_3\,,
\end{align}
which is equivalent to the second line of (\ref{eq:3b.07}) with the generators $P_2$ and $P_3$ exchanged.

The completeness of the description of quantum deformed $\mathfrak{o}(2,1)$ algebras expressed in either $\mathfrak{sl}(2;\mathbbm{R})$ or $\mathfrak{su}(1,1)$ Cartan-Weyl basis has recently been proven in \cite{Lukierski:2017qy}, where it has been explicitly demonstrated that the $\mathfrak{sl}(2;\mathbbm{R})$ and $\mathfrak{su}(1,1)$ bialgebras, determined by the respective classical $r$-matrices, are isomorphic. It turns out that one can conveniently choose the following three non-equivalent basic $r$-matrices for the $D = 3$ Lorentz algebra $\mathfrak{o}(2,1)$ \cite{Lukierski:2017qy}:
\begin{align}\label{eq:3b.09}
r_{st} &= -2i \alpha X_1 \wedge X_2 = \alpha E_+ \wedge E_-\,, \nonumber\\
r'_{st} &= 2 \alpha X_1 \wedge X_3 = -i \alpha E'_+ \wedge E'_-\,, \nonumber\\
r_J &= i \alpha (i X_1 + X_2) \wedge X_3 = \alpha E_+ \wedge H\,,
\end{align}
where the parameter $\alpha \in \mathbbm{R}_+$ (and the prime again denotes generators from the $\mathfrak{su}(1,1)$ basis (\ref{eq:3b.02})). The first two $r$-matrices describe the $q$-analogs of the $\mathfrak{sl}(2;\mathbbm{R})$ and $\mathfrak{su}(1,1)$ algebras, satisfying the following modified YB equation
\begin{align}\label{eq:3b.10}
[[r,r]] = \pm \alpha^2\, \Omega\,, \quad \Omega = -8\, X_1 \wedge X_2 \wedge X_3\,,
\end{align}
where the plus sign corresponds to $r = r'_{st}$ (the standard $r$-matrix of $\mathfrak{su}(1,1)$) and the minus to $r = r_{st}$  (the standard $r$-matrix of $\mathfrak{sl}(2;\mathbbm{R})$). The remaining $r$-matrix $r_J$ provides the Jordanian
deformation of $\mathfrak{sl}(2;\mathbbm{R})$. We should also mention that the choice to employ two particular $r$-matrices in the $\mathfrak{sl}(2;\mathbbm{R})$ Cartan-Weyl basis and one in the $\mathfrak{su}(1,1)$ basis is inspired by the presence of explicit quantization procedure \cite{Majid:1995fy}.

As we already mentioned in the Introduction, the classification of $D = 3$ inhomogeneous Euclidean and Poincar\'{e} $r$-matrices (describing coboundary bialgebras) has been derived in \cite{Stachura:1998ps}. This has been accomplished using the technique of solving the inhomogeneous Yang-Baxter equation for the $\mathfrak{o}(3) \vartriangleright\!\!< {\cal T}^3$ and $\mathfrak{o}(2,1) \vartriangleright\!\!< {\cal T}^{2,1}$ algebras. We rewrite the obtained $r$-matrices in our notation in Sec.~\ref{sec:5.0}.

\subsection{(Pseudo-)orthogonal real forms of $\mathfrak{o}(4;\mathbbm{C})$} \label{sec:3.3}

If we pass from $\mathfrak{o}(3;\mathbbm{C})$ to $\mathfrak{o}(4;\mathbbm{C}) = \mathfrak{o}(3;\mathbbm{C}) \oplus \bar{\mathfrak{o}}(3;\mathbbm{C})$, we need to employ the pair of commuting $\mathfrak{o}(3;\mathbbm{C})$ bases: left-chiral $\{{\bf H},{\bf E}_\pm\}$ and right-chiral $\{\bar {\bf H},\bar {\bf E}_\pm\}$ (cf. (\ref{eq:20.01})). Using the results of Sec.~\ref{sec:3.2} (see also \cite{Borowiec:2016qg,Borowiec:2017ag,Lukierski:2017qy,Borowiec:2017bs}), we can then obtain three $D=4$ (pseudo-)orthogonal real algebras $\mathfrak{o}(4-k,k)$, $k = 0,1,2$ as the appropriate real forms of $\mathfrak{o}(4;\mathbbm{C})$.\footnote{In this paper we do not consider two additional real forms of $\mathfrak{o}(4;\mathbbm{C})$, which are isomorphic to the quaternionic Lie algebra $\mathfrak{o}(2;\mathbbm{H}) \cong \mathfrak{o}(2,1) \oplus \mathfrak{o}(3)$, see \cite{Borowiec:2016qg,Borowiec:2017ag,Borowiec:2017bs}.} The Euclidean algebra $\mathfrak{o}(4) \cong \mathfrak{su}(2) \oplus \bar{\mathfrak{su}}(2)$ is introduced via the following reality conditions for the Cartan-Weyl basis
\begin{align}\label{eq:33.05}
H^\dagger = H\,, \qquad E_\pm^\dagger = E_\mp\,, \qquad \bar H^\dagger = \bar H\,, \qquad \bar E_\pm^\dagger = \bar E_\mp
\end{align}
and the Lorentz algebra $\mathfrak{o}(3,1)$ via the conditions\footnote{On a side, we note that under the automorphism $H \mapsto -H$, $E_\pm \mapsto -E_\mp$ or $\bar H \mapsto -\bar H$, $\bar E_\pm \mapsto -\bar E_\mp$, (\ref{eq:33.06}) transforms into the conditions
\begin{align}
H^\divideontimes = \bar H\,, \qquad E_\pm^\divideontimes = \bar E_\mp\,, \qquad \bar H^\divideontimes = H\,, \qquad \bar E_\pm^\divideontimes = E_\mp\,. \nonumber
\end{align}}
\begin{align}\label{eq:33.06}
H^\ddagger = -\bar H\,, \qquad E_\pm^\ddagger = -\bar E_\pm\,, \qquad \bar H^\ddagger = -H\,, \qquad \bar E_\pm^\ddagger = -E_\pm\,.
\end{align}
Due to the isomorphisms $\mathfrak{o}(2,1) \cong \mathfrak{sl}(2;\mathbbm{R}) \cong \mathfrak{su}(1,1)$, the Kleinian algebra $\mathfrak{o}(2,2) \cong \mathfrak{o}(2,1) \oplus \mathfrak{o}(2,1)$ can be obtained via the following three nonequivalent sets of reality conditions:
\begin{align}\label{eq:33.07}
H^\divideontimes = -H\,, \qquad E_\pm^\divideontimes = -E_\pm\,, \qquad \bar H^\divideontimes = -\bar H\,, \qquad \bar E_\pm^\divideontimes = -\bar E_\pm\,,
\end{align}
corresponding to the real form $\dot{\mathfrak{o}}(2,2) \equiv \mathfrak{sl}(2;\mathbbm{R}) \oplus \bar{\mathfrak{sl}}(2;\mathbbm{R})$;
\begin{align}\label{eq:33.08}
H^\divideontimes = H\,, \qquad E_\pm^\divideontimes = -E_\mp\,, \qquad \bar H^\divideontimes = \bar H\,, \qquad \bar E_\pm^\divideontimes = -\bar E_\mp\,,
\end{align}
corresponding to the real form $\mathfrak{o}'(2,2) \equiv \mathfrak{su}(1,1) \oplus \bar{\mathfrak{su}}(1,1)$;
\begin{align}\label{eq:33.09}
H^\divideontimes = H\,, \qquad E_\pm^\divideontimes = -E_\mp\,, \qquad \bar H^\divideontimes = -\bar H\,, \qquad \bar E_\pm^\divideontimes = -\bar E_\pm\,,
\end{align}
corresponding to the real form $\dot{\mathfrak{o}}'(2,2) \equiv \mathfrak{su}(1,1) \oplus \bar{\mathfrak{sl}}(2;\mathbbm{R})$. The flip $\mathfrak{su}(1,1) \oplus \bar{\mathfrak{sl}}(2;\mathbbm{R}) \mapsto \mathfrak{sl}(2;\mathbbm{R}) \oplus \bar{\mathfrak{su}}(1,1)$ is described by an $\mathfrak{o}(2,2)$ automorphism and therefore does not define an independent real form.

In terms of the chiral Cartesian basis $\{{\bf X}_i,\bar {\bf X}_i\}$ introduced in (\ref{eq:20.02}), the reality conditions (\ref{eq:33.05}-\ref{eq:33.09}) simplify to
\begin{align}\label{eq:33.10}
X_i^\dagger &= -X_i\,, & \bar X_i^\dagger &= -\bar X_i && {\rm for} \quad \mathfrak{o}(4)\,, \nonumber\\
X_i^\ddagger &= -\bar X_i\,, & \bar X_i^\ddagger &= -X_i && {\rm for} \quad \mathfrak{o}(3,1)\,, \nonumber\\
X_i^\# &= (-1)^{i-1} X_i\,, & \bar X_i^\# &= (-1)^{i-1} \bar X_i && {\rm for} \quad \mathfrak{o}(2,2)\,.
\end{align}
(in particular, the above conditions are identical for different decompositions (\ref{eq:33.07}-\ref{eq:33.09}) of $\mathfrak{o}(2,2)$). Similarly, one finds that in terms the orthogonal basis $\{{\bf J}_i,{\bf K}_i\}$ introduced in (\ref{eq:20.05}) the reality conditions (\ref{eq:33.10}) become
\begin{align}\label{eq:33.11}
J_i^\dagger &= -J_i\,, & K_i^\dagger &= -K_i && {\rm for} \quad \mathfrak{o}(4)\,, \nonumber\\
J_i^\ddagger &= -J_i\,, & K_i^\ddagger &= K_i && {\rm for} \quad \mathfrak{o}(3,1)\,, \nonumber\\
J_i^\# &= (-1)^{i-1} J_i\,, & K_i^\# &= (-1)^{i-1} K_i && {\rm for} \quad \mathfrak{o}(2,2)\,.
\end{align}
A natural basis for a real Lie algebra is either purely Hermitian or purely anti-Hermitian. Since choosing the Hermitian convention leads to the appearance of imaginary units in the algebra brackets, we will use the anti-Hermitian bases. For $\mathfrak{o}(4)$ such a basis is simply given in (\ref{eq:33.11}), while in other cases the Hermitian generators have to be Wick-rotated, i.e. appropriately transformed into anti-Hermitian ones (see the next Section). %In the $\mathfrak{o}(3,1)$ algebra, which we associate with the spacetime metric $(1,1,1,-1)$, the $D = 4$ boost generators satisfy the Hermitian reality conditions and have to be transformed (this may be called a Wick rotation) in the following way:
%can be treated as either the (Lorentzian) de Sitter algebra or Euclidean anti-de Sitter algebra (see the next Section).
%\begin{align}\label{eq:33.13a}
%L_i \equiv -i \tilde K_i & {\rm for the IW contraction to} \quad \mathfrak{o}(3) \vartriangleright\!\!< {\cal T}^3\,, \nonumber\\
%{\cal L}_0 \equiv -i \tilde K'_4\,, \qquad {\cal L}_a \equiv \tilde K'_a & {\rm for the IW contraction to} \quad \mathfrak{o}(2,1) \vartriangleright\!\!< {\cal T}^{2,1}\,.
%\end{align}

We have assumed that for the real forms (\ref{eq:33.05}-\ref{eq:33.07}) the relation between the Cartan-Weyl and Cartesian bases is analogous to (\ref{eq:20.02}), namely
\begin{align}\label{eq:33.12}
H = -i X_3\,, \qquad E_\pm = -i X_1 \mp X_2\,, \qquad
\bar H = i \bar X_3\,, \qquad \bar E_\pm = i \bar X_1 \mp \bar X_2\,.
\end{align}
However, as follows from (\ref{eq:3b.04}), in order to reproduce $\mathfrak{o}'(2,2)$ with the reality conditions (\ref{eq:33.10}) we should take instead
\begin{align}\label{eq:30.01a}
H = -i X_2\,, \qquad E_\pm = -i X_1 \pm X_3\,, \qquad
\bar H = i \bar X_2\,, \qquad \bar E_\pm = i \bar X_1 \pm \bar X_3
\end{align}
and for $\dot{\mathfrak{o}}'(2,2)$ we take
\begin{align}\label{eq:30.01b}
H = -i X_2\,, \qquad E_\pm = -i X_1 \pm X_3\,, \qquad
\bar H = i \bar X_3\,, \qquad \bar E_\pm = i \bar X_1 \mp \bar X_2\,.
\end{align}
One can pass from the formulae (\ref{eq:33.12}) to (\ref{eq:30.01a}) by performing two $\frac{\pi}{2}$ rotations $(X_3 \mapsto X_2,X_2 \mapsto -X_3)$ and $(\bar X_3 \mapsto \bar X_2,\bar X_2 \mapsto -\bar X_3)$ in the $(2,3)$ plane, while (\ref{eq:30.01b}) is obtained using only the first of these rotations. Therefore, we have the alternative either to keep three different transformations (\ref{eq:33.12}), (\ref{eq:30.01a}) and (\ref{eq:30.01b}) relating the Cartan-Weyl and Cartesian bases and introduce the same reality conditions for $\dot{\mathfrak{o}}(2,2)$, $\mathfrak{o}'(2,2)$ and $\dot{\mathfrak{o}}'(2,2)$ (see (\ref{eq:33.10})) -- or to assume the same map (\ref{eq:33.12}) for $\dot{\mathfrak{o}}(2,2)$, $\mathfrak{o}'(2,2)$ and $\dot{\mathfrak{o}}'(2,2)$ but at the cost of modifying the reality conditions for $\mathfrak{o}(2,2)$ in (\ref{eq:33.10}-\ref{eq:33.11}), through the substitutions $X_2 \leftrightarrow X_3$, $\bar X_2 \leftrightarrow \bar X_3$ in the case of $\mathfrak{o}'(2,2)$ and $X_2 \leftrightarrow X_3$ in the case of $\dot{\mathfrak{o}}'(2,2)$.

Finally, let us express the reality conditions (\ref{eq:33.11}) in terms of the standard orthogonal basis $\{{\bf M}_{AB}\}$, introduced in (\ref{eq:20.06a}). We obtain
\begin{align}\label{eq:33.13}
M_{ij}^\dagger &= -M_{ij} & M_{i4}^\ddagger &= -M_{i4} && {\rm for} \quad \mathfrak{o}(4)\,, \nonumber\\
M_{ij}^\ddagger &= -M_{ij}\,, & M_{i4}^\ddagger &= M_{i4} && {\rm for} \quad \mathfrak{o}(3,1) \nonumber\\
M_{ij}^\# &= (-1)^{i+j-1} M_{ij}\,, & M_{i4}^\# &= (-1)^{i-1} M_{i4} && {\rm for} \quad \mathfrak{o}(2,2)\,.
\end{align}
For both $\mathfrak{o}(4)$ and $\mathfrak{o}(3,1)$ it shows that $J_i = \epsilon_{ijk} M_{jk}$ generate the $\mathfrak{o}(3)$ algebra, corresponding to the IW contraction chosen along the fourth axis and the spacetime metric $(1,1,1,-1)$ in the latter case.

As it was mentioned at the end of the previous Section, the IW contractions performed along other axes do not always lead to the equivalent result. In the case of the $\mathfrak{o}(3,1)$ real form adjusted to the IW contraction along the third axis (and analogously for the first or second one) we have $J'_p = \frac{1}{2} \epsilon_{pqr} M_{qr}$, $K'_p = M_{p3}$, $p,q,r = 1,2,4$ and therefore (\ref{eq:33.13}) (with the same metric as above) gives the reality conditions
\begin{align}\label{eq:33.14}
{J'_4}^\ddagger = -J'_4\,, \quad {J'_{1,2}}^\ddagger = J'_{1,2}\,, \qquad {K'_4}^\ddagger = K'_4\,, \quad {K'_{1,2}}^\ddagger = -K'_{1,2}\,,
\end{align}
so that $J'_p$ generate the $\mathfrak{o}(2,1)$ algebra. For $\mathfrak{o}(2,2)$ real forms the situation is again different. The reality conditions in (\ref{eq:33.13}) correspond to the metric $(1,-1,1,-1)$ and the algebra (\ref{eq:20.06}) can be decomposed into two copies of $\mathfrak{o}(2,1)$, generated by $M_{ij}$'s and $M_{i4}$'s (the choice adjusted to the IW contraction along the fourth axis) or by $M_{pq}$'s and $M_{p3}$'s (the choice for the IW contraction along the third axis); analogously for the second and first axis.

\section{Deformed $\mathfrak{o}(4-k,k)$ ($k = 0,1,2$) algebras and their inhomogeneous contractions} \label{sec:4.0}

In this Section we shall consider deformations of the real $D = 4$ rotation algebras $\mathfrak{o}(4-k,k)$, $k = 0,1,2$, obtained as (pseudo-)orthogonal real forms of the $\mathfrak{o}(4;\mathbbm{C})$ algebra. The reality conditions for each real form impose some restrictions on the allowed values of deformation parameters of the $\mathfrak{o}(4;\mathbbm{C})$ $r$-matrices (i.e. (\ref{eq:20.09}), (\ref{eq:20.11}) and (\ref{eq:20.10d})), as well as completely exclude certain $r$-matrices. We collect all of this information in Table I. Looking at the $\mathfrak{o}(4;\mathbbm{C})$ $r$-matrices we may also observe that each of them is composed of two types of terms: the ones that describe independent deformations of the $\mathfrak{o}(3;\mathbbm{C})$ and $\bar{\mathfrak{o}}(3;\mathbbm{C})$ subalgebras (for the $\mathfrak{o}(2,1)$ real form such $r$-matrices are given by (\ref{eq:3b.09})); and the terms that mix $\mathfrak{o}(3;\mathbbm{C})$ and $\bar{\mathfrak{o}}(3;\mathbbm{C})$ deformations. It is the latter ones that introduce novel features to the results for $\mathfrak{o}(4;\mathbbm{C})$ (and its real forms) with respect to the already discussed $\mathfrak{o}(3;\mathbbm{C})$ deformations.

\begin{table}[h]
\begin{tabular}{|c|c|c|c|c|c|}
\hline & $r_I$ & $r_{II}$ & $r_{III}$ & $r_{IV}$ & $r_V$ \\
\hline $\mathfrak{o}(4)$ & & & $\gamma, \bar\gamma \in \mathbbm{R}$, $\eta \in i \mathbbm{R}$ & & \\
\hline $\mathfrak{o}(3,1)$ & $\chi \in i \mathbbm{R}$ &  $\bar\chi = \chi \in i \mathbbm{R}$, $\varsigma \in \mathbbm{R}$ & $\bar\gamma = -\gamma^* \in \mathbbm{C}$, $\eta \in \mathbbm{R}$ & $\gamma, \varsigma \in \mathbbm{R}$ & \\
\hline $\dot{\mathfrak{o}}(2,2)$ & $\chi \in i \mathbbm{R}$ & $\chi, \bar\chi, \varsigma \in i \mathbbm{R}$ & $\gamma, \bar\gamma, \eta \in i \mathbbm{R}$ & $\gamma, \varsigma \in i \mathbbm{R}$ & $\gamma, \bar\chi, \rho \in i \mathbbm{R}$ \\
\hline $\mathfrak{o}'(2,2)$ & & & $\gamma, \bar\gamma \in \mathbbm{R}$, $\eta \in i \mathbbm{R}$ & & \\
\hline $\dot{\mathfrak{o}}'(2,2)$ & & & $\gamma, \eta \in \mathbbm{R}$, $\bar\gamma \in i \mathbbm{R}$ & & $\gamma, \rho \in \mathbbm{R}$, $\bar\chi \in i \mathbbm{R}$ \\
\hline
\end{tabular}
\caption{The families of $r$-matrices allowed for (pseudo-)orthogonal real forms of $\mathfrak{o}(4;\mathbbm{C})$ \cite{Borowiec:2017bs}}
\end{table}

When we introduce the cosmological constant, the $\mathfrak{o}(4-k,k)$, $k = 0,1,2$ algebras can appropriately be treated as the $D = 3$ Euclidean and Lorentzian (anti-)de Sitter algebras. Therefore, in such a way we obtain from the results of \cite{Borowiec:2017bs} the complete classification of Hopf-algebraic deformations of the above relativistic symmetry algebras. We will further introduce the unified notation $\{J_i,L_i\}$, $i = 1,2,3$ for the inhomogeneous algebra $\mathfrak{o}(3)$ with curved translations forming $\mathfrak{o}(4)$ or $\mathfrak{o}(3,1)$, and $\{{\cal J}_\mu,{\cal L}_\mu\}$, $\mu = 0,1,2$ for the inhomogeneous algebra $\mathfrak{o}(2,1)$ with curved translations forming $\mathfrak{o}(3,1)$ or $\mathfrak{o}(2,2)$. We also need to mention that in the remaining part of this paper we use the natural system of physical units $c = \hbar = 1$.

\subsection{Deformed $\mathfrak{o}(4)$ contracted to deformed $\mathfrak{o}(3) \vartriangleright\!\!< {\cal T}^3$} \label{sec:4.1}

We begin with the real form $\mathfrak{o}(4)$, which is treated as the $D = 3$ Euclidean de Sitter algebra, with $\Lambda = {\cal R}^{-2} > 0$ (after the IW rescaling analogous to (\ref{eq:20.07})). In the chiral Cartan-Weyl basis it is characterized by the reality conditions (\ref{eq:33.05}). As shown in Table I, there is only one allowed family of Hopf-algebraic deformations of $\mathfrak{o}(4)$, associated with the $r$-matrix (\ref{eq:20.11}). Quantum IW contractions of this $r$-matrix lead to quantum $D = 3$ inhomogeneous Euclidean algebras. It is enough to consider such contractions along the fourth axis of $\mathbbm{R}^4$ (i.e. $A = 4$ in (\ref{eq:20.06})), since the results for other axes differ only by automorphisms. The corresponding basis $\{J_i,\tilde K_i\}$, $i = 1,2,3$ is introduced via the transformation
\begin{align}\label{eq:31.02}
H = -\frac{i}{2} \left(J_3 + {\cal R} \tilde K_3\right)\,, \qquad E_\pm = \frac{1}{2} \left(-i J_1 \mp J_2 - {\cal R}\, (i \tilde K_1 \pm \tilde K_2)\right)\,, \nonumber\\
\bar H = \frac{i}{2} \left(J_3 - {\cal R} \tilde K_3\right)\,, \qquad \bar E_\pm = \frac{1}{2} \left(i J_1 \mp J_2 - {\cal R}\, (i \tilde K_1 \mp \tilde K_2)\right)\,.
\end{align}
%which is equivalent to
%\begin{align}\label{eq:31.02a}
%M_3 = J_3\,, \qquad M_1 = -K_1\,, \qquad M_2 = -K_2\,, \qquad P_3 = -\sqrt{\Lambda}\, K_3\,, \qquad P_1 = -\sqrt{\Lambda}\, J_2\,, \qquad P_2 = \sqrt{\Lambda}\, J_1\,,
%\end{align}
%with the generators satisfying the anti-Hermitian reality conditions $J_i^\divideontimes = -J_i$ and $\tilde K_i^\divideontimes = -\tilde K_i$.
This rescaled orthogonal basis (as well as the analogous bases in Subsec.~\ref{sec:4.2}-\ref{sec:4.3}) will be called the physical basis. In terms of $J_i$ and $L_i \equiv \tilde K_i$, the undeformed brackets of the $\mathfrak{o}(4)$ algebra have the form
\begin{align}\label{eq:31.03}
[J_i,J_j] = \epsilon_{ij}^{\ \ k} J_k\,, \qquad [J_i,L_j] = \epsilon_{ij}^{\ \ k} L_k\,, \qquad [L_i,L_j] = \Lambda \epsilon_{ij}^{\ \ k} J_k\,.
\end{align}

The $r$-matrix (\ref{eq:20.11}) in the physical basis $\{J_i,L_i\}$ becomes
\begin{align}\label{eq:31.04}
r_{III}(\gamma,\bar\gamma,\eta;{\cal R}) = -i \frac{\gamma - \bar\gamma}{2} \left(J_1 \wedge J_2 + {\cal R}^2 L_1 \wedge L_2\right) + {\cal R}\, i \frac{\gamma + \bar\gamma}{2} \left(J_1 \wedge L_2 - J_2 \wedge L_1\right) - {\cal R}\, \frac{\eta}{2}\, J_3 \wedge L_3
\end{align}
and in this case the parameters $\gamma, \bar\gamma \in \mathbbm{R}$, $\eta \in i \mathbbm{R}$. As follows from the analysis in Sec.~\ref{sec:2.0}, (\ref{eq:31.04}) has two inequivalent inhomogeneous contraction limits (if contraction is performed along the first or second axis, obtaining $\hat r_{III}$ requires $\bar\gamma = -\gamma$ instead of $\bar\gamma = \gamma$)
\begin{align}\label{eq:31.04a}
\tilde r_{III}(\tilde\gamma,\tilde{\bar\gamma}) &= -i \frac{\tilde\gamma - \tilde{\bar\gamma}}{2}\, P_1 \wedge P_2\,, \nonumber\\
\hat r_{III}(\hat\gamma,\hat\eta) &= i \hat\gamma \left(J_1 \wedge P_2 - J_2 \wedge P_1\right) - \frac{\hat\eta}{2}\, J_3 \wedge P_3\,,
\end{align}
where $P_i$ denote the Euclidean 3-momenta. %The comparison with the Stachura classification of $r$-matrices for the $D = 3$ inhomogeneous Euclidean algebra will be presented in Subsec.~\ref{sec:5.1}.

\subsection{Deformed $\mathfrak{o}(3,1)$ contracted to deformed $\mathfrak{o}(3) \vartriangleright\!\!< {\cal T}^3$} \label{sec:4.2}

The second real form of $\mathfrak{o}(4;\mathbbm{C})$ that we will consider is $\mathfrak{o}(3,1)$, characterized by the reality conditions (\ref{eq:33.06}), which are in agreement with the spacetime metric $(1,1,1,-1)$ (the convention chosen by us in Subsec.~\ref{sec:3.3}). The algebra $\mathfrak{o}(3,1)$ in $D = 3$ can be treated as either the (Lorentzian) de Sitter algebra (with $\Lambda = {\cal R}^{-2} > 0$) or Euclidean anti-de Sitter algebra (with $\Lambda = -{\cal R}^{-2} < 0$). This corresponds to taking either a space- or timelike direction as the axis associated with the generators rescaled by ${\cal R}^{-1}$, along which the IW contraction (in the limit ${\cal R} \rightarrow \infty$) is later performed.

Let us first pick the fourth (i.e. timelike) axis and consider $\mathfrak{o}(3,1)$ as the Euclidean anti-de Sitter algebra. For the latter algebra there is no possibility to choose another axis. Consequently, the relation between the chiral Cartan-Weyl basis and the basis $\{J_i,\tilde K_i\}$ has the same form as (\ref{eq:31.02}), to wit
\begin{align}\label{eq:32.01a}
H = -\frac{i}{2} \left(J_3 + {\cal R}\, \tilde K_3\right)\,, \qquad E_\pm = \frac{1}{2} \left(-i J_1 \mp J_2 - {\cal R}\, (i \tilde K_1 \pm \tilde K_2)\right)\,, \nonumber\\
\bar H = \frac{i}{2} \left(J_3 - {\cal R}\, \tilde K_3\right)\,, \qquad \bar E_\pm = \frac{1}{2} \left(i J_1 \mp J_2 - {\cal R}\, (i \tilde K_1 \mp \tilde K_2)\right)\,.
\end{align}
Due to the reality conditions (\ref{eq:33.06}), the rotation generators $J_i$ are anti-Hermitian, while $\tilde K_i$'s (which are the $D = 4$ boost generators since $K_i = M_{i4}$) become Hermitian, $\tilde K_i^\divideontimes = \tilde K_i$. As it was described in Subsec.~\ref{sec:3.3}, we need to move to the basis where all generators are anti-Hermitian, using the transformation
\begin{align}\label{eq:32.01h}
L_i \equiv -i \tilde K_i\,.
\end{align}
The undeformed brackets of the $\mathfrak{o}(3,1)$ algebra in this physical basis have exactly the same form as (\ref{eq:31.03}) but with $\Lambda > 0$ being replaced by $\Lambda < 0$.

It is known \cite{Zakrzewski:1994pp,Borowiec:2016qg} that one may define four different families of Hopf-algebraic deformations of the real form $\mathfrak{o}(3,1)$, determined by the $r$-matrices $r_I$, $r_{II}$, $r_{III}$ and $r_{IV}$ \cite{Borowiec:2017bs} (cf. Table I). The first two $r$-matrices in the physical basis can be written as
\begin{align}\label{eq:32.01b}
r_I(\chi;{\cal R}) &= -\chi \left({\cal R}\, J_2 - {\cal R}^2 L_1\right) \wedge L_3\,, \nonumber\\
r_{II}(\chi,\varsigma;{\cal R}) &= -\frac{\chi}{2} \left(\left(J_1 + {\cal R}\, L_2\right) \wedge J_3 + \left({\cal R}\, J_2 - {\cal R}^2 L_1\right) \wedge L_3\right) + \frac{i \varsigma}{2} \left(J_1 + {\cal R}\, L_2\right) \wedge \left(J_2 - {\cal R}\, L_1\right)
&%+ \frac{i \varsigma}{2} \left(J_1 \wedge J_2 - {\cal R}\, J_1 \wedge L_1 - {\cal R}\, J_2 \wedge L_2 + {\cal R}^2 L_1 \wedge L_2\right)\,,
\end{align}
where $\chi \in i \mathbbm{R}$ and $\varsigma \in \mathbbm{R}$, while $\bar\chi$ is eliminated by the relation $\bar\chi = \chi$. Next, $r_{III}$ is given by
\begin{align}\label{eq:32.01c}
r_{III}(\gamma - \bar\gamma,\gamma + \bar\gamma,\eta;{\cal R}) = -i \frac{\gamma - \bar\gamma}{2} \left(J_1 \wedge J_2 - {\cal R}^2 L_1 \wedge L_2\right) + {\cal R}\, \frac{\gamma + \bar\gamma}{2} \left(J_1 \wedge L_2 - J_2 \wedge L_1\right) - {\cal R}\, \frac{i \eta}{2}\, J_3 \wedge L_3\,,
\end{align}
where $\gamma - \bar\gamma = 2 {\rm Re}\gamma \in \mathbbm{R}$, $\gamma + \bar\gamma = 2i {\rm Im}\gamma \in i \mathbbm{R}$ and $\eta \in \mathbbm{R}$ (although for $\mathfrak{o}(3,1)$ we have the relation $\bar\gamma = -\gamma^*$, it will be more convenient not to eliminate $\bar\gamma$ explicitly). The remaining $r_{IV}$ has the form
\begin{align}\label{eq:32.01d}
r_{IV}(\gamma,\varsigma;{\cal R}) = -i \gamma \left(J_1 \wedge J_2 - {\cal R}\, J_3 \wedge L_3 - {\cal R}^2 L_1 \wedge L_2\right) + \frac{i \varsigma}{2} \left(J_1 + {\cal R}\, L_2\right) \wedge \left(J_2 - {\cal R}\, L_1\right)\,,
%{\cal R}\, i \gamma\, J_3 \wedge L_3 - {\cal R}\, \frac{i \varsigma}{2} \left(J_1 \wedge L_1 + J_2 \wedge L_2\right) - i \frac{2\gamma - \varsigma}{2}\, J_1 \wedge J_2 - {\cal R}^2 i \frac{2\gamma + \varsigma}{2}\, L_1 \wedge L_2\,,
\end{align}
where $\gamma,\varsigma \in \mathbbm{R}$. %Let us also mention that at the level of the Hopf algebra, the case of non-vanishing $\varsigma$ is a bit problematic since the corresponding twist deformation is non-unitary.

The possible quantum IW contractions of these real $r$-matrices can be read out from Sec.~\ref{sec:2.0} and they lead to the corresponding quantum $D = 3$ inhomogeneous Euclidean algebras. Namely, $r_I$ and $r_{II}$ have the following inhomogeneous contraction limits
\begin{align}\label{eq:32.01e}
\tilde r_I(\tilde\chi) &= \tilde\chi\, P_1 \wedge P_3\,, \nonumber\\
\tilde r_{II}(\tilde\chi,\tilde\varsigma) &= \frac{\tilde\chi}{2}\, P_1 \wedge P_3 + \frac{i \tilde\varsigma}{2}\, P_1 \wedge P_2\,;
\end{align}
$r_{III}$ has two inequivalent contraction limits
\begin{align}\label{eq:32.01f}
\tilde r_{III}(\tilde\gamma - \tilde{\bar\gamma}) &= i \frac{\tilde\gamma - \tilde{\bar\gamma}}{2}\, P_1 \wedge P_2\,, \nonumber\\
\hat r_{III}(\hat\gamma + \hat{\bar\gamma},\hat\eta) &= \frac{\hat\gamma + \hat{\bar\gamma}}{2} \left(J_1 \wedge P_2 - J_2 \wedge P_1\right) - \frac{i \hat\eta}{2}\, J_3 \wedge P_3
\end{align}
and for $r_{IV}$ we similarly have
\begin{align}\label{eq:32.01g}
\tilde r_{IV}(\tilde\gamma,\tilde\varsigma) &= i \frac{2\tilde\gamma + \tilde\varsigma}{2}\, P_1 \wedge P_2\,, \nonumber\\
\hat r_{IV}(\hat\gamma) &= i \hat\gamma \left(J_1 \wedge P_1 + J_2 \wedge P_2 + J_3 \wedge P_3\right)\,.
\end{align}
%We again refer the reader to Subsec.~\ref{sec:5.1} for the comparison with the known results of Stachura.

\subsection{Deformed $\mathfrak{o}(3,1)$ contracted to deformed $\mathfrak{o}(2,1) \vartriangleright\!\!< {\cal T}^{2,1}$} \label{sec:4.2a}

Let us now take the real form $\mathfrak{o}(3,1)$ as the $D = 3$ de Sitter algebra. To this end we choose the third spatial axis as the direction associated with the rescaled generators. The corresponding basis $\{J'_p,\tilde K'_p\}$, $p = 1,2,4$ is introduced via the following transformation of the chiral Cartan-Weyl basis
\begin{align}\label{eq:32.02}
H = -\frac{i}{2} \left(J'_4 - {\cal R}\, \tilde K'_4\right)\,, \qquad E_\pm = \frac{1}{2} \left(\mp J'_1 + i J'_2 + {\cal R}\, (\pm \tilde K'_1 - i \tilde K'_2)\right)\,, \nonumber\\
\bar H = \frac{i}{2} \left(J'_4 + {\cal R}\, \tilde K'_4\right)\,, \qquad \bar E_\pm = \frac{1}{2} \left(\pm J'_1 + i J'_2 + {\cal R}\, (\pm \tilde K'_1 + i \tilde K'_2)\right)\,,
\end{align}
%or equivalently
%\begin{align}\label{eq:32.02a}
%M = J_3\,, \qquad N_1 = i K_1\,, \qquad N_2 = i K_2\,, \qquad E = i \sqrt{\Lambda}\, K_3\,, \qquad P_1 = -\sqrt{\Lambda}\, J_2\,, \qquad P_2 = \sqrt{\Lambda}\, J_1\,.
%\end{align}
%The formulae (\ref{eq:32.02}) can obviously be obtained from the Euclidean ones (\ref{eq:31.02}) via the Wick rotation $E \equiv -i P_3$, $N_a \equiv -i M_a$, $M \equiv M_3$.
From (\ref{eq:33.14}) we know that the $\tilde K'_{1,2}$ and $J'_4$ generators satisfy the anti-Hermitian reality conditions, while $J'_{1,2}$ and $\tilde K'_4$ (i.e. the $D = 4$ boost generators $M_{24}$, $M_{41}$ and $M_{43}$) are Hermitian. Defining the physical basis, in which all generators are anti-Hermitian, via the transformation
\begin{align}\label{eq:32.02a}
{\cal J}_0 \equiv J'_4\,, \qquad {\cal J}_a \equiv i J'_a\,, \qquad {\cal L}_0 \equiv -i \tilde K'_4\,, \qquad {\cal L}_a \equiv \tilde K'_a\,,
\end{align}
we can write down familiar brackets of the $\mathfrak{o}(3,1)$ algebra:
\begin{align}\label{eq:32.03}
[{\cal J}_\mu,{\cal J}_\nu] = \epsilon_{\mu\nu}^{\ \ \sigma} {\cal J}_\sigma\,, \qquad [{\cal J}_\mu,{\cal L}_\nu] = \epsilon_{\mu\nu}^{\ \ \sigma} {\cal L}_\sigma\,, \qquad [{\cal L}_\mu,{\cal L}_\nu] = -\Lambda \epsilon_{\mu\nu}^{\ \ \sigma} {\cal J}_\sigma
\end{align}
(assuming the convention $\epsilon_{012} = 1$ and rising indices with the Lorentzian metric $(1,1,-1)$).

When $\mathfrak{o}(3,1)$ is treated as the $D = 3$ de Sitter algebra, the list of allowed $r$-matrices naturally remains the same as in (\ref{eq:32.01b}-\ref{eq:32.01d}) but they are expressed in a different physical basis, namely (\ref{eq:32.02a}). $r_I$ and $r_{II}$ are now written as
\begin{align}\label{eq:32.03a}
r_I(\chi;{\cal R}) &= \chi\, {\cal L}_0 \wedge \left({\cal R}\, {\cal J}_2 + {\cal R}^2 {\cal L}_1\right)\,, \nonumber\\
r_{II}(\chi,\varsigma;{\cal R}) &= -\frac{\chi}{2} \left({\cal J}_0 \wedge \left({\cal J}_1 - {\cal R}\, {\cal L}_2\right) + \left({\cal R}\, {\cal J}_2 + {\cal R}^2 {\cal L}_1\right) \wedge {\cal L}_0\right) + \frac{i \varsigma}{2} \left({\cal J}_1 - {\cal R}\, {\cal L}_2\right) \wedge \left({\cal J}_2 + {\cal R}\, {\cal L}_1\right)\,,
%&+ \frac{i \varsigma}{2} \left({\cal J}_1 \wedge {\cal J}_2 + {\cal R}\, {\cal J}_1 \wedge {\cal L}_1 + {\cal R}\, {\cal J}_2 \wedge {\cal L}_2 + {\cal R}^2 {\cal L}_1 \wedge {\cal L}_2\right)\,,
\end{align}
where $\chi \in i \mathbbm{R}$ and $\varsigma \in \mathbbm{R}$. $r_{III}$ acquires the form
\begin{align}\label{eq:32.04}
r_{III}(\gamma - \bar\gamma,\gamma + \bar\gamma,\eta;{\cal R}) = i \frac{\gamma - \bar\gamma}{2} \left({\cal J}_1 \wedge {\cal J}_2 - {\cal R}^2 {\cal L}_1 \wedge {\cal L}_2\right) + {\cal R}\, \frac{\gamma + \bar\gamma}{2} \left({\cal J}_1 \wedge {\cal L}_2 - {\cal J}_2 \wedge {\cal L}_1\right) + {\cal R}\, \frac{i \eta}{2}\, {\cal J}_0 \wedge {\cal L}_0\,,
\end{align}
where $\gamma - \bar\gamma = 2 {\rm Re}\gamma \in \mathbbm{R}$, $\gamma + \bar\gamma = 2i {\rm Im}\gamma \in i \mathbbm{R}$ and $\eta \in \mathbbm{R}$. Finally, $r_{IV}$ is given by
\begin{align}\label{eq:32.04a}
r_{IV}(\gamma,\varsigma;{\cal R}) = i \gamma \left({\cal J}_1 \wedge {\cal J}_2 - {\cal R}\, {\cal J}_0 \wedge {\cal L}_0 - {\cal R}^2 {\cal L}_1 \wedge {\cal L}_2\right) + \frac{i \varsigma}{2} \left({\cal J}_1 - {\cal R}\, {\cal L}_2\right) \wedge \left({\cal J}_2 + {\cal R}\, {\cal L}_1\right)\,,
%-{\cal R}\, i \gamma\, {\cal J}_0 \wedge {\cal L}_0 + {\cal R}\, \frac{i \varsigma}{2} \left({\cal J}_1 \wedge {\cal L}_1 + {\cal J}_2 \wedge {\cal L}_2\right) + i \frac{2\gamma + \varsigma}{2}\, {\cal J}_1 \wedge {\cal J}_2 - {\cal R}^2 i \frac{2\gamma - \varsigma}{2}\, {\cal L}_1 \wedge {\cal L}_2\,,
\end{align}
where $\gamma,\varsigma \in \mathbbm{R}$.

Quantum IW contractions of these $r$-matrices lead to the quantum $D = 3$ Poincar\'{e} algebras. Such inhomogeneous contraction limits of $r_I$ and $r_{II}$ are
\begin{align}\label{eq:32.04b}
\tilde r_I(\tilde\chi) &= \tilde\chi\, {\cal P}_0 \wedge {\cal P}_1\,, \nonumber\\
\tilde r_{II}(\tilde\chi,\tilde\varsigma) &= \frac{\tilde\chi}{2}\, {\cal P}_0 \wedge {\cal P}_1 + \frac{i \tilde\varsigma}{2}\, {\cal P}_1 \wedge {\cal P}_2\,,
\end{align}
where ${\cal P}_\mu$ denote the Lorentzian 3-momenta; $r_{III}$ has the following contraction limits
\begin{align}\label{eq:32.04c}
\tilde r_{III}(\tilde\gamma - \tilde{\bar\gamma}) &= -i \frac{\tilde\gamma - \tilde{\bar\gamma}}{2}\, {\cal P}_1 \wedge {\cal P}_2\,, \nonumber\\
\hat r_{III}(\hat\gamma + \hat{\bar\gamma},\hat\eta) &= \frac{\hat\gamma + \hat{\bar\gamma}}{2} \left({\cal J}_1 \wedge {\cal P}_2 - {\cal J}_2 \wedge {\cal P}_1\right) + \frac{i \hat\eta}{2}\, {\cal J}_0 \wedge {\cal P}_0
\end{align}
and for $r_{IV}$ we have
\begin{align}\label{eq:32.04d}
\tilde r_{IV}(\tilde\gamma,\tilde\varsigma) &= -i \frac{2\tilde\gamma - \tilde\varsigma}{2}\, {\cal P}_1 \wedge {\cal P}_2\,, \nonumber\\
\hat r_{IV}(\hat\gamma) &= -i \hat\gamma \left({\cal J}_0 \wedge {\cal P}_0 - {\cal J}_1 \wedge {\cal P}_1 - {\cal J}_2 \wedge {\cal P}_2\right)\,,
\end{align}
where $\hat r_{IV}$ is obtained for $\varsigma = 2\gamma$ instead of $\varsigma = -2\gamma$, as it was the case in (\ref{eq:32.01g}).
%Furthermore, let us observe that $\hat r_{III}$ in (\ref{eq:31.04a}) can be obtained from $\hat r_{III}$ in (\ref{eq:32.04c}) by the Wick rotation $P_3 = i {\cal P}_0$, $J_a = i {\cal J}_a$, accompanied by the replacement $\hat\gamma \mapsto i \hat\gamma$, $\hat\eta \mapsto -i \hat\eta$.

Furthermore, it turns out that the quantum IW contraction of $\mathfrak{o}(3,1)$ $r$-matrices along the first or second spatial axis leads to the set of $r$-matrices different than the one obtained for the third axis. This is because the form of considered $r$-matrices changes under the ${\cal R}$-rescaled $\mathfrak{o}(3,1)$ automorphisms. We may restrict most of the discussion to the contractions along the first axis (results for the second axis differ only by automorphisms, except in the cases of $r_I$ and $r_{II}$, as we will mention), introducing another anti-Hermitian physical basis
\begin{align}\label{eq:32.04e}
{\cal J}_0 \equiv M_{23}\,, \qquad {\cal J}_{1/2} \equiv -i M_{34/42}\,, \qquad {\cal L}_0 \equiv i {\cal R}^{-1} M_{41}\,, \qquad {\cal L}_{1/2} \equiv {\cal R}^{-1} M_{21/31}\,,
\end{align}
in which the $\mathfrak{o}(3,1)$ brackets preserve their form (\ref{eq:32.03}). Then instead of (\ref{eq:32.03a}-\ref{eq:32.04a}) we obtain the $r$-matrices
\begin{align}\label{eq:32.04f}
r_I^a(\chi;{\cal R}) &= -{\cal R}\, \chi\, {\cal J}_1 \wedge \left({\cal L}_0 - {\cal L}_2\right)\,, \nonumber\\
r_{II}^a(\chi,\varsigma;{\cal R}) &= {\cal R}\, \frac{\chi}{2} \left(({\cal J}_0 - {\cal J}_2) \wedge {\cal L}_1 - {\cal J}_1 \wedge ({\cal L}_0 - {\cal L}_2)\right) - {\cal R}\, \frac{i \varsigma}{2} \left({\cal J}_0 - {\cal J}_2\right) \wedge \left({\cal L}_0 - {\cal L}_2\right)\,, \nonumber\\
r_{III}^a(\gamma - \bar\gamma,\gamma + \bar\gamma,\eta;{\cal R}) &= -{\cal R}\, i \frac{\gamma - \bar\gamma}{2} \left({\cal J}_0 \wedge {\cal L}_2 - {\cal J}_2 \wedge {\cal L}_0\right) - \frac{\gamma + \bar\gamma}{2} \left({\cal J}_0 \wedge {\cal J}_2 - {\cal R}^2 {\cal L}_0 \wedge {\cal L}_2\right) - {\cal R}\, \frac{i \eta}{2}\, {\cal J}_1 \wedge {\cal L}_1\,, \nonumber\\
r_{IV}^a(\gamma,\varsigma;{\cal R}) &= -{\cal R}\, i \gamma \left({\cal J}_0 \wedge {\cal L}_2 - {\cal J}_2 \wedge {\cal L}_0 - {\cal J}_1 \wedge {\cal L}_1\right) - {\cal R}\, \frac{i \varsigma}{2} \left({\cal J}_0 - {\cal J}_2\right) \wedge \left({\cal L}_0 - {\cal L}_2\right)\,.
\end{align}
Although they are equivalent to (\ref{eq:32.03a}-\ref{eq:32.04a}) under the $\mathfrak{o}(3,1)$ automorphism $({\cal J}_{0/1} \mapsto {\cal R}\, {\cal L}_{2/0},{\cal J}_2 \mapsto -{\cal J}_1,{\cal L}_{0/1} \mapsto -{\cal R}^{-1} {\cal J}_{2/0},{\cal L}_2 \mapsto -{\cal L}_1)$, the expressions in (\ref{eq:32.04f}) are rescaled by ${\cal R}$ in a different way. %However, taking the ${\cal R} \rightarrow \infty$ limit of (\ref{eq:32.04f}) leads to results inequivalent to (\ref{eq:32.04b}-\ref{eq:32.04d}).

Performing the rescaling of deformation parameters as it was done for $\tilde r_I$-$\tilde r_{IV}$ in (\ref{eq:32.04b}-\ref{eq:32.04d}),
%\begin{align}\label{eq:32.04f}
%\tilde\gamma \equiv {\cal R}^2 \gamma\,, \qquad \tilde\eta \equiv {\cal R}^2 \eta\,,
%\end{align}
we find that (\ref{eq:32.04f}) have only one contraction limit (equivalent to $\tilde r_{III}$ in (\ref{eq:32.04c}))
\begin{align}\label{eq:32.04g}
\tilde r^a_{III}(\tilde\gamma + \tilde{\bar\gamma}) = \frac{\tilde\gamma + \tilde{\bar\gamma}}{2}\, {\cal P}_0 \wedge {\cal P}_2\,.
\end{align}
On the other hand, the alternative rescaling (under the condition $\gamma + \bar\gamma = 2i {\rm Im}\gamma = 0$ in the case of $r_{III}$)
\begin{align}\label{eq:32.04h}
\hat\chi \equiv {\cal R}\, \chi\,, \qquad \hat\varsigma \equiv {\cal R}\, \varsigma\,, \qquad \hat\gamma \equiv {\cal R}\, \gamma\,, \qquad \hat\eta \equiv {\cal R}\, \eta
\end{align}
leads to the following set of new contraction limits
\begin{align}\label{eq:32.04i}
\hat r^a_I(\hat\chi) &= -\hat\chi\, {\cal J}_1 \wedge \left({\cal P}_0 - {\cal P}_2\right)\,, \nonumber\\
\hat r^a_{II}(\hat\chi,\hat\varsigma) &= \frac{\hat\chi}{2} \left(({\cal J}_0 - {\cal J}_2) \wedge {\cal P}_1 - {\cal J}_1 \wedge ({\cal P}_0 - {\cal P}_2)\right) - \frac{i \hat\varsigma}{2} \left({\cal J}_0 - {\cal J}_2\right) \wedge \left({\cal P}_0 - {\cal P}_2\right)\,, \nonumber\\
\hat r^a_{III}(\hat\gamma - \hat{\bar\gamma},\hat\eta) &= -i \frac{\hat\gamma - \hat{\bar\gamma}}{2} \left({\cal J}_0 \wedge {\cal P}_2 - {\cal J}_2 \wedge {\cal P}_0\right) - \frac{i \hat\eta}{2}\, {\cal J}_1 \wedge {\cal P}_1\,, \nonumber\\
\hat r^a_{IV}(\hat\gamma,\hat\varsigma) &= -i \hat\gamma \left({\cal J}_0 \wedge {\cal P}_2 - {\cal J}_2 \wedge {\cal P}_0 - {\cal J}_1 \wedge {\cal P}_1\right) - \frac{i \hat\varsigma}{2} \left({\cal J}_0 - {\cal J}_2\right) \wedge \left({\cal P}_0 - {\cal P}_2\right)\,.
\end{align}
The quantum IW contraction along the second axis can be performed in the analogous way but $r_I$ will now be expressed only in terms of the ${\cal J}_\mu$ generators and therefore both before and after the contraction (without the necessity of rescaling $\chi$!) we have
\begin{align}\label{eq:32.04j}
r^b_I(\chi) &= \chi \left({\cal J}_0 - {\cal J}_1\right) \wedge {\cal J}_2\,.
\end{align}
Moreover, in this case $\hat r_{II}^b$ does not exist %(since $r_{II}^b$ with $\bar\chi = -\chi$ is forbidden for $\mathfrak{o}(3,1)$) 
but there exists $\tilde r_{II}^b$, equivalent to $\tilde r_{II}$ in (\ref{eq:32.04b}).

%All contraction limits obtained in this Subsection will be compared with the Stachura classification of $r$-matrices for the $D = 3$ Poincar\'{e} algebra in Subsec.~\ref{sec:5.2}.

\subsection{Deformed $\mathfrak{o}(2,2)$ contracted to deformed $\mathfrak{o}(2,1) \vartriangleright\!\!< {\cal T}^{2,1}$} \label{sec:4.3}

Finally, let us investigate the (Kleinian) rotation algebra $\mathfrak{o}(2,2)$, i.e. the $D = 3$ anti-de Sitter algebra, with $\Lambda = -{\cal R}^{-2} < 0$. The corresponding reality conditions (\ref{eq:33.13}) are in agreement with the spacetime metric $(1,-1,1,-1)$ (as discussed in Subsec.~\ref{sec:3.3}). The IW contraction of $\mathfrak{o}(2,2)$ leads to the Poincar\'{e} algebra but in principle there are two distinct possibilities: the contraction can be performed either along a timelike (e.g. the fourth) axis or spacelike (e.g. the third) axis, giving us the $\mathfrak{o}(2,1)$ or $\mathfrak{o}(1,2)$ algebra, respectively. In the absence of a deformation they differ only by a trivial change of the metric signature. As we will show, for deformed algebras it is actually sufficient to consider quantum IW contractions along the fourth and second axis, which lead to deformed Poincar\'{e} algebras with the metric $(-1,1,1)$, as it is also the case in our convention for $\mathfrak{o}(3,1)$ contracted along a spacelike axis.

In the Cartan-Weyl basis, the $\mathfrak{o}(2,2)$ algebra can arise as one of three different real forms of $\mathfrak{o}(4;\mathbbm{C})$, which were presented in (\ref{eq:33.07}-\ref{eq:33.09}). Let us first consider the set of reality conditions (\ref{eq:33.07}) (of the real form denoted as $\mathfrak{o}''(2,2)$ in \cite{Borowiec:2017bs} but in this paper as $\dot{\mathfrak{o}}(2,2)$). We first calculate quantum IW contractions along the fourth axis. For the third axis the results differ only by automorphisms and a change of the metric signature (corresponding to the multiplication of all generators by $-1$). The relation between the chiral Cartan-Weyl basis and the basis $\{J_i,\tilde K_i\}$ has again the form (\ref{eq:31.02}), namely
\begin{align}\label{eq:33.02}
H = -\frac{i}{2} \left(J_3 + {\cal R}\, \tilde K_3\right)\,, \qquad E_\pm = \frac{1}{2} \left(-i J_1 \mp J_2 - {\cal R}\, (i \tilde K_1 \pm \tilde K_2)\right)\,, \nonumber\\
\bar H = \frac{i}{2} \left(J_3 - {\cal R}\, \tilde K_3\right)\,, \qquad \bar E_\pm = \frac{1}{2} \left(i J_1 \mp J_2 - {\cal R}\, (i \tilde K_1 \mp \tilde K_2)\right)\,,
\end{align}
The reality conditions (\ref{eq:33.07}) determine that $J_2$ and $\tilde K_2$ are anti-Hermitian generators, while $J_{1/3}$, $\tilde K_{1/3}$ are Hermitian. The physical basis, in which all generators are anti-Hermitian, is now defined via the transformation\footnote{For the IW rescaling along the third axis it would be instead
\begin{align}
{\cal J}_0 \equiv J'_1\,, \qquad {\cal J}_{1/2} \equiv \pm i J'_{2/3}\,, \qquad {\cal L}_0 \equiv \tilde K'_1\,, \qquad {\cal L}_{1/2} \equiv \pm i \tilde K'_{2/3}\,, \nonumber
\end{align}
where $\pm$ in the formulae for ${\cal J}_{1/2}$ and ${\cal L}_{1/2}$ allows to recover the $(1,-1,-1)$ signature.}
\begin{align}\label{eq:33.02a}
{\cal J}_0 \equiv J_2\,, \qquad {\cal J}_{1/2} \equiv i J_{3/1}\,, \qquad {\cal L}_0 \equiv \tilde K_2\,, \qquad {\cal L}_{1/2} \equiv i \tilde K_{3/1}\,.
\end{align}
The undeformed brackets of the $\mathfrak{o}(2,2)$ algebra in the $\{{\cal J}_\mu,{\cal L}_\mu\}$ basis are identical to (\ref{eq:32.03}) (but with $\Lambda < 0$), irrespective whether we consider $\dot{\mathfrak{o}}(2,2)$, $\mathfrak{o}'(2,2)$ or $\dot{\mathfrak{o}}'(2,2)$. As expected, $\{{\cal J}_\mu\}$ after the IW contraction will consist of one rotation and two boost generators.

It has been shown (cf. Table I) that $\dot{\mathfrak{o}}(2,2)$ is the only real form of $\mathfrak{o}(4;\mathbbm{C})$ that inherits all possible Hopf-algebraic deformations of the latter, given by the $r$-matrices $r_I$, $r_{II}$, $r_{III}$, $r_{IV}$ and $r_V$. Furthermore, in this case all deformation parameters are imaginary, $\chi, \bar\chi, \varsigma, \gamma, \bar\gamma, \eta, \rho \in i \mathbbm{R}$. The first two $r$-matrices in the physical basis (\ref{eq:33.02a}) become
\begin{align}\label{eq:33.03a}
r_I(\chi;{\cal R}) &= -\chi\, {\cal L}_1 \wedge \left({\cal R}\, {\cal J}_0 + {\cal R}^2 {\cal L}_2\right)\,, \nonumber\\
r_{II}(\chi,\bar\chi,\varsigma;{\cal R}) &= \frac{\chi + \bar\chi}{4} \left(\left({\cal R}\, {\cal J}_0 + {\cal R}^2 {\cal L}_2\right) \wedge {\cal L}_1 - {\cal J}_1 \wedge \left({\cal J}_2 + {\cal R}\, {\cal L}_0\right)\right) \nonumber\\
&+ \frac{\chi - \bar\chi}{4} \left(\left({\cal R}\, {\cal J}_2 + {\cal R}^2 {\cal L}_0\right) \wedge {\cal L}_1 - {\cal J}_1 \wedge \left({\cal J}_0 + {\cal R}\, {\cal L}_2\right)\right) - \frac{\varsigma}{2} \left({\cal J}_0 + {\cal R}\, {\cal L}_2\right) \wedge \left({\cal J}_2 + {\cal R}\, {\cal L}_0\right)\,,
%-\left(\frac{\chi + \bar\chi}{4}\, {\cal J}_1 + {\cal R}\, \frac{\chi - \bar\chi}{4}\, {\cal L}_1\right) \wedge \left({\cal J}_2 + {\cal R}\, {\cal L}_0\right) \nonumber\\
%&- \left(\frac{\chi - \bar\chi}{4}\, {\cal J}_1 + {\cal R}\, \frac{\chi + \bar\chi}{4}\, {\cal L}_1\right) \wedge \left({\cal J}_0 + {\cal R}\, {\cal L}_2\right) \nonumber\\
\end{align}
and the remaining three are
\begin{align}\label{eq:33.04}
r_{III}(\gamma,\bar\gamma,\eta;{\cal R}) &= \frac{\gamma - \bar\gamma}{2} \left({\cal J}_0 \wedge {\cal J}_2 + {\cal R}^2 {\cal L}_0 \wedge {\cal L}_2\right) + {\cal R}\, \frac{\gamma + \bar\gamma}{2} \left({\cal J}_0 \wedge {\cal L}_2 - {\cal J}_2 \wedge {\cal L}_0\right) + {\cal R}\, \frac{\eta}{2}\, {\cal J}_1 \wedge {\cal L}_1\,, \nonumber\\
r_{IV}(\gamma,\varsigma;{\cal R}) &= \gamma \left({\cal J}_0 \wedge {\cal J}_2 - {\cal R}\, {\cal J}_1 \wedge {\cal L}_1 + {\cal R}^2 {\cal L}_0 \wedge {\cal L}_2\right) - \frac{\varsigma}{2} \left({\cal J}_0 + {\cal R}\, {\cal L}_2\right) \wedge \left({\cal J}_2 + {\cal R}\, {\cal L}_0\right)\,, \nonumber\\
%&-{\cal R}\, \gamma\, {\cal J}_1 \wedge {\cal L}_1 - {\cal R}\, \frac{\varsigma}{2} \left({\cal J}_0 \wedge {\cal L}_0 - {\cal J}_2 \wedge {\cal L}_2\right) + \frac{2\gamma - \varsigma}{2}\, {\cal J}_0 \wedge {\cal J}_2 + {\cal R}^2 \frac{2\gamma + \varsigma}{2}\, {\cal L}_0 \wedge {\cal L}_2\,, \nonumber\\
r_V(\gamma,\bar\chi,\rho;{\cal R}) &= \frac{\gamma}{2} \left({\cal J}_0 + {\cal R}\, {\cal L}_0\right) \wedge \left({\cal J}_2 + {\cal R}\, {\cal L}_2\right) + \left(\frac{\bar\chi + \rho}{4}\, {\cal J}_1 - {\cal R}\, \frac{\bar\chi - \rho}{4}\, {\cal L}_1\right) \wedge \left({\cal J}_0 - {\cal J}_2 - {\cal R}\, ({\cal L}_0 - {\cal L}_2)\right)\,.
%&+ \frac{\gamma}{2} \left({\cal J}_0 \wedge {\cal J}_2 - {\cal R}\, {\cal J}_0 \wedge {\cal L}_2 + {\cal R}\, {\cal J}_2 \wedge {\cal L}_0 + {\cal R}^2 {\cal L}_0 \wedge {\cal L}_2\right)\,.
\end{align}
Sec.~\ref{sec:2.0} once again shows us what are the possible quantum IW contractions of these $r$-matrices. Namely, $r_I$ and $r_{II}$ have the following inhomogeneous contraction limits
\begin{align}\label{eq:33.04b}
\tilde r_I(\tilde\chi) &= -\tilde\chi\, {\cal P}_1 \wedge {\cal P}_2\,, \nonumber\\
\tilde r_{II}(\tilde\chi,\tilde{\bar\chi},\tilde\varsigma) &= -\frac{\tilde\chi + \tilde{\bar\chi}}{4}\, {\cal P}_1 \wedge {\cal P}_2 + \frac{\tilde\chi - \tilde{\bar\chi}}{4}\, {\cal P}_0 \wedge {\cal P}_1 + \frac{\tilde\varsigma}{2}\, {\cal P}_0 \wedge {\cal P}_2\,;
\end{align}
Each of the remaining $r$-matrices has two independent contraction limits. $r_{III}$ leads to
\begin{align}\label{eq:33.04c}
\tilde r_{III}(\tilde\gamma,\tilde{\bar\gamma}) &= \frac{\tilde\gamma - \tilde{\bar\gamma}}{2}\, {\cal P}_0 \wedge {\cal P}_2\,, \nonumber\\
\hat r_{III}(\hat\gamma,\hat\eta) &= \hat\gamma \left({\cal J}_0 \wedge {\cal P}_2 - {\cal J}_2 \wedge {\cal P}_0\right) + \frac{\hat\eta}{2}\, {\cal J}_1 \wedge {\cal P}_1\,,
\end{align}
$r_{IV}$ to
\begin{align}\label{eq:33.04d}
\tilde r_{IV}(\tilde\gamma,\tilde\varsigma) &= \frac{2\tilde\gamma + \tilde\varsigma}{2}\, {\cal P}_0 \wedge {\cal P}_2\,, \nonumber\\
\hat r_{IV}(\hat\gamma) &= \hat\gamma \left({\cal J}_0 \wedge {\cal P}_0 - {\cal J}_1 \wedge {\cal P}_1 - {\cal J}_2 \wedge {\cal P}_2\right)
\end{align}
and $r_V$ to
\begin{align}\label{eq:33.04e}
\tilde r_V(\tilde\gamma,\tilde{\bar\chi},\tilde\rho) &= \frac{\tilde\gamma}{2}\, {\cal P}_0 \wedge {\cal P}_2 - \frac{\tilde{\bar\chi} - \tilde\rho}{4}\, ({\cal P}_0 - {\cal P}_2) \wedge {\cal P}_1\,, \nonumber\\
\hat r_V(\hat{\bar\chi}) &= -\frac{\hat{\bar\chi}}{2}\, {\cal J}_1 \wedge ({\cal P}_0 - {\cal P}_2)\,.
\end{align}
The only subtlety for contractions along the third axis is that $\hat r_{IV}$ is then obtained under the condition $\varsigma = 2\gamma$.
%The Euclidean counterparts (\ref{eq:32.01e}-\ref{eq:32.01g}) of $\tilde r_I$, $\hat r_{III}$, $\tilde r_{IV}$ and $\hat r_{IV}$ can be straightforwardly obtained by applying to them the Wick rotation $E \equiv -i P_3$, $N_a \equiv -i M_a$, $M \equiv M_3$ together with the maps $\hat\eta \mapsto -i \hat\eta$, $\tilde\varsigma \mapsto i \tilde\varsigma$, $\tilde\gamma \mapsto -i \tilde\gamma$ and $\hat\gamma \mapsto -i \hat\gamma$ (the last one only for $\hat r_{IV}$).

Similarly as it is the case for $\mathfrak{o}(3,1)$, the quantum IW contraction of $\dot{\mathfrak{o}}(2,2)$ $r$-matrices along the first or second spatial axis leads to a different set of $r$-matrices than above. We restrict to the contraction along the second axis (results for the first axis differ only by automorphisms and a change of the metric signature, except the case of $r_I$, as we will mention), introducing the anti-Hermitian physical basis%\footnote{For the IW rescaling along the first axis it would be instead
%\begin{align}
%{\cal J}_0 \equiv M_{42}\,, \qquad {\cal J}_{1/2} \equiv i M_{23/43}\,, \qquad {\cal L}_0 \equiv {\cal R}^{-1} M_{31}\,, \qquad {\cal L}_{1/2} \equiv i {\cal R}^{-1} M_{41/12}\,. \nonumber
%\end{align}}
\begin{align}\label{eq:33.04f}
{\cal J}_0 \equiv M_{13}\,, \qquad {\cal J}_{1/2} \equiv i M_{34/41}\,, \qquad {\cal L}_0 \equiv {\cal R}^{-1} M_{42}\,, \qquad {\cal L}_{1/2} \equiv i {\cal R}^{-1} M_{12/32}\,,
\end{align}
in which the $\mathfrak{o}(2,2)$ brackets (\ref{eq:32.03}) (with $\Lambda < 0$) are preserved. While the form of $r_{III}$ and $r_V$ now remains the same as in (\ref{eq:33.03a}-\ref{eq:33.04}) apart from some irrelevant sign changes, instead of $r_{II}$ and $r_{IV}$ we obtain the following $r$-matrices
\begin{align}\label{eq:33.04g}
r_{II}^a(\chi,\bar\chi,\varsigma;{\cal R}) &= -\frac{\chi + \bar\chi}{4} \left(\left({\cal J}_0 + {\cal J}_2\right) \wedge {\cal J}_1 - {\cal R}^2 {\cal L}_1 \wedge \left({\cal L}_0 + {\cal L}_2\right)\right) \nonumber\\
&- {\cal R}\, \frac{\chi - \bar\chi}{4} \left(\left({\cal J}_0 + {\cal J}_2\right) \wedge {\cal L}_1 - {\cal J}_1 \wedge \left({\cal L}_0 + {\cal L}_2\right)\right) - {\cal R}\, \frac{\varsigma}{2} \left({\cal J}_0 + {\cal J}_2\right) \wedge \left({\cal L}_0 + {\cal L}_2\right)\,, \nonumber\\
%r_{III}^a(\gamma,\bar\gamma,\eta;{\cal R}) &= {\cal R}\, \frac{\gamma - \bar\gamma}{2} \left({\cal J}_0 \wedge {\cal L}_2 - {\cal J}_2 \wedge {\cal L}_0\right) + \frac{\gamma + \bar\gamma}{2} \left({\cal J}_0 \wedge {\cal J}_2 + {\cal R}^2 {\cal L}_0 \wedge {\cal L}_2\right) - {\cal R}\, \frac{\eta}{2}\, {\cal J}_1 \wedge {\cal L}_1\,, \nonumber\\
r_{IV}^a(\gamma,\varsigma;{\cal R}) &= {\cal R}\, \gamma \left({\cal J}_0 \wedge {\cal L}_2 - {\cal J}_2 \wedge {\cal L}_0 + {\cal J}_1 \wedge {\cal L}_1\right) - {\cal R}\, \frac{\varsigma}{2} \left({\cal J}_0 + {\cal J}_2\right) \wedge \left({\cal L}_0 + {\cal L}_2\right)\,. %\nonumber\\
%r_V^a(\gamma,\bar\chi,\rho;{\cal R}) &= \frac{\gamma}{2} \left({\cal J}_0 + {\cal R}\, {\cal L}_0\right) \wedge \left({\cal J}_2 + {\cal R}\, {\cal L}_2\right) + \left(\frac{\bar\chi - \rho}{4}\, {\cal J}_1 - {\cal R}\, \frac{\bar\chi + \rho}{4}\, {\cal L}_1\right) \wedge \left({\cal J}_0 + {\cal J}_2 - {\cal R}\, ({\cal L}_0 + {\cal L}_2)\right)\,.
\end{align}
%$r_{III}^a$ becomes identical to $r_{III}$ after the sign reversal of parameters $\bar\gamma \mapsto -\bar\gamma$, $\eta \mapsto -\eta$. 
They are equivalent to $r_{II}$ and $r_{IV}$ under the $\mathfrak{o}(2,2)$ automorphism $({\cal J}_0 \mapsto -{\cal J}_0,{\cal J}_{1/2} \mapsto \pm {\cal R}\, {\cal L}_{1/2},{\cal L}_0 \mapsto -{\cal L}_0,{\cal L}_{1/2} \mapsto \pm {\cal R}^{-1} {\cal J}_{1/2})$ but the expressions in (\ref{eq:33.04g}) are rescaled by ${\cal R}$ in a different way. %However, taking the ${\cal R} \rightarrow \infty$ limit of (\ref{eq:33.04g}) leads to some results inequivalent to (\ref{eq:33.04b}-\ref{eq:33.04e}).

If we perform the rescaling of deformation parameters as it was done for $\tilde r_{II}$ and $\tilde r_{IV}$ in (\ref{eq:33.04b},\ref{eq:33.04e}),
%\begin{align}\label{eq:33.04g}
%\tilde\chi \equiv |\Lambda|^{-1} \chi\,, \qquad \tilde{\bar\chi} \equiv |\Lambda|^{-1} \bar\chi\,, \qquad \tilde\gamma \equiv |\Lambda|^{-1} \gamma\,, \qquad \tilde{\bar\gamma} \equiv |\Lambda|^{-1} \bar\gamma\,, \qquad \tilde\eta \equiv |\Lambda|^{-1} \eta\,, \qquad \tilde\rho \equiv |\Lambda|^{-1} \rho\,,
%\end{align}
we find that (\ref{eq:33.04g}) have one non-vanishing contraction limit %($\tilde r^a_{III}$ is equivalent to $\tilde r_{III}$ and $\tilde r^a_V$ to $\tilde r_V$)
\begin{align}\label{eq:33.04h}
\tilde r^a_{II}(\tilde\chi,\tilde{\bar\chi}) &= -\frac{\tilde\chi + \tilde{\bar\chi}}{4} \left({\cal P}_0 + {\cal P}_2\right) \wedge {\cal P}_1\,, %\nonumber\\
%\tilde r^a_{III}(\tilde\gamma,\tilde{\bar\gamma}) &= \frac{\tilde\gamma + \tilde{\bar\gamma}}{2}\, {\cal P}_0 \wedge {\cal P}_2\,, \nonumber\\
%\tilde r^a_V(\tilde\gamma,\tilde{\bar\chi},\tilde\rho) &= \frac{\tilde\gamma}{2}\, {\cal P}_0 \wedge {\cal P}_2 - \frac{\tilde{\bar\chi} + \tilde\rho}{4}\, \left({\cal P}_0 + {\cal P}_2\right) \wedge {\cal P}_1\,.
\end{align}
The alternative rescaling of parameters, performed under the condition $\bar\chi = -\chi$ for $r_{II}$, %$\bar\gamma = -\gamma$ for $r_{III}$ and $\gamma = 0$, $\rho = -\bar\chi$ for $r_V$,
\begin{align}\label{eq:33.04i}
\hat\chi \equiv {\cal R}\, \chi\,, \qquad \hat\varsigma \equiv {\cal R}\, \varsigma\,, \qquad \hat\gamma \equiv {\cal R}\, \gamma\,, %\qquad \hat\eta \equiv {\cal R}\, \eta\,, \qquad \hat{\bar\chi} \equiv {\cal R}\, \bar\chi\,,
\end{align}
leads to the additional two contraction limits %($\hat r^a_{III}$ is equivalent to $\hat r_{III}$ and $\hat r^a_V$ to $\hat r_V$)
\begin{align}\label{eq:33.04j}
\hat r^a_{II}(\hat\chi,\hat\varsigma) &= -\frac{\hat\chi}{2} \left(({\cal J}_0 + {\cal J}_2) \wedge {\cal P}_1 - {\cal J}_1 \wedge ({\cal P}_0 + {\cal P}_2)\right) - \frac{\hat\varsigma}{2} \left({\cal J}_0 + {\cal J}_2\right) \wedge \left({\cal P}_0 + {\cal P}_2\right)\,, \nonumber\\
%\hat r^a_{III}(\hat\gamma,\hat\eta) &= \hat\gamma \left({\cal J}_0 \wedge {\cal P}_2 - {\cal J}_2 \wedge {\cal P}_0\right) - \frac{\hat\eta}{2}\, {\cal J}_1 \wedge {\cal P}_1\,, \nonumber\\
\hat r^a_{IV}(\hat\gamma,\hat\varsigma) &= \hat\gamma \left({\cal J}_0 \wedge {\cal P}_2 - {\cal J}_2 \wedge {\cal P}_0 + {\cal J}_1 \wedge {\cal P}_1\right) - \frac{\hat\varsigma}{2} \left({\cal J}_0 + {\cal J}_2\right) \wedge \left({\cal P}_0 + {\cal P}_2\right)\,. %\nonumber\\
%\hat r^a_V(\hat{\bar\chi}) &= -\frac{\hat{\bar\chi}}{2}\, {\cal J}_1 \wedge \left({\cal P}_0 + {\cal P}_2\right)\,.
\end{align}
Finally, $r_I$ in the considered basis (\ref{eq:33.04f}) is expressed only in terms of the ${\cal J}_\mu$ generators and both before and after the (quantum) IW contraction is given by
\begin{align}\label{eq:33.04k}
r^a_I(\chi) = -\chi \left({\cal J}_0 + {\cal J}_2\right) \wedge {\cal J}_1\,.
\end{align}
The quantum IW contractions of $r_I$ along the first axis lead instead to $r$-matrices equivalent to (\ref{eq:33.04e}).\newline

The second real form of $\mathfrak{o}(4;\mathbbm{C})$ that corresponds to the Kleinian algebra $\mathfrak{o}(2,2)$ is $\mathfrak{o}'(2,2)$ (in \cite{Borowiec:2017bs} simply denoted as $\mathfrak{o}(2,2)$), specified by the reality conditions (\ref{eq:33.08}). In this case the transformation from the chiral Cartan-Weyl basis to the orthogonal basis rescaled along the fourth axis is given by (cf. (\ref{eq:30.01a}))
\begin{align}\label{eq:35.02}
H = -\frac{i}{2} \left(J_2 + {\cal R}\, \tilde K_2\right)\,, \qquad E_\pm = \frac{1}{2} \left(-i J_1 \pm J_3 - {\cal R}\, (i \tilde K_1 \mp \tilde K_3)\right)\,, \nonumber\\
\bar H = \frac{i}{2} \left(J_2 - {\cal R}\, \tilde K_2\right)\,, \qquad \bar E_\pm = \frac{1}{2} \left(i J_1 \pm J_3 - {\cal R}\, (i \tilde K_1 \pm \tilde K_3)\right)\,.
\end{align}

The only possible Hopf-algebraic deformation of $\mathfrak{o}'(2,2)$ is associated with the $r$-matrix $r_{III}$ (\cite{Borowiec:2017ag}). In the physical basis (\ref{eq:33.02a}) introduced for (\ref{eq:35.02}) it now becomes
\begin{align}\label{eq:35.03}
r_{III}(\gamma,\bar\gamma,\eta;{\cal R}) = i \frac{\gamma - \bar\gamma}{2} \left({\cal J}_1 \wedge {\cal J}_2 + {\cal R}^2 {\cal L}_1 \wedge {\cal L}_2\right) + {\cal R}\, i \frac{\gamma + \bar\gamma}{2} \left({\cal J}_1 \wedge {\cal L}_2 - {\cal J}_2 \wedge {\cal L}_1\right) - {\cal R}\, \frac{\eta}{2}\, {\cal J}_0 \wedge {\cal L}_0\,,
\end{align}
where $\gamma, \bar\gamma \in \mathbbm{R}$ and $\eta \in i \mathbbm{R}$. Sec.~\ref{sec:2.0} shows us that two $D = 3$ Poincar\'{e} $r$-matrices obtained via the quantum IW contraction of (\ref{eq:35.03}) are
\begin{align}\label{eq:35.04}
\tilde r_{III}(\tilde\gamma,\tilde{\bar\gamma}) &= i \frac{\tilde\gamma - \tilde{\bar\gamma}}{2}\, {\cal P}_1 \wedge {\cal P}_2, \nonumber\\
\hat r_{III}(\hat\gamma,\hat\eta) &= i \hat\gamma \left({\cal J}_1 \wedge {\cal P}_2 - {\cal J}_2 \wedge {\cal P}_1\right) - \frac{\hat\eta}{2}\, {\cal J}_0 \wedge {\cal P}_0\,.
\end{align}
Furthermore, if we choose instead the second axis, using the anti-Hermitian basis (\ref{eq:33.04e}) introduced for (\ref{eq:35.02}), we still obtain the same contraction limits as (\ref{eq:35.04}) (up to some signs, which can be changed via automorphisms). %i.e.
%\begin{align}\label{eq:35.05}
%\tilde r^a_{III}(\tilde\gamma,\tilde{\bar\gamma}) &= -i \frac{\tilde\gamma - \tilde{\bar\gamma}}{2}\, {\cal P}_1 \wedge {\cal P}_2\,, \nonumber\\
%\hat r^a_{III}(\hat\gamma,\hat\eta) &= -i \hat\gamma \left({\cal J}_1 \wedge {\cal P}_2 - {\cal J}_2 \wedge {\cal P}_1\right) - \frac{\hat\eta}{2}\, {\cal J}_0 \wedge {\cal P}_0\,.
%\end{align}
These results are also equivalent to what is obtained for the third or first axis, although in both these cases the $r$-matrix $\hat r_{III}$/$\hat r^a_{III}$ requires satisfying the relation $\bar\gamma = -\gamma$ instead of $\bar\gamma = \gamma$. \newline

The last pseudo-orthogonal real form of $\mathfrak{o}(4;\mathbbm{C})$ is $\dot{\mathfrak{o}}'(2,2)$ (in \cite{Borowiec:2017bs} denoted as $\mathfrak{o}'(2,2)$), characterized by the reality conditions (\ref{eq:33.09}). The transformation from the chiral Cartan-Weyl basis is introduced by (cf. (\ref{eq:30.01b}))
\begin{align}\label{eq:36.02}
H = -\frac{i}{2} \left(J_2 + {\cal R}\, \tilde K_2\right)\,, \qquad E_\pm = \frac{1}{2} \left(-i J_1 \pm J_3 - {\cal R}\, (i \tilde K_1 \mp \tilde K_3)\right)\,, \nonumber\\
\bar H = \frac{i}{2} \left(J_3 - {\cal R}\, \tilde K_3\right)\,, \qquad \bar E_\pm = \frac{1}{2} \left(i J_1 \mp J_2 - {\cal R}\, (i \tilde K_1 \mp \tilde K_2)\right)\,.
\end{align}

$\dot{\mathfrak{o}}'(2,2)$ has two possible Hopf-algebraic deformations, given by the $r$-matrices $r_{III}$ and $r_V$ \cite{Borowiec:2017ag}. In the physical basis (\ref{eq:33.02a}) introduced for (\ref{eq:36.02}) they acquire the following form
\begin{align}\label{eq:36.03}
r_{III}(\gamma,\bar\gamma,\eta;{\cal R}) &= \frac{i \gamma}{2} \left({\cal J}_1 + {\cal R}\, {\cal L}_1\right) \wedge \left({\cal J}_2 + {\cal R}\, {\cal L}_2\right) - \frac{\bar\gamma}{2} \left({\cal J}_0 - {\cal R}\, {\cal L}_0\right) \wedge \left({\cal J}_2 - {\cal R}\, {\cal L}_2\right) \nonumber\\
&- \frac{i \eta}{4} \left({\cal J}_0 + {\cal R}\, {\cal L}_0\right) \wedge \left({\cal J}_1 - {\cal R}\, {\cal L}_1\right)\,, \nonumber\\
r_V(\gamma,\bar\chi,\rho;{\cal R}) &= \frac{i \gamma}{2} \left({\cal J}_1 + {\cal R}\, {\cal L}_1\right) \wedge \left({\cal J}_2 + {\cal R}\, {\cal L}_2\right) + \frac{\bar\chi}{4} \left({\cal J}_1 - {\cal R}\, {\cal L}_1\right) \wedge \left({\cal J}_0 - {\cal J}_2 - {\cal R}\, ({\cal L}_0 - {\cal L}_2)\right) \nonumber\\
&+ \frac{i \rho}{4} \left({\cal J}_0 + {\cal R}\, {\cal L}_0\right) \wedge \left({\cal J}_0 - {\cal J}_2 - {\cal R}\, ({\cal L}_0 - {\cal L}_2)\right)\,.
\end{align}
where $\gamma, \eta, \rho \in \mathbbm{R}$ and $\bar\gamma, \bar\chi \in i \mathbbm{R}$. For this particular real form there is an essential difference with respect to the contractions of $\mathfrak{o}(4;\mathbbm{C})$ $r$-matrices along the fourth axis discussed in Sec.~\ref{sec:2.0}. Namely, both $r_{III}$ and $r_V$ from (\ref{eq:36.03}) have only one contraction limit
\begin{align}\label{eq:36.04}
\tilde r_{III}(\tilde\gamma,\tilde{\bar\gamma},\tilde\eta) &= \frac{i \tilde\gamma}{2}\, {\cal P}_1 \wedge {\cal P}_2 - \frac{\tilde{\bar\gamma}}{2}\, {\cal P}_0 \wedge {\cal P}_2 + \frac{i \tilde\eta}{4}\, {\cal P}_0 \wedge {\cal P}_1\,, \nonumber\\
\tilde r_V(\tilde\gamma,\tilde{\bar\chi},\tilde\rho) &= \frac{2i \tilde\gamma - \tilde{\bar\chi}}{4}\, {\cal P}_1 \wedge {\cal P}_2 - \frac{\tilde{\bar\chi}}{4}\, {\cal P}_0 \wedge {\cal P}_1 + \frac{i \tilde\rho}{4}\, {\cal P}_0 \wedge {\cal P}_2\,.
\end{align}
As one can notice, we obtain no $r$-matrices that depend on the ${\cal J}_i$ generators. Furthermore, if we perform the contraction along the second axis, using the anti-Hermitian basis (\ref{eq:33.04e}) introduced for (\ref{eq:36.02}), it leads to identical contraction limits as (\ref{eq:36.04}) (up to some irrelevant sign changes). %i.e.
%\begin{align}\label{eq:36.05}
%\tilde r^a_{III}(\tilde\gamma,\tilde{\bar\gamma},\tilde\eta) &= -\frac{i \tilde\gamma}{2}\, {\cal P}_1 \wedge {\cal P}_2 + \frac{\tilde{\bar\gamma}}{2}\, {\cal P}_0 \wedge {\cal P}_2 + \frac{i \tilde\eta}{4}\, {\cal P}_0 \wedge {\cal P}_1\,, \nonumber\\
%\tilde r^a_V(\tilde\gamma,\tilde{\bar\chi},\tilde\rho) &= -\frac{2i \tilde\gamma - \tilde{\bar\chi}}{4}\, {\cal P}_1 \wedge {\cal P}_2 - \frac{\tilde{\bar\chi}}{4} {\cal P}_0 \wedge {\cal P}_1 - \frac{i \tilde\rho}{4} {\cal P}_0 \wedge {\cal P}_2\,.
%\end{align}
The situation is the same for contractions along the third or first axis.

%We again refer the reader to Subsec.~\ref{sec:5.2} for the comparison of results from this Subsection with the known classification of $D = 3$ Poincar\'{e} $r$-matrices.

\subsection{Summary of the contraction results} \label{sec:4.4}

As we already explained for complex $r$-matrices (\ref{eq:20.10ca}), (\ref{eq:20.16a}) and (\ref{eq:20.22}), the most general quantum IW contraction limits for $r_{II}$, $r_{III}$ and $r_V$ are combinations of $r$-matrices of the type $\hat r$ and $\tilde r$. Therefore, the most general results for deformed $\mathfrak{o}(3) \vartriangleright\!\!< {\cal T}^3$ contractions can be written down in the following way (here we absorb imaginary units $i$ into the parameters, so that they all become imaginary-valued):
\begin{itemize}
\item deformation of $\mathfrak{o}(4)$ leads to (cf. Subsec.~\ref{sec:4.1})
\begin{align}\label{eq:37.01}
\hat r_{III}(\hat\gamma + \hat{\bar\gamma},\hat\eta) + \tilde r_{III}(\tilde\gamma - \tilde{\bar\gamma}) = \frac{\hat\gamma + \hat{\bar\gamma}}{2} \left(J_1 \wedge P_2 - J_2 \wedge P_1\right) - \frac{\hat\eta}{2}\, J_3 \wedge P_3 - \frac{\tilde\gamma - \tilde{\bar\gamma}}{2}\, P_1 \wedge P_2\,;
\end{align}
\item deformations of $\mathfrak{o}(3,1)$ lead to (cf. Subsec.~\ref{sec:4.2})
\begin{align}\label{eq:37.02}
\tilde r_I(\tilde\chi) &= \tilde\chi\, P_1 \wedge P_3\,, \nonumber\\
\tilde r_{II}(\tilde\chi,\tilde\varsigma) &= \frac{\tilde\chi}{2}\, P_1 \wedge P_3 + \frac{\tilde\varsigma}{2}\, P_1 \wedge P_2\,, \nonumber\\
\hat r_{III}(\hat\gamma + \hat{\bar\gamma},\hat\eta) + \tilde r_{III}(\tilde\gamma - \tilde{\bar\gamma}) &= \frac{\hat\gamma + \hat{\bar\gamma}}{2} \left(J_1 \wedge P_2 - J_2 \wedge P_1\right) - \frac{\hat\eta}{2}\, J_3 \wedge P_3 + \frac{\tilde\gamma - \tilde{\bar\gamma}}{2}\, P_1 \wedge P_2\,, \nonumber\\
\hat r_{IV}(\hat\gamma) &= \hat\gamma \left(J_1 \wedge P_1 + J_2 \wedge P_2 + J_3 \wedge P_3\right)\,, \nonumber\\
\tilde r_{IV}(\tilde\gamma,\tilde\varsigma) &= \frac{2\tilde\gamma + \tilde\varsigma}{2}\, P_1 \wedge P_2\,.
\end{align}
\end{itemize}

Similarly, the most general results for deformed $\mathfrak{o}(2,1) \vartriangleright\!\!< {\cal T}^{2,1}$ contractions are:
\begin{itemize}
\item deformations of $\mathfrak{o}(3,1)$ lead to (cf. Subsec.~\ref{sec:4.2a})
\begin{align}\label{eq:37.03}
\tilde r_I(\tilde\chi) &= \tilde\chi\, {\cal P}_0 \wedge {\cal P}_1\,, \nonumber\\
\tilde r_{II}(\tilde\chi,\tilde\varsigma) &= \frac{\tilde\chi}{2}\, {\cal P}_0 \wedge {\cal P}_1 + \frac{\tilde\varsigma}{2}\, {\cal P}_1 \wedge {\cal P}_2\,, \nonumber\\
\hat r_{III}(\hat\gamma + \hat{\bar\gamma},\hat\eta) + \tilde r_{III}(\tilde\gamma - \tilde{\bar\gamma}) &= \frac{\hat\gamma + \hat{\bar\gamma}}{2} \left({\cal J}_1 \wedge {\cal P}_2 - {\cal J}_2 \wedge {\cal P}_1\right) + \frac{\hat\eta}{2}\, {\cal J}_0 \wedge {\cal P}_0 - \frac{\tilde\gamma - \tilde{\bar\gamma}}{2}\, {\cal P}_1 \wedge {\cal P}_2\,, \nonumber\\
\hat r_{IV}(\hat\gamma) &= -\hat\gamma \left({\cal J}_0 \wedge {\cal P}_0 - {\cal J}_1 \wedge {\cal P}_1 - {\cal J}_2 \wedge {\cal P}_2\right)\,, \nonumber\\
\tilde r_{IV}(\tilde\gamma,\tilde\varsigma) &= -\frac{2\tilde\gamma - \tilde\varsigma}{2}\, {\cal P}_1 \wedge {\cal P}_2
\end{align}
and
\begin{align}\label{eq:37.04}
r^b_I(\chi) &= \chi \left({\cal J}_0 - {\cal J}_1\right) \wedge {\cal J}_2\,, \nonumber\\
\hat r^a_I(\hat\chi) &= -\hat\chi\, {\cal J}_1 \wedge \left({\cal P}_0 - {\cal P}_2\right)\,, \nonumber\\
\hat r^a_{II}(\hat\chi,\hat\varsigma) &= \frac{\hat\chi}{2} \left(({\cal J}_0 - {\cal J}_2) \wedge {\cal P}_1 - {\cal J}_1 \wedge ({\cal P}_0 - {\cal P}_2)\right) - \frac{\hat\varsigma}{2} \left({\cal J}_0 - {\cal J}_2\right) \wedge \left({\cal P}_0 - {\cal P}_2\right)\,, \nonumber\\
\hat r_{III}(\hat\gamma - \hat{\bar\gamma},\hat\eta) + \tilde r_{III}(\tilde\gamma + \tilde{\bar\gamma}) &= - \frac{\hat\gamma - \hat{\bar\gamma}}{2} \left({\cal J}_0 \wedge {\cal P}_2 - {\cal J}_2 \wedge {\cal P}_0\right) - \frac{\hat\eta}{2}\, {\cal J}_1 \wedge {\cal P}_1 + \frac{\tilde\gamma + \tilde{\bar\gamma}}{2}\, {\cal P}_0 \wedge {\cal P}_2\,, \nonumber\\
\hat r^a_{IV}(\hat\gamma,\hat\varsigma) &= -\hat\gamma \left({\cal J}_0 \wedge {\cal P}_2 - {\cal J}_2 \wedge {\cal P}_0 - {\cal J}_1 \wedge {\cal P}_1\right) - \frac{\hat\varsigma}{2} \left({\cal J}_0 - {\cal J}_2\right) \wedge \left({\cal P}_0 - {\cal P}_2\right)\,;
\end{align}
\item deformations of $\dot{\mathfrak{o}}(2,2)$ lead to (cf. Subsec.~\ref{sec:4.3})
\begin{align}\label{eq:37.05}
\tilde r_I(\tilde\chi) &= -\tilde\chi\, {\cal P}_1 \wedge {\cal P}_2\,, \nonumber\\
\tilde r_{II}(\tilde\chi,\tilde{\bar\chi},\tilde\varsigma) &= -\frac{\tilde\chi + \tilde{\bar\chi}}{4}\, {\cal P}_1 \wedge {\cal P}_2 + \frac{\tilde\chi - \tilde{\bar\chi}}{4}\, {\cal P}_0 \wedge {\cal P}_1 + \frac{\tilde\varsigma}{2}\, {\cal P}_0 \wedge {\cal P}_2\,, \nonumber\\
\hat r_{III}(\hat\gamma + \hat{\bar\gamma},\hat\eta) + \tilde r_{III}(\tilde\gamma - \tilde{\bar\gamma}) &= \frac{\hat\gamma + \hat{\bar\gamma}}{2} \left({\cal J}_0 \wedge {\cal P}_2 - {\cal J}_2 \wedge {\cal P}_0\right) + \frac{\hat\eta}{2}\, {\cal J}_1 \wedge {\cal P}_1 + \frac{\tilde\gamma - \tilde{\bar\gamma}}{2}\, {\cal P}_0 \wedge {\cal P}_2\,, \nonumber\\
\hat r_{IV}(\hat\gamma) &= \hat\gamma \left({\cal J}_0 \wedge {\cal P}_0 - {\cal J}_1 \wedge {\cal P}_1 - {\cal J}_2 \wedge {\cal P}_2\right)\,, \nonumber\\
\tilde r_{IV}(\tilde\gamma,\tilde\varsigma) &= \frac{2\tilde\gamma + \tilde\varsigma}{2}\, {\cal P}_0 \wedge {\cal P}_1\,, \nonumber\\
\hat r_V(\hat{\bar\chi} + \hat\rho) + \tilde r_V(\tilde\gamma,\tilde{\bar\chi} - \tilde\rho) &= -\frac{\hat{\bar\chi} + \hat\rho}{4}\, {\cal J}_1 \wedge ({\cal P}_0 - {\cal P}_2) + \frac{\tilde\gamma}{2}\, {\cal P}_0 \wedge {\cal P}_2 - \frac{\tilde{\bar\chi} - \tilde\rho}{4}\, ({\cal P}_0 - {\cal P}_2) \wedge {\cal P}_1
\end{align}
and
\begin{align}\label{eq:37.06}
r^a_I(\chi) &= -\chi \left({\cal J}_0 + {\cal J}_2\right) \wedge {\cal J}_1\,, \nonumber\\
\hat r_{II}^a(\hat\chi - \hat{\bar\chi},\hat\varsigma) + \tilde r_{II}^a(\tilde\chi + \tilde{\bar\chi}) &= -\frac{\hat\chi - \hat{\bar\chi}}{4} \left(({\cal J}_0 + {\cal J}_2) \wedge {\cal P}_1 - {\cal J}_1 \wedge ({\cal P}_0 + {\cal P}_2)\right) - \frac{\hat\varsigma}{2} \left({\cal J}_0 + {\cal J}_2\right) \wedge \left({\cal P}_0 + {\cal P}_2\right) \nonumber\\
&-\frac{\tilde\chi + \tilde{\bar\chi}}{4} \left({\cal P}_0 + {\cal P}_2\right) \wedge {\cal P}_1\,, \nonumber\\
\hat r^a_{IV}(\hat\gamma,\hat\varsigma) &= \hat\gamma \left({\cal J}_0 \wedge {\cal P}_2 - {\cal J}_2 \wedge {\cal P}_0 + {\cal J}_1 \wedge {\cal P}_1\right) - \frac{\hat\varsigma}{2} \left({\cal J}_0 + {\cal J}_2\right) \wedge \left({\cal P}_0 + {\cal P}_2\right)\,;
\end{align}
\item deformation of $\mathfrak{o}'(2,2)$ leads to (cf. Subsec.~\ref{sec:4.3})
\begin{align}\label{eq:37.07}
\hat r_{III}(\hat\gamma + \hat{\bar\gamma},\hat\eta) + \tilde r_{III}(\tilde\gamma - \tilde{\bar\gamma}) &= \frac{\hat\gamma + \hat{\bar\gamma}}{2} \left({\cal J}_1 \wedge {\cal P}_2 - {\cal J}_2 \wedge {\cal P}_1\right) - \frac{\hat\eta}{2}\, {\cal J}_0 \wedge {\cal P}_0 + \frac{\tilde\gamma - \tilde{\bar\gamma}}{2}\, {\cal P}_1 \wedge {\cal P}_2\,;
\end{align}
\item deformations of $\dot{\mathfrak{o}}'(2,2)$ lead to (cf. Subsec.~\ref{sec:4.3})
\begin{align}\label{eq:37.08}
\tilde r_{III}(\tilde\gamma,\tilde{\bar\gamma},\tilde\eta) &= \frac{\tilde\gamma}{2}\, {\cal P}_1 \wedge {\cal P}_2 - \frac{\tilde{\bar\gamma}}{2}\, {\cal P}_0 \wedge {\cal P}_2 + \frac{\tilde\eta}{4}\, {\cal P}_0 \wedge {\cal P}_1\,, \nonumber\\
\tilde r_V(\tilde\gamma,\tilde{\bar\chi},\tilde\rho) &= \frac{2\tilde\gamma - \tilde{\bar\chi}}{4}\, {\cal P}_1 \wedge {\cal P}_2 - \frac{\tilde{\bar\chi}}{4}\, {\cal P}_0 \wedge {\cal P}_1 + \frac{\tilde\rho}{4}\, {\cal P}_0 \wedge {\cal P}_2\,.
\end{align}
\end{itemize}

\section{$D = 3$ inhomogeneous contractions compared to Stachura classification} \label{sec:5.0}

\subsection{$D = 3$ inhomogeneous Euclidean deformations} \label{sec:5.1}

In the case of $D = 3$ inhomogeneous Euclidean algebra $\mathfrak{o}(3) \vartriangleright\!\!< {\cal T}^3$, we can identify the following relation between the notation of \cite{Stachura:1998ps} and ours:
\begin{align}\label{eq:A0.01}
e_i = P_i\,, \qquad k_i = J_i\,,
\end{align}
as well as we replace the names of parameters $\alpha$ and $\rho$ by $\beta$ and $\varrho$, respectively. Then we rewrite the complete classification (up to an algebra automorphism) of Hermitian $r$-matrices for the $D = 3$ inhomogeneous Euclidean algebra from Subsec.~3.2 of \cite{Stachura:1998ps}, which includes
\begin{align}\label{eq:A0.02}
r_1 &= \beta\, (J_1 \wedge P_2 - J_2 \wedge P_1) - \varrho\, J_3 \wedge P_3 + \theta\, P_1 \wedge P_2\,, \nonumber\\
r_2 &= J_1 \wedge P_1 + J_2 \wedge P_2 + J_3 \wedge P_3\,, \nonumber\\
r_3 &= \theta^{ij} P_i \wedge P_j\,,
\end{align}
where $\beta \in \{0,1\}$, $\varrho \geq 0$, $\beta = 0 \Leftrightarrow \varrho \neq 0$ and $\theta,\theta_{ij} = -\theta_{ji} \in \mathbbm{R}$. Let us note that the part $c \in {\cal L} \wedge {\cal L}$ of a classical $r$-matrix of a Lie algebra ${\cal L} \vartriangleright\!\!< {\cal T}$ vanishes for all $r$-matrices (\ref{eq:A0.02}).

\begin{table}[!h]
\begin{tabular}{|c|c|c|}
\hline Stachura class & contractions of $\mathfrak{o}(4)$ & contractions of $\mathfrak{o}(3,1)$ \\
\hline $r_1$ & $\hat r_{III} + \tilde r_{III}$ (\ref{eq:37.01}) & $\hat r_{III} + \tilde r_{III}$ (\ref{eq:37.02}) \\
\hline $r_2$ &  & $\hat r_{IV}$ (\ref{eq:37.02}) \\
\hline $r_3$ & $\tilde r_{III}$ & $\tilde r_{I-IV}$ \\
\hline
\end{tabular}
\caption{Comparison of our results from Subsec. \ref{sec:4.1}-\ref{sec:4.2} with the classification (\ref{eq:A0.02}).}
\end{table}

A comparison with the results of Subsec.~\ref{sec:4.1} and \ref{sec:4.2} shows (cf. Table II):
\begin{itemize}
\item the $r$-matrices $\tilde r_I$, $\tilde r_{II}$, $\tilde r_{III}$ and $\tilde r_{IV}$ in (\ref{eq:37.02}), as well as $\tilde r_{III}$ in (\ref{eq:37.01}), depend only on the translation generators and therefore they all belong to the type $r_3$ above;
\item $\hat r_{IV}$ in (\ref{eq:37.02}) is proportional to $r_2$;
\item $\hat r_{III} + \tilde r_{III}$ in (\ref{eq:37.01}) and (\ref{eq:37.02}) are equivalent to $r_1$ (up to the automorphism $(J_1 \mapsto -J_1,J_3 \mapsto -J_3,P_2 \mapsto -P_2)$ or $(J_2 \mapsto -J_2,J_3 \mapsto -J_3,P_1 \mapsto -P_1)$).
\end{itemize}
Therefore, our (parametrized families of) $r$-matrices (\ref{eq:37.01}-\ref{eq:37.02}) can be obtained by multiplying an appropriate expression from (\ref{eq:A0.02}) by an imaginary parameter and using the automorphism in the $r_1$ case.

\subsection{$D = 3$ Poincar\'{e} deformations} \label{sec:5.2}

On the other hand, for $D = 3$ Poincar\'{e} algebra $\mathfrak{o}(2,1) \vartriangleright\!\!< {\cal T}^{2,1}$, we can identify the following relation between the notation of \cite{Stachura:1998ps} and ours:
\begin{align}\label{eq:A0.03}
e_1 = -{\cal P}_0\,, \qquad e_{2/3} = {\cal P}_{1/2}\,, \qquad k_i = {\cal J}_{i-1}\,,
\end{align}
as well as we replace the names of parameters $\alpha$, $\rho$ by $\beta$, $\varrho$, respectively. Then we rewrite the complete classification (up to an algebra automorphism) of Hermitian $r$-matrices for $D = 3$ Poincar\'{e} algebra ${\cal L} \vartriangleright\!\!< {\cal T}$ from Subsec.~3.1 of \cite{Stachura:1998ps}, which includes one $r$-matrix with non-vanishing part $c \in {\cal L} \wedge {\cal L}$ (i.e. an extension of a $\mathfrak{o}(2,1)$ $r$-matrix),
\begin{align}\label{eq:A0.04}
r_1 &= \frac{1}{\sqrt{2}}\, ({\cal J}_0 + {\cal J}_1) \wedge {\cal J}_2 + \beta\, ({\cal J}_0 \wedge {\cal P}_0 - {\cal J}_1 \wedge {\cal P}_1 - {\cal J}_2 \wedge {\cal P}_2)\,,
\end{align}
where $\beta \in \{0,1\}$; three $r$-matrices of the form
\begin{align}\label{eq:A0.05}
r_2 &= \varrho\, {\cal J}_2 \wedge {\cal P}_2 + \beta\, ({\cal J}_0 \wedge {\cal P}_1 - {\cal J}_1 \wedge {\cal P}_0) + \theta\, {\cal P}_0 \wedge {\cal P}_1 + \mathbbm{1}_{\varrho = \beta}\, \theta'\, ({\cal P}_0 + {\cal P}_1) \wedge {\cal P}_2\,, \nonumber\\
r_3 &= -\varrho\, {\cal J}_0 \wedge {\cal P}_0 + \beta\, ({\cal J}_1 \wedge {\cal P}_2 - {\cal J}_2 \wedge {\cal P}_1) + \theta\, {\cal P}_1 \wedge {\cal P}_2\,, \nonumber\\
r_4 &= \varrho\, ({\cal J}_0 + {\cal J}_1) \wedge ({\cal P}_0 + {\cal P}_1) - \frac{\beta}{\sqrt{2}} \left(({\cal J}_0 + {\cal J}_1) \wedge {\cal P}_2 - {\cal J}_2 \wedge ({\cal P}_0 + {\cal P}_1)\right) + \theta\, ({\cal P}_0 + {\cal P}_1) \wedge {\cal P}_2\,,
\end{align}
where $\beta \in \{0,1\}$, $\varrho \geq 0$, $\beta = 0 \Leftrightarrow \varrho \neq 0$ and $\theta,\theta' \in \mathbbm{R}$; two $r$-matrices
\begin{align}\label{eq:A0.06}
r_5 &= \frac{1}{\sqrt{2}}\, {\cal J}_2 \wedge ({\cal P}_0 + {\cal P}_1) + \theta^{\mu\nu} {\cal P}_\mu \wedge {\cal P}_\nu\,, \nonumber\\
r_6 &= ({\cal J}_0 + {\cal J}_1) \wedge ({\cal P}_0 + {\cal P}_1) - \varrho\, ({\cal J}_0 \wedge {\cal P}_1 - {\cal J}_1 \wedge {\cal P}_0 + {\cal J}_2 \wedge {\cal P}_2) + \theta^{\mu\nu} {\cal P}_\mu \wedge {\cal P}_\nu\,,
\end{align}
where $\varrho \in \mathbbm{R}\backslash\{0\}$; and two $r$-matrices
\begin{align}\label{eq:A0.07}
r_7 &= {\cal J}_0 \wedge {\cal P}_0 - {\cal J}_1 \wedge {\cal P}_1 - {\cal J}_2 \wedge {\cal P}_2\,, \nonumber\\
r_8 &= \theta^{\mu\nu} {\cal P}_\mu \wedge {\cal P}_\nu\,.
\end{align}
The parameters $\theta_{\mu\nu} = -\theta_{\nu\mu} \in \mathbbm{R}$, $\mu,\nu = 0,1,2$ and can be further restricted via automorphisms. %Let us note that $r_8$ included as the last term in the $r$-matrices $r_2$-$r_6$ is of limited relevance, e.g. it does not generate inhomogeneity in the corresponding Yang-Baxter equations.
\newline

\begin{table}[!h]
\begin{tabular}{|c|c|c|c|c|}
\hline Stachura class & contractions of $\mathfrak{o}(3,1)$ & contractions of $\dot{\mathfrak{o}}(2,2)$ & contractions of $\mathfrak{o}'(2,2)$ & contractions of $\dot{\mathfrak{o}}'(2,2)$ \\
\hline $r_1$ & $r^b_I$ (\ref{eq:37.04}) & $r^a_I$ (\ref{eq:37.06}) & & \\
\hline $r_2$ & $\hat r^a_{III} + \tilde r^a_{III}$ (\ref{eq:37.04}) & $\hat r_{III} + \tilde r_{III}$ (\ref{eq:37.05}) & & \\
\hline $r_3$ & $\hat r_{III} + \tilde r_{III}$ (\ref{eq:37.03}) & & $\hat r_{III} + \tilde r_{III}$ (\ref{eq:37.07}) & \\
\hline $r_4$ & $\hat r^a_{II}$ (\ref{eq:37.04}) & $\hat r^a_{II} + \tilde r^a_{II}$ (\ref{eq:37.06}) & & \\
\hline $r_5$ & $\hat r^a_I$ (\ref{eq:37.04}) & $\hat r_V + \tilde r_V$ (\ref{eq:37.05}) & & \\
\hline $r_6$ & $\hat r^a_{IV}$ (\ref{eq:37.04}) & $\hat r^a_{IV}$ (\ref{eq:37.06}) & & \\
\hline $r_7$ & $\hat r_{IV}$ (\ref{eq:37.03}) & $\hat r_{IV}$ (\ref{eq:37.05}) & & \\
\hline $r_8$ & $\tilde r_{I-IV}$, $\tilde r^a_{III}$ & $\tilde r_{I-V}$, $\tilde r^a_{II}$ & $\tilde r_{III}$ & $\tilde r_{III}$, $\tilde r_V$ \\
\hline
\end{tabular}
\caption{Comparison of our results from Subsec. \ref{sec:4.2a}-\ref{sec:4.3} with the classification (\ref{eq:A0.04}-\ref{eq:A0.07}).}
\end{table}

Comparing the above classification with our results, we observe that (cf. Table III and the formulae (\ref{eq:60.01})):
\begin{itemize}
\item the $r$-matrices $\tilde r_I$, $\tilde r_{II}$/$\tilde r^a_{II}$, $\tilde r_{III}$/$\tilde r^a_{III}$, $\tilde r_{IV}$ and $\tilde r_V$ in (\ref{eq:37.03}-\ref{eq:37.08}) (which depend only on the translation generators) belong to the type $r_8$;
\item $\hat r_{IV}$ in (\ref{eq:37.03}) and (\ref{eq:37.05}) is proportional to $r_7$;
\item $\hat r^a_{III} + \tilde r^a_{III}$ in (\ref{eq:37.04}) and $\hat r_{III} + \tilde r_{III}$ in (\ref{eq:37.05}) are equivalent to $r_2$ (with $\theta' = 0$ and up to the automorphism $({\cal J}_0 \mapsto -{\cal J}_0,{\cal J}_2 \mapsto -{\cal J}_2,{\cal P}_1 \mapsto -{\cal P}_1)$) but to see this one has to act on $r_2$ with an algebra automorphism $({\cal J}_{1/2} \mapsto \pm {\cal J}_{2/1},{\cal P}_{1/2} \mapsto \pm {\cal P}_{2/1})$ or $({\cal J}_{1/2} \mapsto \mp {\cal J}_{2/1},{\cal P}_{1/2} \mapsto \mp {\cal P}_{2/1})$;
\item $\hat r_{III} + \tilde r_{III}$ in (\ref{eq:37.03}) and (\ref{eq:37.07}) is equivalent to $r_3$ (up to the automorphism $({\cal J}_0 \mapsto -{\cal J}_0,{\cal J}_1 \mapsto -{\cal J}_1,{\cal P}_2 \mapsto -{\cal P}_2)$ or $({\cal J}_0 \mapsto -{\cal J}_0,{\cal J}_2 \mapsto -{\cal J}_2,{\cal P}_1 \mapsto -{\cal P}_1)$);
\item $\hat r^a_{II}$ in (\ref{eq:37.04}) and $\hat r^a_{II} + \tilde r^a_{II}$ in (\ref{eq:37.06}) can be obtained from $r_4$ (up to the automorphism $({\cal J}_0 \mapsto -{\cal J}_0,{\cal J}_1 \mapsto -{\cal J}_1,{\cal P}_2 \mapsto -{\cal P}_2)$ and with $\theta = 0$ in the (\ref{eq:37.04}) case) via the respective automorphisms $({\cal J}_{1/2} \mapsto \mp {\cal J}_{2/1},{\cal P}_{1/2} \mapsto \mp {\cal P}_{2/1})$ and $({\cal J}_{1/2} \mapsto \pm {\cal J}_{2/1},{\cal P}_{1/2} \mapsto \pm {\cal P}_{2/1})$;
\item $\hat r^a_{IV}$ in (\ref{eq:37.04}) and (\ref{eq:37.06}) can be obtained from $r_6$ (with $\theta^{\mu\nu} = 0$) via the respective automorphisms $({\cal J}_{1/2} \mapsto \mp {\cal J}_{2/1},{\cal P}_{1/2} \mapsto \mp {\cal P}_{2/1})$ and $({\cal J}_{1/2} \mapsto \pm {\cal J}_{2/1},{\cal P}_{1/2} \mapsto \pm {\cal P}_{2/1})$;
\item $\hat r^a_I$ in (\ref{eq:37.04}) and $\hat r_V + \tilde r_V$ in (\ref{eq:37.05}) can be both obtained from $r_5$ (with $\theta_{\mu\nu} = 0$ in the (\ref{eq:37.05}) case) via the automorphism $({\cal J}_{1/2} \mapsto \mp {\cal J}_{2/1},{\cal P}_{1/2} \mapsto \mp {\cal P}_{2/1})$;
\item $r^b_I$ in (\ref{eq:37.04}) and $r^a_I$ in (\ref{eq:37.06}) can be obtained from $r_1$ (with $\beta = 0$) via the respective automorphisms $({\cal J}_0 \mapsto -{\cal J}_0,{\cal J}_2 \mapsto -{\cal J}_2,{\cal P}_1 \mapsto -{\cal P}_1)$ and $({\cal J}_{1/2} \mapsto \pm {\cal J}_{2/1},{\cal P}_{1/2} \mapsto \pm {\cal P}_{2/1})$.
\end{itemize}
Our (parametrized families of) $r$-matrices (\ref{eq:37.03}-\ref{eq:37.08}) are constructed by multiplying an appropriate expression from (\ref{eq:A0.05}-\ref{eq:A0.07}) by an imaginary parameter and acting on it with the automorphisms described above.

\section{$D = 3$ classical $r$-matrices and 3D (quantum) gravity} \label{sec:6.0}

\subsection{3D gravity as Chern-Simons theory} \label{sec:6.1}

The major reason why the (pseudo-)orthogonal groups considered in this paper are of physical interest is that they play the role of gauge groups in 3D gravity. Namely, in the Chern-Simons formulation of gravity in 2+1 dimensions, the local gauge group describes local isometries of spacetime and is given by the $D = 3$ Poincar\'{e} or (anti-)de Sitter group for vanishing, negative or positive cosmological constant, respectively. The above formalism extends to the Euclidean version of the theory, where the corresponding gauge groups are the inhomogeneous Euclidean and Euclidean (anti-)de Sitter groups \cite{Achucarro:1987vz,Witten:1988hc}. However, in this Section we will restrict ourselves to the physically more important Lorentzian signature. %(as can be seen from Subsec.~\ref{sec:5.1}, the situation in the Euclidean case is much simpler).

In order to formulate the Chern-Simons theory of classical gravity, we first introduce the gauge field $A$ that is a Cartan connection with values in the appropriate local isometry (Lie) algebra. In terms of the physical basis $\{{\cal J}_\mu,{\cal L}_\mu\}$ used in the previous Sections (in the Poincar\'{e} case ${\cal L}_\mu$ become ${\cal P}_\mu$), the gauge field is constructed as follows
\begin{align}\label{grav.1}
A = e^\mu {\cal L}_\mu + \omega^\mu {\cal J}_\mu\,,
\end{align}
where $e^\mu$ and $\omega^\mu$ are the dreibein and spin connection one-forms, respectively. The Chern-Simons action\footnote{We note that in 2+1 dimensions the Newton's constant $G^{(3)}$ has the dimension of inverse mass. It is actually the Planck mass, which is a quantum gravity related concept in 3+1 dimensions but here appears already at the classical level and also plays the fundamental role in quantization, leading to noncommutativity of spacetime geometry.}
\begin{align}\label{grav.2}
S = \frac{1}{16\pi G^{(3)}}\, \int \left(A \wedge dA\right) + \frac{1}{3} \left(A \wedge A \wedge A\right)
\end{align}
is equivalent to the Einstein-Hilbert action of general relativity in 2+1 dimensions\footnote{In the sense that both produce the same set of solutions of vacuum Einstein equations, which in 2+1 dimensions is the set of all torsionless and Riemannian flat spin connections (after subtracting the cosmological constant term).} if the Ad-invariant bilinear form $(.,.)$ is defined in terms of the gauge algebra generators as %\footnote{The most general Ad-invariant bilinear form is a linear combination of \eqref{grav3} and the `diagonal' one but it leads to unphysical results for particles coupled to the theory \cite{Meusburger:2008dc}.}
\cite{Witten:1988hc}
\begin{align}\label{grav3}
({\cal J}_\mu,{\cal J}_\nu) = ({\cal L}_\mu,{\cal L}_\nu) = 0\,, \quad ({\cal J}_\mu,{\cal L}_\nu) = -\eta_{\mu\nu}\,.
\end{align}
%see \cite{Osei:2017ybk} and references therein for a detailed discussion.

The Chern-Simons theory is a topological theory and thus gravity in 2+1 dimensions does not have any dynamical degrees of freedom. In the Hamiltonian picture, after singling out the time direction, the action (\ref{grav.2}) contains two terms: the kinematical one, defining the symplectic structure, which is directly related to the bilinear form (\ref{grav3}), and the constraint taking the form of the requirement that the curvature of the connection $A$ vanishes on constant time surfaces $\Sigma$,
\begin{align}\label{grav.4}
F(A)\big|_\Sigma = 0
\end{align}
(even for non-zero cosmological constant $\Lambda$, since this is not the Riemannian curvature of $\Sigma$). It follows that the action (\ref{grav.2}) describes a theory of flat connections on a two-dimensional manifold (Riemann surface) $\Sigma$. Punctures on the Riemann surface are interpreted as point particles, each labeled by its mass and spin. If such punctures are present, the right hand side of (\ref{grav.4}) becomes the sum of delta functions at the positions of particles, each one multiplied by a gauge algebra element parametrized by the particle's mass and spin. Another way to introduce (topological) degrees of freedom in the theory is via nontrivial topology of the Riemann surface, with some number of handles, which do not modify (\ref{grav.4}) but imply additional continuity conditions on $A$ (see \cite{Matschull:1997du} and \cite{Meusburger:2003ta,Meusburger:2003hc,Meusburger:2006pe,Meusburger:2008dc} for details).

In order to quantize such a theory, we need to know the Poisson (or symplectic) structure of its phase space. The symplectic structure can be derived from \eqref{grav.2} and has the form
\begin{align}\label{grav.a}
\omega \sim \int \delta A \wedge \delta A\,.
\end{align}
Since this symplectic structure is invariant under gauge transformations
\begin{align}\label{grav.b}
A \rightarrow A' = g^{-1} A g + g^{-1} dg\,,
\end{align}
one has to compute the symplectic form not on the space of flat connections $A$ but on the space of their gauge-equivalent classes. At this point the classical $r$-matrices associated with the gauge group become relevant.

The celebrated Fock-Rosly construction provides the auxiliary Poisson structure in the case when spacetime has the topology of $\mathbb{R} \times S$, where the space $S$ is an oriented, closed two-dimensional manifold (the spatial infinity can be added as a distinguished puncture, at least when $\Lambda = 0$ \cite{Meusburger:2006pe}). After gauge fixing, the auxiliary Poisson structure becomes the Poisson structure on the moduli space of flat connections, i.e., the gauge equivalent classes of solutions of (\ref{grav.4}) for a generic Riemann surface with $g$ handles and $n$ punctures. Such auxiliary Poisson structures are defined on the direct product of $n+2g$ copies of the gauge group and are characterized by the Fock-Rosly (FR) $r$-matrices $r_{FR}$, which contain not only the antisymmetric terms (like $r$-matrices we considered so far) but also the symmetric ones.\footnote{The $r$-matrices with symmetric terms were introduced in \cite{Belavin:1982os} and are sometimes called the Belavin-Drinfeld forms.} We have the following Fock-Rosly conditions, restricting the possible form of $r_{FR}$ and ensuring the consistency between the phase space structure and Chern-Simons action (\ref{grav.2}):

\begin{enumerate}
\item $r_{FR}$ satisfies the classical Yang-Baxter equation;
\item the symmetric part of $r_{FR}$ corresponds to the inner product (\ref{grav3}) used in the construction of the Chern-Simons action for gravity.
\end{enumerate}

In the following two subsections, we will discuss separately the cases of vanishing and non-vanishing cosmological constant $\Lambda$.

\subsection{Fock-Rosly-compatible classical $r$-matrices for $\Lambda = 0$} \label{sec:6.2}

$r$-matrices satisfying the Fock-Rosly conditions for $\Lambda = 0$ have the form (we recall that in the Poincar\'{e} case ${\cal L}_\mu$ becomes ${\cal P}_\mu$)
\begin{align}\label{grav5}
r_{FR} = r_A + r_S\,, \quad r_S = \alpha \left({\cal J}^\mu \otimes {\cal P}_\mu + {\cal P}_\mu \otimes {\cal J}^\mu\right)\,,
\end{align}
where $\alpha$ is a non-zero real number and $r_S$ describes the split bilinear Casimir. Using the identity
\begin{align}
[[r_{FR},r_{FR}]] = [[r_A,r_A]] + [[r_S,r_S]]\,,
\end{align}
which holds due to the Ad-invariance of $r_S$, we obtain the explicit condition
\begin{align}\label{grav6}
[[r_A,r_A]] = -[[r_S,r_S]] = - \frac{\alpha^2}{2} \epsilon^{\mu\nu\sigma} {\cal J}_\mu \wedge {\cal P}_\nu \wedge {\cal P}_\sigma
\end{align}
(let us stress that $\epsilon^{012} = -1$). The question that we need to answer is which of the antisymmetric $r$-matrices $r_A$ listed in the previous Section are compatible with the Fock-Rosly construction (FR-compatible) in the sense of (\ref{grav6}). In particular, it is clear that all triangular $r$-matrices (i.e. satisfying the classical Yang-Baxter equation) are not FR-compatible and can not be employed in defining the quantum 3D gravity models.

It is actually sufficient to consider Poincar\'{e} $r$-matrices without Abelian terms (i.e. belonging to seven Stachura classes $r_1$-$r_7$ with $\theta,\theta',\theta_{\mu\nu} = 0$, cf. (\ref{eq:A0.04}-\ref{eq:A0.07})). Following the discussion from Subsec.~\ref{sec:5.2}, one can show that all such $r$-matrices that we derived via quantum IW contractions (cf. (\ref{eq:37.03}-\ref{eq:37.07})) can be transformed by the appropriate algebra automorphisms into the following ones:\footnote{For certain values of the parameters, some $r$-matrices may be related to other ones, e.g. $\hat r_6(\hat\gamma,\hat\varsigma = 0) = \hat r_2(\hat\gamma,\hat\eta = 2\hat\gamma)$.}
\begin{align}\label{eq:60.01}
\hat r_1(\chi) &= \chi\, ({\cal J}_0 + {\cal J}_1) \wedge {\cal J}_2\,, \nonumber\\
\hat r_2(\hat\gamma,\hat\eta) &= \hat\gamma \left({\cal J}_0 \wedge {\cal P}_2 - {\cal J}_2 \wedge {\cal P}_0\right) + \frac{\hat\eta}{2}\, {\cal J}_1 \wedge {\cal P}_1\,, \nonumber\\
\hat r_3(\hat\gamma,\hat\eta) &= \hat\gamma \left({\cal J}_1 \wedge {\cal P}_2 - {\cal J}_2 \wedge {\cal P}_1\right) + \frac{\hat\eta}{2}\, {\cal J}_0 \wedge {\cal P}_0\,, \nonumber\\
\hat r_4(\hat\chi,\hat\varsigma) &= \frac{\hat\chi}{2} \left(({\cal J}_0 + {\cal J}_2) \wedge {\cal P}_1 - {\cal J}_1 \wedge ({\cal P}_0 + {\cal P}_2)\right) - \frac{\hat\varsigma}{2}\, ({\cal J}_0 + {\cal J}_2) \wedge ({\cal P}_0 + {\cal P}_2)\,, \nonumber\\
\hat r_5(\hat{\bar\chi}) &= \frac{\hat{\bar\chi}}{2}\, {\cal J}_1 \wedge ({\cal P}_0 + {\cal P}_2)\,, \nonumber\\
\hat r_6(\hat\gamma,\hat\varsigma) &= \hat\gamma \left({\cal J}_0 \wedge {\cal P}_2 - {\cal J}_2 \wedge {\cal P}_0 - {\cal J}_1 \wedge {\cal P}_1\right) - \frac{\hat\varsigma}{2}\, ({\cal J}_0 + {\cal J}_2) \wedge ({\cal P}_0 + {\cal P}_2)\,, \nonumber\\
\hat r_7(\hat\gamma) &= \hat\gamma \left({\cal J}_0 \wedge {\cal P}_0 - {\cal J}_1 \wedge {\cal P}_1 - {\cal J}_2 \wedge {\cal P}_2\right)\,,
\end{align}
where we have to identify $\hat{\bar\chi}/2 = \hat\chi$ to obtain $\hat r_5$ from $\hat r_I^a$ in (\ref{eq:37.04}). In the context of 3D gravity, we take deformation parameters to be real-valued, since we need $r$-matrices that are Hermitian, in contrast to anti-Hermitian ones considered in the previous Sections (as introduced in (\ref{eq:30.01a})).

The antisymmetric $r$-matrices (\ref{eq:60.01}) are known to be associated with particular quantum Hopf-algebraic deformations of the $D = 3$ Poincar\'{e} algebra. $\hat r_3$, $\hat r_2$ and $\hat r_4$ describe respectively the (twisted) time-, space- and lightlike $\kappa$-deformations (see \cite{Borowiec:2006jt} for the lightlike deformation and \cite{Lukierski:2005tn} for the twist), while $\hat r_6$ is a special combination of the twisted space- and lightlike $\kappa$-deformations; $\hat r_1$ and $\hat r_5$ are quasi-Jordanian deformations (cf. \cite{Lukierski:2017qy}) and $\hat r_7$ is obtained from a Drinfeld double of the $D = 3$ Lorentz algebra (see  Subsec.~\ref{sec:6.4}). Furthermore, the $r$-matrices (\ref{eq:60.01}) satisfy the following set of Yang-Baxter equations
\begin{align}\label{eq:60.02}
[[\hat r_1,\hat r_1]] = [[\hat r_4,\hat r_4]] = [[\hat r_5,\hat r_5]] &= 0\,, \nonumber\\
[[\hat r_3,\hat r_3]] &= \hat\gamma^2 \epsilon^{\mu\nu\sigma} {\cal J}_\mu \wedge {\cal P}_\nu \wedge {\cal P}_\sigma\,, \nonumber\\
[[\hat r_2,\hat r_2]] = [[\hat r_6,\hat r_6]] = [[\hat r_7,\hat r_7]] &= -\hat\gamma^2 \epsilon^{\mu\nu\sigma} {\cal J}_\mu \wedge {\cal P}_\nu \wedge {\cal P}_\sigma\,.
\end{align}
One can observe that $\hat r_2$, $\hat r_6$ and $\hat r_7$ are FR-compatible\footnote{In contrast to $\hat r_1$, the Stachura $r$-matrix $r_1$ multiplied by $\gamma$ satisfies the same Yang-Baxter equation as $\hat r_7$.} (when $\hat\gamma \neq 0$), with $\alpha = \sqrt{2}\, \hat\gamma$ in the formulae (\ref{grav5}) and (\ref{grav6}), while $\hat r_3$ is FR-compatible only if $\alpha = i \sqrt{2}\, \hat\gamma$, as it has recently been considered in \cite{Rosati:2017dt} (it remains to be verified whether choosing $\alpha$ to be imaginary leads to a physically meaningful theory). The presence of Abelian terms of the type $\tilde r$ (cf. (\ref{eq:37.03}-\ref{eq:37.07})) in $\hat r_2$ or $\hat r_3$ would change the form of inhomogeneity in the equations (\ref{eq:60.02}), which is why we have discarded such terms here. Let us also note that quantum IW contractions of $r_I$ and $r_{II}$, which satisfy the classical Yang-Baxter equation, lead to the $r$-matrices $\hat r_1$, $\hat r_5$ and $\hat r_4$ or the type $\tilde r$, also satisfying the classical Yang-Baxter equation. The case of $r_V$ is more peculiar: it satisfies the modified Yang-Baxter equation, while its contractions $\hat r_5$ and $\tilde r$ satisfy the classical one (cf. Table III and the equations (\ref{eq:60.01a},\ref{eq:60.01c},\ref{eq:60.01f}) in the next Subsection). $\hat r_4$ and $\hat r_5$ satisfy the classical Yang-Baxter equation even when terms of the type $\tilde r$ are included.

\subsection{Fock-Rosly-compatible classical $r$-matrices for $\Lambda \neq 0$} \label{sec:6.3}

For non-vanishing cosmological constant, the inner product \eqref{grav5} can actually be generalized to a two-parameter family of such products \cite{Witten:1988hc}, corresponding to the symmetric part of the $r$-matrix generalized to (cf. (\ref{grav5}))
\begin{equation}\label{grav5lambda}
r_S = \alpha \left({\cal J}^\mu \otimes {\cal L}_\mu + {\cal L}_\mu \otimes {\cal J}^\mu\right) + \beta\left(\Lambda {\cal J}^\mu \otimes {\cal J}_\mu - {\cal L}^\mu \otimes {\cal L}_\mu\right)\,,
\end{equation}
where $\alpha,\beta \in \mathbbm{R}$ and $\alpha \neq 0$ or $\beta \neq 0$. In this case the Schouten bracket (\ref{grav6}) takes the form  \cite{Meusburger:2008dc}
\begin{align}\label{eq:60.00}
[[r_A,r_A]] = -[[r_S,r_S]] &= -(\alpha^2 -\Lambda\beta^2) \left( \Lambda {\cal J}_0 \wedge {\cal J}_1 \wedge {\cal J}_2 + \frac{1}{2}\, \epsilon^{\mu\nu\sigma}\, {\cal J}_\mu \wedge {\cal L}_\nu \wedge {\cal L}_\sigma \right) \nonumber\\
&- 2\alpha \beta \left( \frac{1}{2}\, \Lambda\, \epsilon^{\mu\nu\sigma}\, {\cal J}_\mu \wedge {\cal J}_\nu \wedge {\cal L}_\sigma + {\cal L}_0 \wedge {\cal L}_1 \wedge {\cal L}_2 \right)\,.
\end{align}

The relevant (Hermitian) $r$-matrices for positive cosmological constant $\Lambda = {\cal R}^{-2} > 0$ are $\mathfrak{o}(3,1)$ $r$-matrices listed in Subsec.~\ref{sec:4.2a} but with all parameters set to be real (let us remind that the combinations $\gamma - \bar\gamma$ and $\gamma + \bar\gamma$ are actual parameters in the case of $r_{III}$ / $r_{III}^a$). They satisfy the following Yang-Baxter equations (the results are the same for $r_I$ and $r_I^a$, $r_{III}$ and $r_{III}^a$, etc.)
\begin{align}\label{eq:60.01a}
[[r_I,r_I]] = [[r_{II},r_{II}]] &= 0\,, \nonumber\\
[[r_{III},r_{III}]] &= -\tfrac{1}{2} {\cal R}^2 \left((\gamma - \bar\gamma)^2 - (\gamma + \bar\gamma)^2\right) \left(\Lambda {\cal J}_0 \wedge {\cal J}_1 \wedge {\cal J}_2 + \tfrac{1}{2} \epsilon^{\mu\nu\sigma} {\cal J}_\mu \wedge {\cal L}_\nu \wedge {\cal L}_\sigma\right) \nonumber\\
&+ {\cal R}^3 (\gamma - \bar\gamma) (\gamma + \bar\gamma) \left(\tfrac{1}{2} \Lambda \epsilon^{\mu\nu\sigma} {\cal J}_\mu \wedge {\cal J}_\nu \wedge {\cal L}_\sigma + {\cal L}_0 \wedge {\cal L}_1 \wedge {\cal L}_2\right)\,, \nonumber\\
[[r_{IV},r_{IV}]] &= -2{\cal R}^2 \gamma^2 \left(\Lambda {\cal J}_0 \wedge {\cal J}_1 \wedge {\cal J}_2 + \tfrac{1}{2} \epsilon^{\mu\nu\sigma} {\cal J}_\mu \wedge {\cal L}_\nu \wedge {\cal L}_\sigma\right)\,.
\end{align}
It follows that $r_I$ and $r_{II}$ are not FR-compatible. The $r$-matrix $r_{III}$ is FR-compatible for the most general form of $r_S$ \eqref{grav5lambda}, while $r_{IV}$ is FR-compatible only in the case $\beta = 0$. The explicit form of FR-compatible $r$-matrices are (up to automorphisms that do not mix ${\cal J}_\mu$ with ${\cal L}_\mu$)
\begin{align}\label{eq:60.01b}
%r_{III}({\rm Re}\gamma,\eta;{\cal R}) &= {\rm Re}\gamma \left({\cal J}_1 \wedge {\cal J}_2 - {\cal R}^2 {\cal L}_1 \wedge {\cal L}_2\right) + {\cal R}\, \frac{\eta}{2}\, {\cal J}_0 \wedge {\cal L}_0\,, \nonumber\\
%r_{III}^a({\rm Re}\gamma,\eta;{\cal R}) &= {\cal R}\, {\rm Re}\gamma \left({\cal J}_0 \wedge {\cal L}_1 - {\cal J}_1 \wedge {\cal L}_0\right) + {\cal R}\, \frac{\eta}{2}\, {\cal J}_2 \wedge {\cal L}_2\,, \nonumber\\
r_{III}(\gamma - \bar\gamma,\gamma + \bar\gamma,\eta;{\cal R}) &= \frac{\gamma - \bar\gamma}{2} \left({\cal J}_1 \wedge {\cal J}_2 - {\cal R}^2 {\cal L}_1 \wedge {\cal L}_2\right) + {\cal R}\, \frac{\gamma + \bar\gamma}{2} \left({\cal J}_1 \wedge {\cal L}_2 - {\cal J}_2 \wedge {\cal L}_1\right) + {\cal R}\, \frac{\eta}{2}\, {\cal J}_0 \wedge {\cal L}_0\,, \nonumber\\
r_{III}^a(\gamma - \bar\gamma,\gamma + \bar\gamma,\eta;{\cal R}) &= {\cal R}\, \frac{\gamma - \bar\gamma}{2} \left({\cal J}_0 \wedge {\cal L}_2 - {\cal J}_2 \wedge {\cal L}_0\right) + \frac{\gamma + \bar\gamma}{2} \left({\cal J}_0 \wedge {\cal J}_2 - {\cal R}^2 {\cal L}_0 \wedge {\cal L}_2\right) + {\cal R}\, \frac{\eta}{2}\, {\cal J}_1 \wedge {\cal L}_1\,, \nonumber\\
r_{IV}(\gamma,\varsigma;{\cal R}) &= \gamma \left({\cal J}_1 \wedge {\cal J}_2 - {\cal R}\, {\cal J}_0 \wedge {\cal L}_0 - {\cal R}^2 {\cal L}_1 \wedge {\cal L}_2\right) + \frac{\varsigma}{2} \left({\cal J}_1 - {\cal R}\, {\cal L}_2\right) \wedge \left({\cal J}_2 + {\cal R}\, {\cal L}_1\right)\,, \nonumber\\
r_{IV}^a(\gamma,\varsigma;{\cal R}) &= {\cal R}\, \gamma \left({\cal J}_0 \wedge {\cal L}_1 - {\cal J}_1 \wedge {\cal L}_0 - {\cal J}_2 \wedge {\cal L}_2\right) + {\cal R}\, \frac{\varsigma}{2} \left({\cal J}_0 - {\cal J}_1\right) \wedge \left({\cal L}_0 - {\cal L}_1\right),
\end{align}
corresponding to $\alpha = \pm\sqrt{2}^{-1} {\cal R}\, (\gamma - \bar\gamma)$, $\beta = \mp\sqrt{2}^{-1} {\cal R}^2 (\gamma + \bar\gamma)$ in the first two lines and $\alpha = \pm\sqrt{2} {\cal R}\, \gamma$ in the latter two. In particular, $r_{III}^a$ describes the (twisted) spacelike $\kappa$-de Sitter deformation if $\beta = 0$ and $r_{III}$ describes the (twisted) timelike $\kappa$-de Sitter deformation if $\alpha = 0$ (see \cite{Ballesteros:2014ts} for a discussion of these deformations). As can also be seen from the previous Sections, $r_{III}$ does not have FR-compatible quantum IW contraction limits.
\newline

Similarly, the relevant $r$-matrices in the case of negative cosmological constant $\Lambda = -{\cal R}^{-2} < 0$ are $\dot{\mathfrak{o}}(2,2)$, $\mathfrak{o}'(2,2)$ and $\dot{\mathfrak{o}}'(2,2)$ $r$-matrices from Subsec.~\ref{sec:4.3}, with parameters set to be real. In particular, $\dot{\mathfrak{o}}(2,2)$ $r$-matrices satisfy the Yang-Baxter equations (again, the results are the same for $r_I$ and $r_I^a$, $r_{III}$ and $r_{III}^a$, etc.)
\begin{align}\label{eq:60.01c}
[[r_I,r_I]] = [[r_{II},r_{II}]] &= 0\,, \nonumber\\
[[r_{III},r_{III}]] &= -{\cal R}^2 \left(\gamma^2 + \bar\gamma^2\right) \left(\Lambda {\cal J}_0 \wedge {\cal J}_1 \wedge {\cal J}_2 + \tfrac{1}{2} \epsilon^{\mu\nu\sigma} {\cal J}_\mu \wedge {\cal L}_\nu \wedge {\cal L}_\sigma\right) \nonumber\\
&+ {\cal R}^3 \left(\gamma^2 - \bar\gamma^2\right) \left(\tfrac{1}{2} \Lambda \epsilon^{\mu\nu\sigma} {\cal J}_\mu \wedge {\cal J}_\nu \wedge {\cal L}_\sigma + {\cal L}_0 \wedge {\cal L}_1 \wedge {\cal L}_2\right)\,, \nonumber\\
[[r_{IV},r_{IV}]] &= -2{\cal R}^2 \gamma^2 \left(\Lambda {\cal J}_0 \wedge {\cal J}_1 \wedge {\cal J}_2 + \tfrac{1}{2} \epsilon^{\mu\nu\sigma} {\cal J}_\mu \wedge {\cal L}_\nu \wedge {\cal L}_\sigma\right)\,, \nonumber\\
[[r_V,r_V]] &= -{\cal R}^2 \gamma^2 \left(\Lambda {\cal J}_0 \wedge {\cal J}_1 \wedge {\cal J}_2 + \tfrac{1}{2} \epsilon^{\mu\nu\sigma} {\cal J}_\mu \wedge {\cal L}_\nu \wedge {\cal L}_\sigma\right. \nonumber\\
&\left.+ \tfrac{1}{2} {\cal R}^{-1} \epsilon^{\mu\nu\sigma} {\cal J}_\mu \wedge {\cal J}_\nu \wedge {\cal L}_\sigma - {\cal R}\, {\cal L}_0 \wedge {\cal L}_1 \wedge {\cal L}_2\right)\,.
\end{align}
We again observe that $r_I$ and $r_{II}$ are not FR-compatible, while $r_{IV}$ is FR-compatible only if $\beta = 0$. The complete list of FR-compatible $r$-matrices is (up to automorphisms that do not mix ${\cal J}_\mu$ with ${\cal L}_\mu$)
\begin{align}\label{eq:60.01d}
%r_{III}(\gamma = -\bar\gamma,\eta;{\cal R}) &= \gamma \left({\cal J}_0 \wedge {\cal J}_1 + {\cal R}^2 {\cal L}_0 \wedge {\cal L}_1\right) + {\cal R}\, \frac{\eta}{2}\, {\cal J}_2 \wedge {\cal L}_2\,, \nonumber\\
%r_{III}(\gamma = \bar\gamma,\eta;{\cal R}) &= {\cal R}\, \gamma \left({\cal J}_0 \wedge {\cal L}_1 - {\cal J}_1 \wedge {\cal L}_0\right) + {\cal R}\, \frac{\eta}{2}\, {\cal J}_2 \wedge {\cal L}_2\,, \nonumber\\
r_{III}(\gamma,\bar\gamma,\eta;{\cal R}) &= \frac{\gamma - \bar\gamma}{2} \left({\cal J}_0 \wedge {\cal J}_2 + {\cal R}^2 {\cal L}_0 \wedge {\cal L}_2\right) + {\cal R}\, \frac{\gamma + \bar\gamma}{2} \left({\cal J}_0 \wedge {\cal L}_2 - {\cal J}_2 \wedge {\cal L}_0\right) + {\cal R}\, \frac{\eta}{2}\, {\cal J}_1 \wedge {\cal L}_1\,, \nonumber\\
r_{IV}(\gamma,\varsigma;{\cal R}) &= \gamma \left({\cal J}_0 \wedge {\cal J}_1 - {\cal R}\, {\cal J}_2 \wedge {\cal L}_2 + {\cal R}^2 {\cal L}_0 \wedge {\cal L}_1\right) - \frac{\varsigma}{2} \left({\cal J}_0 + {\cal R}\, {\cal L}_1\right) \wedge \left({\cal J}_1 + {\cal R}\, {\cal L}_0\right)\,, \nonumber\\
r_{IV}^a(\gamma,\varsigma;{\cal R}) &= {\cal R}\, \gamma \left({\cal J}_0 \wedge {\cal L}_1 - {\cal J}_1 \wedge {\cal L}_0 + {\cal J}_2 \wedge {\cal L}_2\right) - {\cal R}\, \frac{\varsigma}{2} \left({\cal J}_0 + {\cal J}_1\right) \wedge \left({\cal L}_0 + {\cal L}_1\right)\,, \nonumber\\
r_V(\gamma,\bar\chi,\rho;{\cal R}) &= \frac{\gamma}{2} \left({\cal J}_0 + {\cal R}\, {\cal L}_0\right) \wedge \left({\cal J}_2 + {\cal R}\, {\cal L}_2\right) + \left(\frac{\bar\chi + \rho}{4}\, {\cal J}_1 - {\cal R}\, \frac{\bar\chi - \rho}{4}\, {\cal L}_1\right) \wedge \left({\cal J}_0 - {\cal J}_2 - {\cal R}\, ({\cal L}_0 - {\cal L}_2)\right)\,,
\end{align}
where $r_{IV}$ and $r_{IV}^a$ correspond to $\alpha = \sqrt{2} {\cal R}\, \gamma$; $r_{III}$ corresponds to $\alpha = \pm\sqrt{2}^{-1} {\cal R}\, (\gamma + \bar\gamma)$, $\beta = \mp\sqrt{2}^{-1} {\cal R}^2 (\gamma - \bar\gamma)$ or $\alpha = \pm\sqrt{2}^{-1} {\cal R}\, (\gamma - \bar\gamma)$, $\beta = \mp\sqrt{2}^{-1} {\cal R}^2 (\gamma + \bar\gamma)$; and $r_V$ corresponds to $\alpha = \pm\sqrt{2}^{-1} {\cal R}\, \gamma$, $\beta = \mp\sqrt{2}^{-1} {\cal R}^2 \gamma$ (here both $\alpha$ and $\beta$ have to be nonzero). In particular, $r_{III}(\gamma = \bar\gamma)$ describes the (twisted) spacelike $\kappa$-anti-de Sitter deformation, which fits either $\beta = 0$ or $\alpha = 0$ (see \cite{Ballesteros:2014ts} for a discussion of this deformation). As we already mentioned, $r_V$ does not have FR-compatible quantum IW contraction limits.

The only $\mathfrak{o}'(2,2)$ $r$-matrix satisfies the equation
\begin{align}\label{eq:60.01e}
[[r_{III},r_{III}]] &= {\cal R}^2 \left(\gamma^2 + \bar\gamma^2\right) \left(\Lambda {\cal J}_0 \wedge {\cal J}_1 \wedge {\cal J}_2 + \tfrac{1}{2} \epsilon^{\mu\nu\sigma} {\cal J}_\mu \wedge {\cal L}_\nu \wedge {\cal L}_\sigma\right) \nonumber\\
&- {\cal R}^3 \left(\gamma^2 - \bar\gamma^2\right) \left(\tfrac{1}{2} \Lambda \epsilon^{\mu\nu\sigma} {\cal J}_\mu \wedge {\cal J}_\nu \wedge {\cal L}_\sigma + {\cal L}_0 \wedge {\cal L}_1 \wedge {\cal L}_2\right)\,,
\end{align}
while $\dot{\mathfrak{o}}'(2,2)$ $r$-matrices satisfy
\begin{align}\label{eq:60.01f}
[[r_{III},r_{III}]] &= {\cal R}^2 \left(\gamma^2 - \bar\gamma^2\right) \left(\Lambda {\cal J}_0 \wedge {\cal J}_1 \wedge {\cal J}_2 + \tfrac{1}{2} \epsilon^{\mu\nu\sigma} {\cal J}_\mu \wedge {\cal L}_\nu \wedge {\cal L}_\sigma\right) \nonumber\\
&- {\cal R}^3 \left(\gamma^2 + \bar\gamma^2\right) \left(\tfrac{1}{2} \Lambda \epsilon^{\mu\nu\sigma} {\cal J}_\mu \wedge {\cal J}_\nu \wedge {\cal L}_\sigma + {\cal R}\, {\cal L}_0 \wedge {\cal L}_1 \wedge {\cal L}_2\right)\,, \nonumber\\
[[r_V,r_V]] &= {\cal R}^2 \gamma^2 \left(\Lambda {\cal J}_0 \wedge {\cal J}_1 \wedge {\cal J}_2 + \tfrac{1}{2} \epsilon^{\mu\nu\sigma} {\cal J}_\mu \wedge {\cal L}_\nu \wedge {\cal L}_\sigma\right. \nonumber\\
&\left.+ \tfrac{1}{2} {\cal R}^{-1} \epsilon^{\mu\nu\sigma} {\cal J}_\mu \wedge {\cal J}_\nu \wedge {\cal L}_\sigma - {\cal R}\, {\cal L}_0 \wedge {\cal L}_1 \wedge {\cal L}_2\right)
\end{align}
which means that $r_{III}$ in (\ref{eq:60.01e}) and $r_V$ in (\ref{eq:60.01f}) could be FR-compatible only if both $\alpha$ and $\beta$ are imaginary. We leave it to future investigations to check whether such a choice of parameters is of use in 3D quantum gravity models. %However, such a choice of these parameters would contradict the assumption that the Chern-Simons action \eqref{grav.1} is real and therefore is excluded.
Meanwhile, $r_{III}(\gamma = 0)$ in (\ref{eq:60.01f}), i.e.
\begin{align}\label{eq:60.01g}
r_{III}(\bar\gamma,\eta;{\cal R}) &= \frac{\bar\gamma}{2} \left({\cal J}_0 - {\cal R}\, {\cal L}_0\right) \wedge \left({\cal J}_2 - {\cal R}\, {\cal L}_2\right) + \frac{\eta}{4} \left({\cal J}_0 + {\cal R}\, {\cal L}_0\right) \wedge \left({\cal J}_1 - {\cal R}\, {\cal L}_1\right)
\end{align}
is FR-compatible with $\alpha = \pm\sqrt{2}^{-1} {\cal R}\, \bar\gamma$, $\beta = \mp\sqrt{2}^{-1} {\cal R}^2 \bar\gamma$ (i.e. both $\alpha$ and $\beta$ have to be nonzero) and hence can not have FR-compatible quantum IW contraction limits.

It turns out that the terms of Poincar\'{e} $r$-matrices $\hat r_2$, $\hat r_3$ and $\hat r_6$ proportional to $\hat\eta$ and $\hat\varsigma$, as well as terms of (anti-)de Sitter $r$-matrices (\ref{eq:60.01b}), (\ref{eq:60.01d}) and (\ref{eq:60.01g}) proportional to $\eta$ and $\varsigma$, are not involved in generating inhomogeneity -- necessary for the FR-compatibility -- in the Yang-Baxter equations. Therefore, the values of $\hat\eta$, $\hat\varsigma$, $\eta$ and $\varsigma$ can be arbitrary. Our list of FR-compatible (anti-)de Sitter $r$-matrices agrees with the most complete previous classification, which was given in \cite{Osei:2017ybk}; however, it would require more work to verify that the latter does not contain any additional results with respect to ours (it is clear that at least some of their $r$-matrices differ from ours just by automorphisms). Meanwhile, as we showed in Sec.~\ref{sec:5.0}-\ref{sec:6.0}, the FR-compatible $D = 3$ Poincar\'{e} $r$-matrices that can be obtained via quantum IW contractions form a subset within the Stachura classification, which should be contained within the classification of \cite{Osei:2017ybk}.

\subsection{Drinfeld double $r$-matrices} \label{sec:6.4}

Let us now discuss a special class of the classical $r$-matrices, called the Drinfeld double $r$-matrices, which all satisfy the Fock-Rosly conditions \cite{Ballesteros:2013dy,Ballesteros:2018te}. The Drinfeld double algebra is a $2d$-dimensional extension of a $d$-dimensional Lie algebra, with a basis $(Y_1,\ldots,Y_d;y^1,\ldots,y^d)$ such that the brackets
\begin{align}\label{dd1}
[Y_i, Y_j] = c^k_{ij}\, Y_k\,,\quad [y^i, y^j] = f_k^{ij}\, y^k\,,\quad [y^i, Y_j] = c^i_{jk}\, y^k - f_j^{ik}\, Y_k\,,
\end{align}
where $c^k_{ij}$ and $f_k^{ij}$ are structure constants. We note that to a given Lie algebra one can in principle attach many inequivalent Drinfeld double structures. For example, for all 3-dimensional real Lie algebras (i.e. 11 Bianchi algebras) there exist 22 inequivalent Drinfeld doubles \cite{Snobl:2002kq}.

The structure of a Drinfeld double allows to define the non-degenerate, symmetric, Ad-invariant bilinear form
\begin{align}\label{dd2}
(Y_i,Y_j) = 0\,, \quad (y^i,y^j) = 0\,, \quad (y^i,Y_j) = \delta^i_j\,,
\end{align}
hence the Lie algebra $\mathfrak{g}^*$ generated by $\{y^i\}$ becomes the dual of $\mathfrak{g}$ generated by $\{Y_i\}$. The unique canonical $r$-matrix associated with a Drinfeld double has the form
\begin{align}\label{dd3}
r = \sum_i y^i \otimes Y_i = \frac{1}{2} \sum y^i \wedge Y_i + \Omega\,, \qquad \Omega = \frac{1}{2} \left(y^i \otimes Y_i + Y_i \otimes y^i \right)
\end{align}
where $\Omega$ is the quadratic split Casimir of the universal enveloping algebra of (\ref{dd1}). It can be shown by an explicit calculation that (\ref{dd3}) satisfies the Fock-Rosly conditions.

The $r$-matrices associated with all possible Drinfeld double structures on the $D = 3$ Poincar\'{e} algebra were recently classified in \cite{Ballesteros:2018te,Gutierrez-Sagredo:2018zog}, where it was found that there are eight inequivalent $D = 3$ Poincar\'{e} Drinfeld doubles. In terms of the Stachura classification from Subsec.~\ref{sec:5.2}, four of them lead to the $r$-matrices of the type $r_6$ with $\varrho = \pm 1$ (and appropriate $\theta^{\mu\nu}$), two have the $r$-matrix of the type $r_2$ with $\varrho = \beta = 1$ (and appropriate $\theta,\theta'$), and the remaining two Drinfeld double $r$-matrices are of the types $r_1$ with $\beta = 1$ and $r_7$, respectively.

In the set of $r$-matrices that we derived via quantum IW contractions, $r_1$ (i.e. Case 1 in \cite{Ballesteros:2018te}) is replaced by $\hat r_1$, which is not associated with a Drinfeld double (and also not FR-compatible, cf. (\ref{eq:60.02})). %Further, if we take $\hat r_2$ with $\hat\eta \neq \hat\gamma$ (or even $\hat\eta = 0$), as well as $\hat r_6$ with $\hat\varsigma \neq \pm 2\hat\gamma$, it violates the Drinfeld double structure but not the FR-compatibility.
$r_2$ and $r_6$ are respectively replaced by $\hat r_2 + \tilde\gamma\, {\cal P}_0 \wedge {\cal P}_2$ and $\hat r_6$, corresponding to $\theta' = 0$ in the former case and $\theta^{\mu\nu} = 0$ in the latter. Therefore, Cases 3, 4 and 5 from \cite{Ballesteros:2018te} also can not be obtained via quantum IW contractions. %The most peculiar case is $\hat r_3$, which can not be associated with a Drinfeld double for any set of values of its parameters but in principle \cite{Rosati:2017dt} may be FR-compatible (i.e. satisfy the condition (\ref{grav6})), although in the literature there is no agreement about the latter claim \cite{Osei:2017ybk}.

The situation is simpler for the relation between the known list \cite{Ballesteros:2013dy,Ballesteros:2018te} of Drinfeld double $r$-matrices in the case of $D = 3$ (anti-)de Sitter algebra (i.e. $\Lambda \neq 0$) and the list of all FR-compatible $r$-matrices associated with these algebras, provided by us in (\ref{eq:60.01b}) and (\ref{eq:60.01d},\ref{eq:60.01g}), respectively. If we use the notation of \cite{Ballesteros:2018te}, de Sitter $r$-matrices $r'_A$ and $r'_B$ are equivalent (under certain automorphisms that do not mix ${\cal J}_\mu$ with ${\cal L}_\mu$) to our $r_{IV}$ with $\varsigma = \pm 2\gamma = 1$, while $r'_C$ and $r'_D$ are equivalent to $r^a_{III}$ with $\eta = 2{\rm Re}\gamma = 1$, ${\rm Im}\gamma = 0$ in the former case and ${\rm Re}\gamma = \frac{\mu^2 - 1}{2\mu}$, ${\rm Im}\gamma = 1$, $\eta = \frac{\mu^2 + 1}{\mu}$, $\mu > 0$ in the latter case. We see that $r'_A$, $r'_B$ and $r'_C$ are FR-compatible only for $\beta = 0$ and $r'_D$ only for $\beta \neq 0$.

Similarly, (in the notation of \cite{Ballesteros:2018te}) the anti-de Sitter $r$-matrix $r'_E$ is equivalent (under certain automorphisms that do not mix ${\cal J}_i$ with ${\cal L}_i$) to our $r_{IV}$ with $\varsigma = -2\gamma = -1$, while $r'_F$ and $r'_G$ are equivalent to $r_{III}$ with $\eta = 2\gamma = 2\bar\gamma = 1$ in the former case and $\gamma = 1/2$, $\bar\gamma = \rho^2/2$, $\eta = \rho/2$, $-1 < \rho < 1$ in the latter case. $r'_E$ is FR-compatible only for $\beta = 0$ and $r'_F$ only for $\beta = 0$ or $\alpha = 0$ (the latter possibility was not noticed in \cite{Ballesteros:2013dy}). %$r'_G(\rho = 1)$ is equivalent to $r_{III}(\gamma = \bar\gamma)$ with $\pm\eta = \gamma = 1/2$.

\section{Conclusions and outlook} \label{sec:7.0}

%Our reference point in this paper are the (undeformed) local isometry algebras of 3D gravity with non-vanishing cosmological constant $\Lambda$. In general, we have four cases, corresponding to the Euclidean or Lorentzian metric, with positive or negative $\Lambda$. A familiar form of the brackets of each algebra can be obtained from (\ref{eq:20.01}) by an appropriate transformation to the basis that we call the physical basis. There are several such possible transformations (mainly due to the freedom of choice of the contraction axis) and therefore the choice of one of them is a convention. This choice is actually restricted by imposing the set of reality conditions on the algebra generators, which introduces the correspondence between a given local isometry algebra and one of the real forms of $\mathfrak{o}(4;\mathbbm{C})$. However, we have two subtleties here. Firstly, the Euclidean isometry algebra with $\Lambda < 0$ corresponds to the same real form as the Lorentzian isometry algebra with $\Lambda > 0$, although the brackets in the physical basis are different. The second special case is the Lorentzian isometry algebra with $\Lambda < 0$, which corresponds to three possible real forms but each of them is associated with a different kind of transformation to the physical basis.

The main aim of this paper was to check whether all classical $r$-matrices (up to an automorphism) for $D = 3$ inhomogeneous Euclidean and $D = 3$ Poincar\'{e} algebra, derived by Stachura as solutions of the homogeneous or modified Yang-Baxter equation \cite{Stachura:1998ps}, can be obtained as well via quantum IW contractions of classical $r$-matrices for the real (pseudo-)orthogonal algebras $\mathfrak{o}(4-k,k)$, $k = 0,1,2$. It turns out that this is almost true -- the caveats in the Poincar\'{e} case (cf. Subsec.~\ref{sec:5.2}) are that we recover $r_1$ only with $\beta = 0$ (i.e. without the term proportional to $r_7$), $r_2$ with $\theta' = 0$ and $r_6$ with $\theta^{\mu\nu} = 0$. In Sec.~\ref{sec:5.0}, we explicitly write down the $\mathfrak{o}(3) \vartriangleright\!\!< {\cal T}^3$ and $\mathfrak{o}(2,1) \vartriangleright\!\!< {\cal T}^{2,1}$ automorphisms that are necessary in order to connect our formulae with the ones provided by Stachura. Let us stress that the success of the current work stems from the recently obtained complete classification of $r$-matrices for $\mathfrak{o}(4-k,k)$ ($k = 0,1,2$), in the unified setting of the $\mathfrak{o}(4;\mathbbm{C})$ algebra \cite{Borowiec:2017bs}. We also recall that the IW contraction parameter ${\cal R}$ in the context of classical and quantum 3D gravity (with the Lorentzian or Euclidean metric signature) directly corresponds to the cosmological constant. As we discussed in detail in the previous Section, our classification of $r$-matrices for $D = 3$ Poincar\'{e} and (A)dS algebras %(as well as for the $D = 3$ inhomogeneous Euclidean and Euclidean (A)dS algebras) 
contains a confirmation of earlier results \cite{Osei:2017ybk} for $r$-matrices compatible with 3D gravity.

There are the following problems that are worth to be explored in future investigations.
\begin{itemize}
\item The most obvious next step is to describe the quantizations of symmetries that are determined by all $D = 3$ classical $r$-matrices (\ref{eq:60.01}), i.e. to find the corresponding Hopf algebras. Since all $\mathfrak{o}(4-k,k)$ $r$-matrices have already been quantized in the Cartan-Weyl basis in \cite{Borowiec:2017bs}, the deformed $D = 3$ inhomogeneous Euclidean and $D = 3$ Poincar\'{e} Hopf algebras can be obtained simply by introducing physical bases of the algebras and performing suitable quantum IW contractions of the Hopf-algebraic formulae presented in that paper.
\item It is still not completely clear what is the actual role of different $r$-matrices in 3D gravity: whether they lead to different physical theories or, effectively, reduce to the only one model when expressed in terms of suitably chosen gauge invariant observables. In particular, in the approach studied in \cite{Cianfrani:2016ss}, the algebra of symmetries of quantum 3D spacetime was derived with the help of the Loop Quantum Gravity quantization scheme. This algebra is introduced as a commutator algebra of the appropriately smeared quantum constraint operators, obtained from the classical Poisson algebra of Poincar\'{e} symmetries of flat spacetime. It turns out that such an algebra leads to a Hopf algebra, whose non-trivial coproducts are derived from a quantum $R$-matrix. The quantum $R$-matrix has been calculated within the particular regularization scheme, employed in the process of loop quantization, and it is the one corresponding to the classical $r$-matrix \eqref{eq:31.04} for $\bar\gamma = -\gamma$, $\eta = 0$ (describing the $D = 3$ Euclidean $\kappa$-de Sitter deformation). In this framework the presence of different $r$-matrices may lead to different Loop Quantum Gravity quantizations, associated with different regularization schemes for the non-commutative connections.
\item Let us also mention that in \cite{Kowalski:2014dy,Trzesniewski:2018ey} a different type of contraction of the $D = 3$ de Sitter symmetries has been considered, using a local decomposition of the gauge group ${\rm O}(3,1) \cong {\rm SL}(2;\mathbbm{C}) = {\rm SL}(2;\mathbbm{R}) \vartriangleright\!\!\vartriangleleft {\rm AN}(2)$ and making the ${\rm SL}(2;\mathbbm{R})$ generators Abelian via the IW contraction procedure. It could be interesting to apply such a type of contraction to the Hopf-algebraic deformations of symmetries, characterized by classical $r$-matrices.
\item On the other hand, in order to study quantum symmetries of 4D quantum gravity models, one needs the classical $r$-matrices for $\mathfrak{o}(5-k,k)$ ($k = 0,1,2$) algebras as well as their quantum IW contractions to the $\mathfrak{o}(4) \vartriangleright\!\!< {\cal T}^4$ and $\mathfrak{o}(3,1) \vartriangleright\!\!< {\cal T}^{3,1}$ $r$-matrices. Unfortunately, the complete classification of $r$-matrices describing deformed $D = 5$ rotations is still unknown (for some effort in this direction see \cite{Lucas:2017as}). As we already mentioned, for the $D = 4$ Poincar\'{e} algebra the most extensive list has been given a long time ago in \cite{Zakrzewski:1994pp} (see also \cite{Tolstoy:2008ts}). Moreover, we refer the reader to \cite{Ballesteros:2015tt} for an example of how quantum symmetries relevant in 3D gravity could generalize to 4D. %We also recall that the first derivation of the $\kappa$-Poincar\'{e} algebra \cite{Lukierski:1991qa,Lukierski:1992ny} has been done via the quantum IW contraction of the $q$-deformed $\mathfrak{o}(3,2)$ algebra.
\item Another potentially interesting direction of research is to consider the $D = 3$ and $D = 4$ supersymmetric classical $r$-matrices, which determine quantum deformations of $D = 3$ and $D = 4$ (A)dS supersymmetries, and to study their quantum IW contractions. The best known superextension of the $\mathfrak{o}(5-k,k)$ algebra is provided for $k = 2$, as the $D = 4$ AdS superalgebra $\mathfrak{osp}(N;4)$ ($\mathfrak{sp}(4) \cong \mathfrak{o}(3,2)$); for $k = 0,1$ one should consider the quaternionic superalgebras: $\mathfrak{uu}_\alpha(2;N|\mathbbm{H})$ for $k = 0$ ($\mathfrak{u}(2|\mathbbm{H}) \equiv \mathfrak{usp}(4) \cong \mathfrak{o}(5)$) and $\mathfrak{uu}_\alpha(1,1;N|\mathbbm{H})$ for $k = 1$ ($\mathfrak{u}(1,1|\mathbbm{H}) \equiv \mathfrak{usp}(2,2) \cong \mathfrak{o}(4,1)$) -- see \cite{Lukierski:1986qs}. So far, only particular cases of quantum IW contractions of the above mentioned superalgebras has been studied (see e.g. \cite{Borowiec:2005js}, where the contraction of $\mathfrak{osp}(1;4)$ $r$-matrix is shown to lead to the $D = 3$ lightcone $\kappa$-Poincar\'{e} algebra; see also \cite{Borowiec:2016rs}).
\end{itemize}

Concluding, we hope that the broader knowledge of quantum-deformed $D = 3$ and $D = 4$ (super)symmetries will be helpful in the construction of consistent and physically reliable quantum 3D and 4D (super)gravity models.

\section*{Acknowledgements}

We thank A. Borowiec and G. Rosati for some valuable comments. The work of JKG and JL was partially supported by the Polish National Science Centre projects no. 2017/27/B/ST2/01902, while TT was supported by the Polish National Science Centre project no. DEC-2017/26/E/ST2/00763. JKG would also like to acknowledge the contribution of the COST Action CA18108.

\end{document}